\documentclass[]{jfm}
\usepackage{amsmath,amssymb,amsbsy,graphicx,subfigure,epstopdf}
\usepackage[a4paper, left=2cm, right=2cm, top=2cm, bottom=2cm]{geometry}
\usepackage{natbib}

\pdfoutput=1

\newcommand{\diff}[3][]{\dfrac{\mathrm{d}^{#1}#2}{\mathrm{d}{#3}^{#1}}}
\title[Capillary Breakup]
{Capillary Breakup of a Liquid Bridge:\\ Identifying Regimes and Transitions}

\author[Y. Li and J.E. Sprittles]
{Y\ls U\ls A\ls N \ns L\ls I\footnote{E-mail: y.li.2@bham.ac.uk} \and J\ls A\ls M\ls E\ls S\ns E.\ns S\ls P\ls R\ls I\ls
T\ls T\ls L\ls E\ls S\footnote{E-mail: J.E.Sprittles@warwick.ac.uk} }

\affiliation{School of Mathematics, University of Birmingham, Birmingham, B15 2TT, UK,\newline 
Mathematics Institute, University of Warwick, Coventry, CV4 7AL, UK.}

\begin{document}
\label{firstpage} \maketitle

\begin{abstract}

Computations of the breakup of a liquid bridge are used to establish the limits of applicability of similarity solutions derived for different breakup regimes.  These regimes are based on particular viscous-inertial balances, that is different limits of the Ohnesorge number $Oh$. To accurately establish the transitions between regimes, the minimum bridge radius is resolved through four orders of magnitude using a purpose-built multiscale finite element method.  This allows us to construct a quantitative phase diagram for the breakup phenomenon which includes the appearance of a recently discovered low-$Oh$ viscous regime.  The method used to quantify the accuracy of the similarity solutions allows us to identify a number of previously unobserved features of the breakup, most notably an oscillatory convergence towards the viscous-inertial similarity solution.  Finally, we discuss how the new findings open up a number of challenges for both theoretical and experimental analysis.

\end{abstract}

\section{Introduction} \label{intro}

The breakup of liquid volumes is a process that is ubiquitous throughout industry and nature.  Recently, this process has attracted significant attention due to its importance for the functioning of a range of microfluidic technologies where one would like to be able to control the generation of uniform sized droplets which then become building blocks, as in 3D printers \citep{derby10}, or modes of transport for reagents, as in lab-on-a-chip devices \citep{stone04}.  In order to optimise the functioning of these technologies it is  key to be able to understand the physical mechanisms controlling the breakup process and, ideally, to have a tool capable of predicting the response of the system to alterations in operating conditions.

From a theoretical perspective, the breakup phenomenon has attracted interest due to both its technological importance as well as its status as a free-surface flow which exhibits finite time singularities \citep{eggers97}.  In the former case, motivated by a desire to describe engineering-scale systems, the focus has been on capturing the global dynamics of the process using multiphysics computational fluid dynamics (CFD) codes, with the actual breakup usually under-resolved due to the inherently multiscale nature of the problem.  In contrast, in the latter case most interest has revolved around using asymptotic methods to study the micromechanics of the breakup process, with the larger scale flow often not considered. Potentially, the advances made in understanding the micromechanics could be fed into large-scale CFD codes in order to exploit the advantages of both approaches, but this is yet to be achieved.  

The failure of CFD packages to reliably capture breakup phenomena has been highlighted in \cite{fawehinmi05} for drop formation, where the appearance of satellite drops depended on grid resolution, and \cite{hysing09} for a 2D-rising bubble in a regime where breakup was anticipated.  In \cite{hysing09}, six different codes were tested for the same problem and all gave different results ranging from no observed breakup through to multiple bubble detachments.  In other words, the codes were unreliable for this process.  The conclusion of the authors provides motivation for the current work: ``Although the obtained benchmark quantities were in the same ranges, they did not agree on the point of break up or even what the bubble should look like afterwards, rendering these results rather inconclusive. To establish reference benchmark solutions including break up and coalescence will clearly require much more intensive efforts by the research community.''

Computational approaches cannot, in principle, resolve the thinning of a liquid volume down to the point at which breakup occurs, when the thread is infinitesimally small. Instead, at some point the pre-breakup process must be terminated, by `cutting' the liquid thread joining two volumes, and a post-breakup state must be initiated based on the final solution in the pre-breakup phase. Whilst some computational codes do this process `automatically', notably those using Eulerian meshes such as Volume-of-Fluid, there is no guarantee that this approach accurately represents the physical reality.  Theoretical approaches to this problem have been considered in different regimes and are described in \cite{eggers14}.  To implement an appropriate post-breakup solution, one must identify which pre-breakup regime the process is in, based on the scale for the cut-off.  Therefore, knowing what regime a given breakup process is in, and determining the transitions between these regimes, is critical for the topological change to be correctly handled. The complexity of this procedure has been revealed in \cite{castrejon15}, see \S\ref{S:multiple} for details, where multiple regimes were discovered for each breakup event.

Here, by computing the pre-breakup solution to the smallest possible scales that our simulation will allow, we will measure the accuracy of various similarity solutions proposed in the literature and, importantly, establish their limits of applicability.  To do so, rather than just identifying power-law behaviour, methods will be developed to determine the regions of phase space where similarity solutions proposed for the different `regimes' accurately approximate the full solution.  The procedure is intended to leave no ambiguity as to when a certain regime is encountered.  This attempt at a more rigorous approach will lead us to discover previously unobserved features of the breakup process that open-up several new avenues of enquiry.

\section{Regimes and Transitions in the Breakup Phenomenon}

\begin{figure}
     \centering
\includegraphics[scale=0.6,viewport=-100 500 800 750]{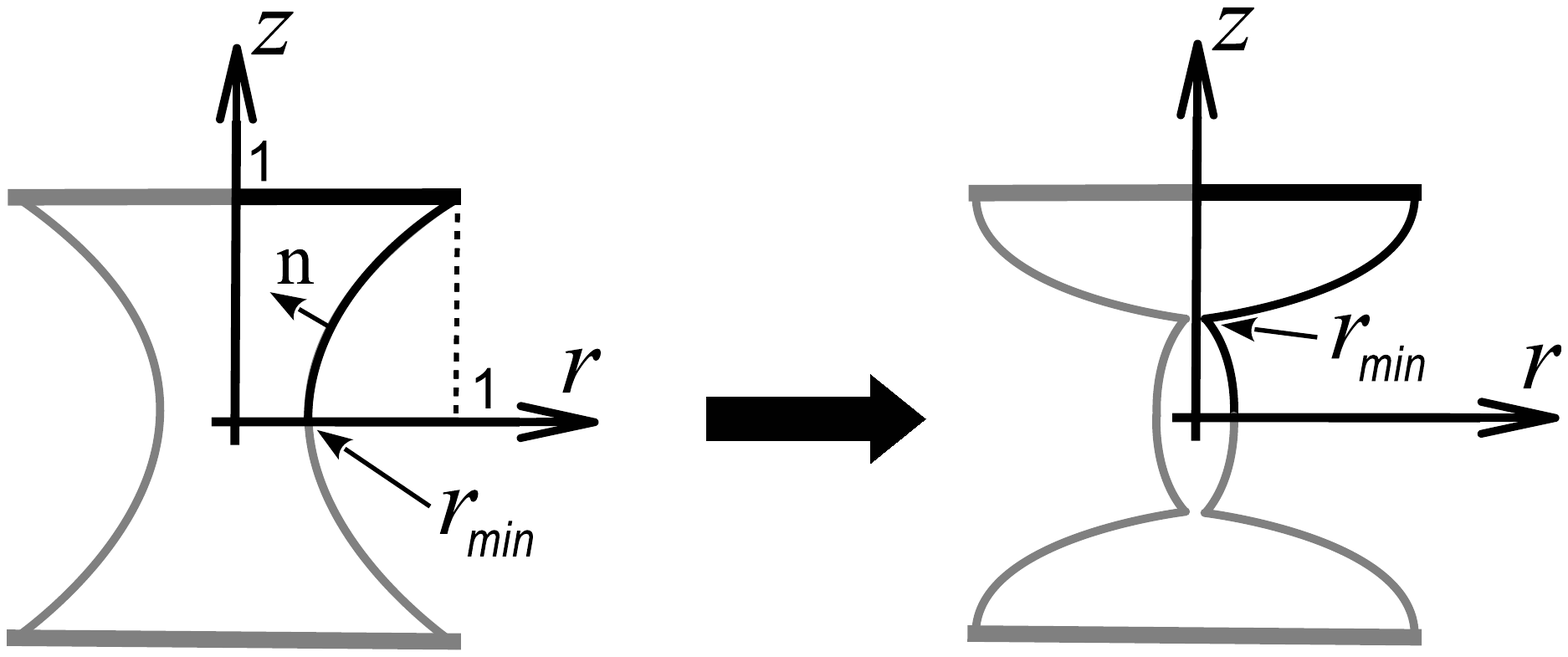}
\caption{Illustration of breakup in the liquid bridge geometry using the (dimensionless) coordinate system $(r,z)$, with the black lines showing the computational domain. The minimum radius of the thread is $r_{min}$, which can change vertical position as the breakup proceeds if satellite drops are formed (right hand side), and the inward normal to the free-surface is $\mathbf{n}$.}
\label{F:sketch}
\end{figure}

This article focuses on the breakup of incompressible Newtonian liquids with constant viscosity $\mu$ and density $\rho$, which have a constant surface tension $\sigma$ at their liquid-gas interface and are surrounded by a dynamically passive gas.  The breakup is considered axisymmetric with the axis of the thread along the $z$-axis of a cylindrical coordinate system ($r,z$), as shown in Figure~\ref{F:sketch} for a liquid bridge geometry, with characteristic length scale $R$.  
In general, the flow is governed by the Navier-Stokes equations, but when inertial forces are weak the dominant balance of viscous and capillary forces gives a characteristic speed of breakup of $U=\sigma/\mu$ so that the capillary number $Ca=\mu U/\sigma=1$. Then the appropriate dimensionless number characterising breakup is the Ohnesorge number $Oh$ which is obtained by substituting $U$ into the Reynolds number $Re=\rho U R/\sigma = \rho \sigma R /\mu^2 = Oh^{-2}$.  The Ohnesorge number is the square-root of the ratio of a viscous length scale $\ell_{\mu}=\mu^2/(\rho\sigma)$ to the characteristic scale of the system $R$. Then $Oh^{-1}=0$ is Stokes flow and $Oh=0$ gives inviscid flow. 

The scales used so far are based only on global quantities. To analyse the breakup process locally it is useful to introduce a time-dependent local Reynolds number $Re_{local}(t) =\rho U_z L_z/\mu$ based on instantaneous axial-scales $U_z(t)$ and $L_z(t)$ (which will be larger than, or comparable to, radial scales). The practicalities of calculating this quantity are discussed in \S\ref{S:analysis}.  This local parameter will indicate the dominant forces during a breakup event and hence should identify the flow regime.  

\subsection{Regimes of Breakup}\label{S:regimes}

Near breakup much research has focussed on finding similarity solutions for the different regimes, all of which are categorised in \S4 of \cite{eggers08}.   A summary of the key results is provided here with particular attention paid to the dependence of the minimum thread radius $r_{min}$, the maximum axial velocity $w_{max}$, and the local Reynolds number $Re_{local}$ on the time from breakup $\tau=t_b-t$, where $t_b$ is the breakup time.  Quantities have been made dimensionless with characteristic scales for lengths, velocities and time of $R$, $\sigma/\mu$ and $\mu R/\sigma$. 

\subsubsection{The Inertial Regime}

In the `inertial regime', henceforth the I-regime, where there is a balance between inertial and capillary forces, dimensional analysis \citep{keller83,brenner97} of inviscid flow ($Oh=0$) gives as $\tau\to0$ that
\begin{equation}\label{inertial}
r_{min}= A_I (Oh~\tau)^{2/3},  \qquad Re_{local} \sim \left(\tau/Oh^2\right)^{1/3}, \qquad w_{max}\sim (Oh^2/\tau)^{1/3},
\end{equation}
where $A_I$ is a constant of proportionality, given in \cite{eggers08} as $A_I\approx 0.7$.  Notably, computations performed in the inertial regime have never listed $A_I$ and so it will be of interest to determine this value so that (\ref{inertial}) is precisely determined and becomes a predictive tool.  It will be useful to note, in terms of $r_{min}$, that $Re_{local} \sim Oh^{-1}~r_{min}^{1/2}$ and $w_{max}\sim Oh~r_{min}^{-1/2}$

A feature of the inertial regime is the `overturning' of the free-surface, observed in both experiments \citep{chen02} and simulations in this regime \citep{schulkes94}, which invalidates attempts at a slender description of the thread.  In \cite{day98} it was predicted that at breakup the free-surface forms a double-cone shape with angles of $18.1^\circ$ and $112.8^\circ$ with the $z$-axis.  This phenomenon has been observed experimentally in \cite{castrejon12} and computationally in \cite{wilkes99}. 

\subsubsection{The Viscous Regime}
 
The solutions derived for both the Viscous and Viscous-Inertial regimes, henceforth referred to as the V-regime and the VI-regime, were derived in a one-dimensional approximation of the Navier-Stokes equations which relies on the thread remaining `slender' as the breakup is approached.  For a free-surface represented by $r=r(z,t)$ this means that the gradient of the free-surface must remain small $\frac{\partial r}{\partial z} \ll1$.
 
For the V-regime, the similarity solution for Stokes flow ($Oh^{-1}=0$) was derived in \cite{papageorgiou95a,papageorgiou95} and is given by
\begin{equation}\label{papa}
r_{min}=0.0709~\tau, \qquad Re_{local} \sim \tau^{2\beta -1}/Oh^2, ,\qquad w_{max}\sim \tau^{\beta-1},
\end{equation}
where $\beta\approx0.175$. Recently, \cite{eggers12} showed that this is the only similarity solution of infinitely many possible ones which remains stable to small perturbations.  The free surface profile associated with this similarity solution remains symmetric about the pinch point and the local profile depends on an external length scale (i.e. it is not `universal'), so that it is a self-similar problem of the second kind \citep{barenblatt96}.  The persistence of the V-regime for high-viscosity liquids has been observed experimentally in \cite{mckinley00}.

\subsubsection{The Viscous-Inertial Regime}

As $\tau\to0$, from (\ref{inertial}) and (\ref{papa}) we find in the I-regime that $Re_{local}\to0$ and in the V-regime that $Re_{local}\to\infty$.  This contradicts the initial assumptions behind their derivation \citep{lister98}.  Therefore, neither the I- or V-regimes are valid right up to breakup and a regime where viscous and inertial effects are in balance, so that $Re_{local}\sim 1$, must be considered. In this VI-regime, the Navier-Stokes equations are required. It was shown in \cite{eggers93} that for the VI-regime a `universal' set of exponents exist in which
\begin{equation}\label{eggers}
r_{min}=0.0304~\tau,\qquad Re_{local}\sim 1, \qquad w_{max}=1.8~Oh~\tau^{-1/2},
\end{equation}
with a highly asymmetric free surface shape joining a thin thread to a much steeper `drop-like' profile.  It was later demonstrated in \cite{brenner96} that this solution is the most favourable of a countably infinite set of similarity solutions. Notably $w_{max}=\max(w)$ is the maximum velocity out of the thin thread ($w>0$ in what follows), rather than $\max |w|$ which scales in the same way but has a pre-factor $3.1$ \citep{eggers93}. In terms of $r_{min}$ this gives $w_{max}= 0.3~Oh~r_{min}^{-1/2}$.

\subsection{Transitions Between Regimes}

It has been established that whichever regime the breakup process starts in, it will eventually end up in the VI-regime. However, it is possible that the transition from either the V- or I-regimes will occur at such small scales that this regime is irrelevant, at least from a practical perspective.  For example, experiments in \cite{burton04} for mm-sized mercury drops show that the similarity solution in the I-regime adequately describes the breakup down to the nanoscale . Therefore, determining the transitions between the regimes is an important part of understanding the breakup process. 

Phase diagrams which identify the regions of $(Oh,r_{min})$ space associated with different regimes have been constructed for the related problem of drop coalescence in \citep{paulsen12,paulsen13} and \citep{sprittles14_jfm2} \footnote{Intriguingly, these publications disagree on the number of regimes.}, whereas for the breakup phenomenon an equivalent diagram is currently lacking.  Most progress in this direction has been made in \cite{castrejon15}, where sketches of two characteristic phase space trajectories were presented (in their Figure 1E).  Here, we provide a systematic exploration of $(Oh,r_{min})$ space in order to construct a phase diagram for breakup which is equivalent to those obtained for coalescence. To do so, the scalings proposed for the transitions between the various regimes as $r_{min},\tau\to0$ are considered.

\subsubsection{Viscous to Viscous-Inertial Transition (V$\to$VI)}

When $Oh\gg1$, the V-regime adequately describes the initial stages of breakup.  However, the inertial term in the Navier-Stokes equations does not remain negligible as  $r_{min}\to0$ and so a transition to the VI-regime occurs at a bridge radius $r^{V\rightarrow VI}_{min}$.  In \cite{basaran02} and \cite{eggers05}, by balancing the solutions in the V- and VI-regimes, i.e.\ by taking $Re_{local}\sim1$ in (\ref{papa}), it was shown that this should occur when 
\begin{equation}\label{V-VI}
r^{V\to VI}_{min}\sim Oh^{2/(2\beta-1)}\sim Oh^{-3.1},
\end{equation}
However, experimental evidence in \cite{rothert03} appears to contradict this result, finding instead that $r^{V\rightarrow VI}_{min}$ is constant, i.e.\ independent of $Oh$.  Investigating these regimes computationally should provide new insight into the transition.  

\subsubsection{Inertial to Viscous-Inertial Transition (I$\to$VI)}

When $Oh\ll1$ the Euler equations accurately capture the initial stages of breakup until the local Reynolds number in (\ref{inertial}) drops to $Re_{local}\sim 1$, which occurs when $\tau\sim Oh^2$ so that 
\begin{equation}\label{I-VI}
r^{I\rightarrow VI}_{min}\sim Oh^{2}.
\end{equation}
This crossover was first confirmed computationally in \cite{notz01} and experimentally in \cite{chen02} for the case of a water-glycerol mixture with $Oh=0.16$, but has never been studied systematically across a range of $Oh$.

\subsubsection{The Discovery of Multiple Regime Transitions in \cite{castrejon15}}\label{S:multiple}

\begin{figure}
     \centering
\subfigure[Expected phase diagram]{\includegraphics[scale=0.26]{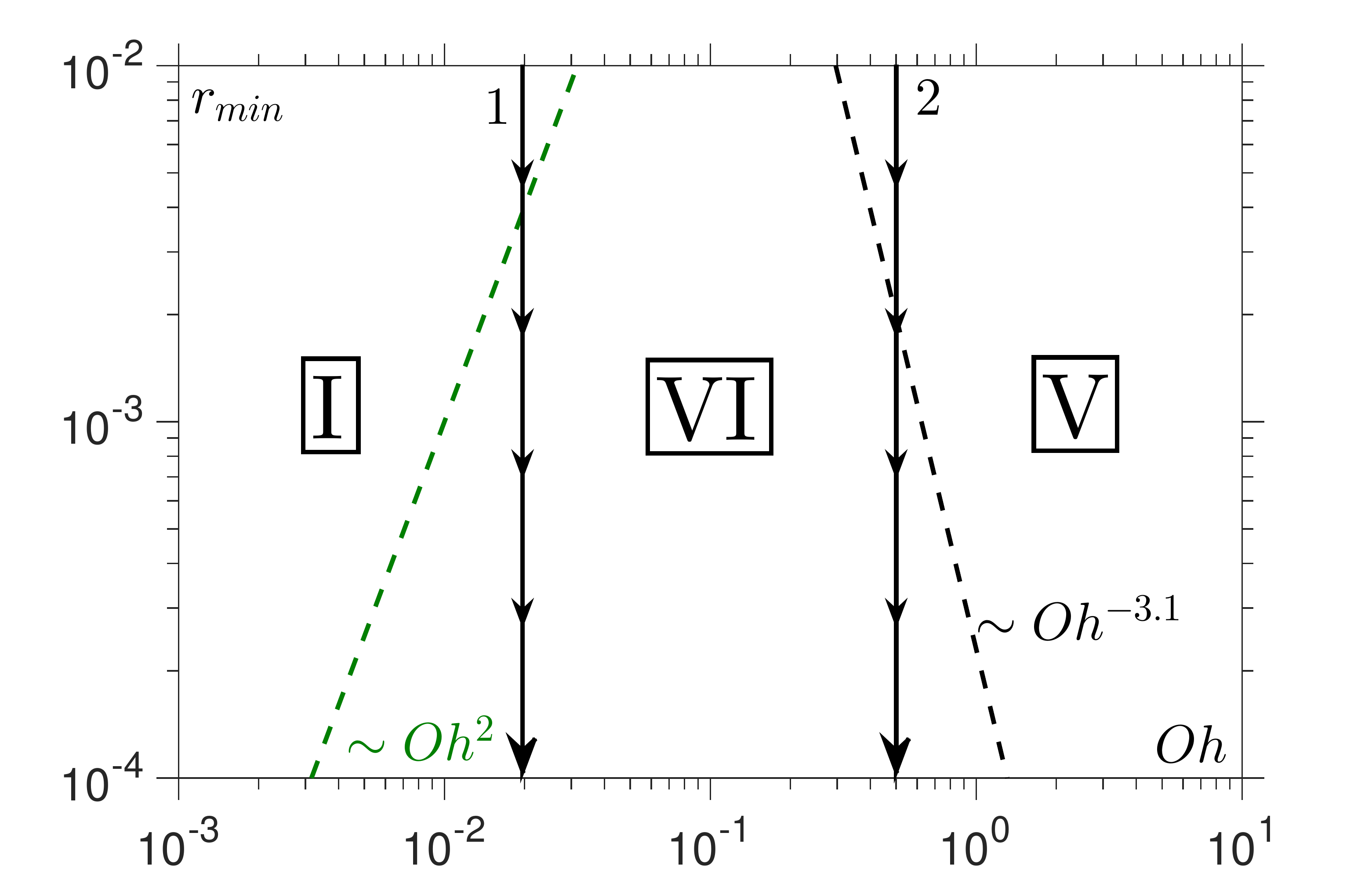}\label{F:expected}}
\subfigure[Computed phase diagram]{\includegraphics[scale=0.26]{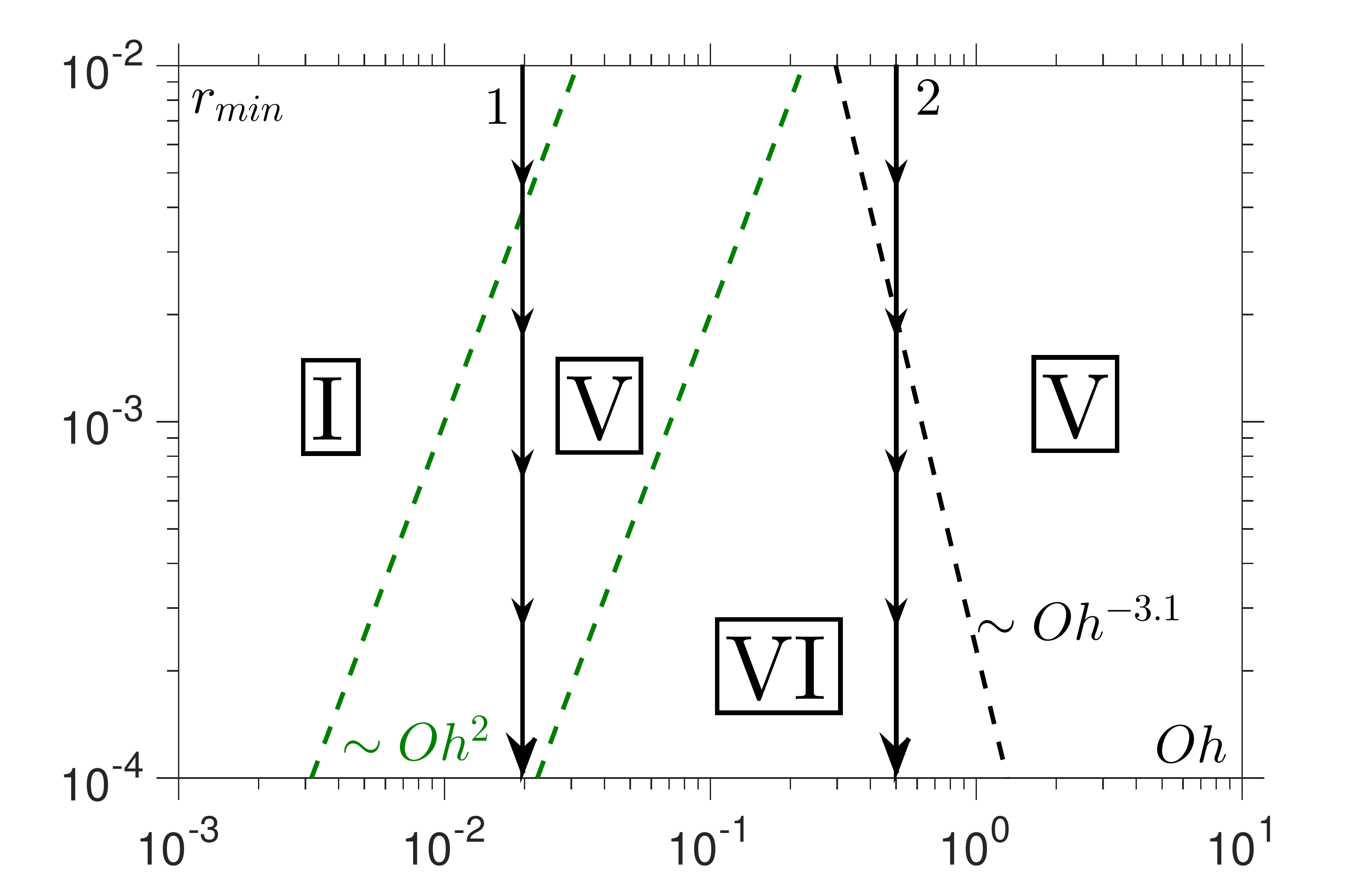}\label{F:actual}}
\caption{Phase diagrams of Ohnesorge number $Oh$ against minimum bridge radius $r_{min}$. The path of typical breakup events are given by arrowed lines 1 and 2, which show how different regimes are encountered as $r_{min}$ decreases.   At higher $Oh$ (path 2) a V$\to$VI transition is both expected and computed, whilst at lower $Oh$ (path 1), a single I$\to$VI transition was expected but a more complex behaviour was computed, with a low-$Oh$ V-regime encountered before the VI-regime, so that I$\to$V$\to$VI transitions are found, as discovered in \cite{castrejon15}.  The dashed lines show the scaling of the transitions between the different regimes as $r_{min}\to0$, with I$\to$V and V$\to$VI transitions scaling as $r_{min}\sim Oh^2$ and V$\to$VI occurring when $r_{min}\sim Oh^{-3.1}$.}
\end{figure}

From the predicted transitions of (\ref{V-VI}) and (\ref{I-VI}) the phase diagram may be expected to look qualitatively like Figure~\ref{F:expected}.  In this Figure, to see which regimes are encountered during breakup at a given $Oh$, one follows a vertical line (at the given $Oh$) from the top axis downwards (through decreasing $r_{min}$), e.g.\ see paths 1 and 2.  For small $Oh$ (path 1) one has an I-regime crossing into a VI-regime when $r_{min}\sim Oh^2$ whilst for large $Oh$ (path 2) one has a V-regime crossing into the VI-regime at $r_{min}\sim Oh^{-3.1}$.   

However, a recent publication by \cite{castrejon15} shows that the picture can be more complex: the breakup can pass transiently through multiple different regimes. For example, \cite{castrejon15} show that, counter-intuitively, at $Oh=0.23$ there is no I$\to$VI transition but rather an I$\to$V$\to$VI, so that a new unexpected low-$Oh$ V-regime is encountered.  Similarly, a high-$Oh$ I-regime is observed.  This behaviour is sketched out \emph{qualitatively} in a phase diagram (their Figure 1E) showing how $Re_{local}$ varies as breakup is approach for different values of $Oh$.

In this work, we will use computations to build the first \emph{fully computed and quantitatively constructed} phase diagram for the breakup so that the nature of the transitions can be established.  To orientate the reader over the forthcoming sections, Figure~\ref{F:actual} gives a preview of the computed phase diagram, in a form that is only asymptotically accurate as $r_{min}\to 0$ but serves to neatly illustrate the computed flow transitions. The result shows clearly how the new phase diagram differs from the expected behaviour (Figure~\ref{F:expected}) due to the appearance of the low-$Oh$ V-regime sandwiched between the I- and VI-regimes, so that path 1 now first encounters an I$\to$V transition. In contrast to \cite{castrejon15}, no evidence for the high-$Oh$ I-regime can be seen. These features will be considered in further detail over the forthcoming sections.

In \S\ref{S:V-regime}--\S\ref{S:I-regime} we will show how the phase diagram in Figure~\ref{F:actual} was actually constructed, but before doing so we must specify the problem considered.

\section{Problem Formulation}\label{S:form}

A liquid bridge geometry (Figure~\ref{F:sketch}) is used with the liquid trapped between two stationary solid discs of (dimensional) radius $R$ a distance $2R$ apart and surrounded by a dynamically passive gas.  This setup allows us to isolate the breakup dynamics, limit the elongation of the domain (e.g.\ in contrast to dripping phenomena) and retain an experimentally realisable setup. A future work will consider the effects of geometry through elongation by allowing the plates to move apart. The contact line where the liquid-gas free-surface meets the solid remains pinned at the disc's edge throughout.  Assuming gravitational effects are negligible allows us to consider a plane of symmetry at $z=0$ of a cylindrical polar coordinate system $(r,z,\theta)$ so that, using $R$ as a characteristic scale for lengths, the solid is located at dimensionless position $z=1$ and the free-surface is pinned at $(r,z)=(1,1)$.  The extension to include gravity is not a difficult one, but it does not add significant value to a study focused on the small-scale breakup.

Using $U=\sigma/\mu$ as a scale for velocity; $T=R/U$ as a scale for time; and $\mu U/R$ for pressure, the (dimensionless) Navier-Stokes equations are
\begin{equation}\label{ns}
\nabla\cdot\mathbf{u} = 0,\qquad \frac{\partial\mathbf{u}}{\partial t}+\mathbf{u}\cdot\nabla\mathbf{u} = Oh^2~\nabla\cdot \mathbf{P},
\end{equation}
where the stress tensor is
\begin{equation}\label{stresst}
\mathbf{P} = -p\mathbf{I} + \left[\nabla\mathbf{u}+\left(\nabla\mathbf{u}\right)^T\right] .
\end{equation}
Here, $\mathbf{u}$ and $p$ are, respectively, the velocity and pressure in the liquid, and the Ohnesorge number is $Oh=\mu/\sqrt{\rho\sigma R}$.  

On the free-surface, whose location $f(r,z,t)=0$ must be obtained as part of the solution, the kinematic equation 
\begin{equation}\label{fs}
\frac{\partial f}{\partial t} + \mathbf{u}\cdot\nabla f =0,
\end{equation}
is applied alongside the usual balance of fluid stresses with capillarity in the directions tangential and normal to the free-surface
\begin{equation}\label{fs}
\mathbf{n}\cdot\mathbf{P}\cdot\left(\mathbf{I}-\mathbf{n}\mathbf{n}\right) = \mathbf{0},  \qquad \mathbf{n}\cdot\mathbf{P}\cdot\mathbf{n} = \nabla\cdot\mathbf{n},
\end{equation}
where $\mathbf{n}$ is the inward normal, $\mathbf{I}$ is the metric tensor of the coordinate system and $p$ is taken relative to the (constant) gas pressure. 

At the liquid-solid boundary, no-slip and impermeability are applied $\mathbf{u}=0$ and the free surface is pinned at the contact line $f(1,1,t)=0$. 

\subsection{Initial Conditions and Initiation of Breakup}

The dynamics of breakup are studied by generating a thread shape that is taken to the edge of its Rayleigh-Plateau stability limit \citep{slobozhanin93,lowry95}, so that  small perturbations will trigger breakup. The movie `FullBreakup' in the Supplementary Material shows a typical breakup event resulting from this approach. In this way, the breakup dynamics can be computed independently of a driving force such as the flux from a nozzle, as in jetting, or the pull of gravity, as in dripping \citep{rubio13}.  This reduces the dimensionality of the system to just one  controlling parameter $Oh$ so that constructing a phase diagram becomes a tractable task.

The required initial thread shape can be generated by either solving the Young-Laplace equation for the free-surface position \citep{fordham48,brown80,thoroddsen05} or using the finite element code, described below, in a quasi-static mode, to generate this shape by gradually removing fluid from the thread in a manner similar to the experimental setup described in \cite{meseguer85}.  The latter method is used, as this also enables the shape on the stable-unstable boundary to be established. The result is the first profile in Figure~\ref{F:Oh10}(a).  The instability can be triggered by either continuing to gradually remove fluid from the thread or by very slowly moving the plates apart, and both approaches were shown to produce indistinguishable results, thus confirming that the only control parameter is $Oh$.

\section{Multiscale Finite Element Computations}\label{S:computations}

The mathematical model formulated in \S\ref{S:form} is for an unsteady free-surface flow influenced by the forces of inertia, viscosity and capillarity.  To study this system over the entire range of $Oh$ requires computational methods.

\subsection{Previous Computational Studies}
 
The breakup of liquid volumes has been studied computationally using the Stokes \citep{gaudet96,pozrikidis99}, Euler \citep{day98,schulkes94}, and Navier-Stokes equations for both Newtonian \citep{popinet09,ashgriz95} and Non-Newtonian \citep{bhat10,li03} liquids.  Considerable progress has been made by the Group at Purdue University using finite element methods for axisymmetric flows such as dripping \citep{ambravaneswaran00}, jetting \citep{ambravaneswaran04}, drop-on-demand \citep{chen02a} and liquid bridge breakup \citep{suryo06} with an array of different physical effects such as non-Newtonian fluids \citep{yildrim01}, surfactant dynamics \citep{mcgough06} and electric field effects \citep{collins08}.  These algorithms were the first to demonstrate a number of experimental features such as overturning of the free-surface for a viscous fluid \citep{wilkes99}, transitions between scaling regimes \citep{chen02}, and the identification of multiple regime transitions \citep{castrejon15}.  
 
Notably, it has been shown that with sufficient care, the results of sharp interface methods, such as those implemented at Purdue, can sometimes be recovered by the diffuse interface approach, in which topological changes are easier to handle computationally.  In particular, the diffuse interface method developed in \cite{yue04} was shown in \cite{zhou06} to accurately recover the sharp-interface results of \cite{wilkes99}. Furthermore, in \cite{zhou06}, the flexibility of the diffuse interface method was demonstrated by computing compound drop formation, where additional complexity arises from the need to track two free surfaces.  Despite these advances, we will implement a sharp interface approach (see \S\ref{S:comp}) whose reliability and accuracy has been repeatedly confirmed.
 
Despite the numerous successes of computational schemes in accurately predicting many of the global features of breakup, there have been fewer investigations of the local dynamics and their comparison with the similarity solutions. One of the most impressive works in this direction is \cite{suryo06}, where scales comparable to those resolved in the current work ($r_{min}<10^{-4}$) were captured and the scaling behaviour of both Newtonian and power-law liquids was investigated.  Other progress includes results in \cite{notz01} and \cite{chen02}, where I$\to$VI transitions were shown, and \cite{castrejon15}, in which multiple transitions were observed (see \S\ref{S:multiple}).  Our work will build on these previous investigations by systematically quantifying the accuracy of the similarity solutions across the entire parameter space (of Newtonian liquids) in order to build the first quantitatively constructed phase diagram for breakup.

In principle, computations are the ideal tool with which to map a phase diagram for the breakup process and to investigate in which regions each of the similarity solutions is accurate; a similar procedure was used for the coalescence phenomenon in \cite{sprittles14_jfm2}.  The difficulty of this procedure is that many decades of $r_{min}$ are required in order to reliably determine the scaling behaviour of different regime boundaries (if they exist).    Although one may hope to compute $r_{min}$ until its behaviour falls into the VI-regime, in practise this may occur at scales below the possible computational resolution, e.g.\ $r_{min}<10^{-5}$ \citep{burton04}.  Therefore, establishing the range of applicability of the similarity solutions using computations and then using these solutions to carry the dynamics to scales below realisable computational resolution seems to be a promising strategy for capturing breakup.

\subsection{Computational Approach}\label{S:comp}

A focus of this work is to resolve the spatial and temporal dynamics of the pinch-off process to the smallest scales possible in order to compare with the similarity solutions proposed in the literature for the final stages of breakup.   The approach used is based on the finite element framework originally developed in \cite{sprittles12_ijnmf,sprittles12_jcp} to capture dynamic wetting problems and subsequently used to study the coalescence of liquids drops \citep{sprittles12_pof,sprittles14_pre}, including two-phase calculations \citep{sprittles14_jfm2}; drop impact phenomena \citep{sprittles12_pof}; the detachment of bubbles from an orifice \citep{simmons15}; and dynamic wetting in a Knudsen gas \citep{sprittles15_jfm}.    These flow configurations are all inherently multiscale, either due to the disparity of length scales in the problem formulation or because of the dynamics of the process itself, which generates small scales during its evolution, as is the case for breakup phenomena.   The framework allows the global flow in these problems to be resolved alongside localised smaller-scales, without relying on particular limits of $Oh$.

As a step-by-step user-friendly guide to the implementation of this computational framework has been provided in \cite{sprittles12_ijnmf}, only the main details will be recapitulated here alongside some aspects which are specific to the current work.  The code uses the arbitrary Lagrangian Eulerian scheme, based on the method of spines \citep{ruschak80,kistler83},  to capture the evolution of the free surface in two-dimensional or three-dimensional axisymmetric flows.  In order to keep the problem computationally tractable, the mesh is graded so that small elements can be used near the pinch-off region whilst larger elements are used where scales associated with the global flow are present.  

The mesh starts with $\approx4000$ triangular elements ($\approx500$ surface nodes) and then adaptively refines as the breakup progresses by adding spines whenever elements become too deformed.  At the end of the computation, the number of elements is in the range $7000$--$50000$ ($800$--$7000$ surface nodes), depending on the particular breakup requirements.  For example, for a computation at $Oh = 10^{-3}$, the mesh is refined due to the formation of a `corner' in the free surface (see Figure~\ref{F:Oh0p001}), resulting in 869 surface nodes with a minimum surface element length of $3\times10^{-5}$.  In contrast, at $Oh=0.16$, refinement is required due to the formation of a thin thread of liquid (Figure~\ref{F:Oh0p16}) which requires 5263 surface nodes with surface elements in this region all having length $\approx10^{-4}$. Overturning of the free surface is permitted by using angled spines (i.e. not only horizontal) whose slope varies along the thread. 

Refinement of the mesh is restricted by controlling the smallest permitted element size $h_{min}$, which affects the final number of elements.  Reducing $h_{min}$ allows the code to converge to smaller $r_{min}$ so that more of the breakup is recovered.  By reducing $h_{min}$ from $10^{-2}$ down to $10^{-5}$, convergence of the scheme under spatial refinement has been established, with each decrease in $h_{min}$ revealing more of the solution (smaller $r_{min}$) but producing curves which are graphically indistinguishable from those obtained  on cruder meshes, where both solutions exist.
%
%
%The overturning has been allowed by the designed mesh alignment in our codes. We divide the whole domain into two parts by a straight line: z = k*r + c, and use geometric progression to distribute the spines and directors from this line. If the slope is sufficient large and a group of directors in the small neighborhood around this line could cover the pinchoff region which also have the similar slope to k, then the mesh is capable of capturing the overturning of the free surface using method of spines. For all computations we choose k = -0.8. c could be decided by a quick preliminary computation for each Oh. For the convenience but without loss of accuracy, for any Oh>0.01, c = 0.6; for 0.005<Oh =< 0.01, c = 0.8; for oh =<0.005, c = 0.85.

The result of the spatial discretisation is a system of non-linear differential algebraic equations of index two \citep{lotstedt86} which are solved using the second-order backward differentiation formula, whose application to the Navier-Stokes equations is described in detail in \cite{gresho2}.  The time step automatically adapts during a simulation to capture the appropriate temporal scale at each instant.

In the problem considered, the smallest scale that needs resolving is the minimum neck radius $r_{min}$ and eventually this becomes so small that accurate converged solutions can no longer be obtained. For the entire range of $Oh$ considered computations remain accurate down to $r_{min}=10^{-4}$ and in certain cases they can go as far as $r_{min}=10^{-5}$.  

\section{Quantitatively Identifying Regimes}\label{S:quant_regimes}

Computations are unable to resolve down to $r_{min}=0$, i.e.\ the point at which the topological change occurs, so that the precise time $t=t_b$ of pinch-off is unknown.  This can be problematic when comparing to scaling laws of the form $r_{min}=A\tau^B+C$, where $\tau=t_b-t$ requires the time of break up, as small changes in $t_b$ can have a large effect on a curve of $r_{min}$ vs $\tau$ which is then used to determine which regime a breakup process is in.  Similar conclusions were reached for coalescence in \cite{thoroddsen05}.  In the final regime one has $C=0$, otherwise for `transitional' or `transient' regimes $C$ must be determined as well. In general, there is no easy method to overcome these issues without assuming a particular functional form for the breakup.   

However, in the V- and VI-regimes it is possible to exploit $r_{min}$ being a linear function of $\tau$ ($B=1$) so that the speed at which the minimum thread radius evolves is constant, i.e.\ independent of both $t_b$ and $C$.  Therefore, to identify these regimes the `speed of breakup' given by $\dot{r}_{min}=\diff{r_{min}}{t}$ can be compared to the predictions of (\ref{papa}) that $\dot{r}_{min}=-0.071$ and (\ref{eggers}) that $\dot{r}_{min}=-0.030$ as a function of $r_{min}$ without needing to know $t_b$.  This has the further advantage that $C$ doesn't have to be fitted when a regime is observed transiently, as in \cite{castrejon15}.
%
%The same approach cannot be used to identify the inertial regime where $r_{min}=A_I(Oh~\tau)^{B_I}+C_I$, as $B_I=2/3$, so that the speed of breakup is no longer constant. Here, finding a quantity which is independent of both $B_I$ and $C_I$ is possible, but far messier and without any obvious physical interpretation.  In particular, the quantity $M_{min}(t)$ defined by 
%
%\begin{equation}\label{M_{min}} 
%M_{min} =(\frac{-9\dot{r}_{min}^4}{8\ddot{r}_{min} Oh^2})^{1/3}
%\end{equation}
%
%satisfies $M_{min}=A_I$ in the I-regime, i.e.\ a time-independent constant which is independent of both $t_b$ and $C_I$.  

The same approach cannot be used to identify the inertial regime where $B=2/3$, so that the speed of breakup is no longer constant.  Therefore, we try a similar idea to that above and determine an $Oh$-independent quantity which should be linear in the inertial regime.  This relies on $C\approx 0$, so that this regime must be close to breakup.  Then, given that  $r_{min}=A_I(Oh~\tau)^{2/3}$, we introduce $l_{min}=Oh^{-1} r_{min}^{3/2}$ which will satisfy $l_{min}=A_I^{3/2}\tau$ if the expected scaling holds.  Differentiating this quantity with respect to time gives $\dot{l}_{min}=\diff{l_{min}}{t}=-A_I^{3/2}$, i.e.\ a time-independent constant.  

A key aspect of this work is to develop techniques to \emph{quantify} when the breakup dynamics are in a certain regime \citep{sprittles14_jfm2}.  To do so, we must define what it means to be `in a regime'.  We choose to define a breakup process to be in the V- or VI-regimes when the speed of breakup $\dot{r}_{min}$ is within $0.015$ of that predicted by (\ref{papa}) or (\ref{eggers}), respectively.  In other words, speeds in the range $\dot{r}_{min}=-0.071\pm0.015$ indicate V-regime dynamics whilst $\dot{r}_{min}=-0.030\pm0.015$ for the VI-regime. In a similar way, the inertial regime is defined by points at which $\dot{l}_{min}$ is within 0.15 of $A_{I}^{3/2}$, where we will later see that $A_{I}^{3/2}=0.5$.  Choosing smaller (larger) margins to define each regime will shrink (expand) the area of a given regime in the phase diagram but should not alter the qualitative picture.  The particular values chosen avoided regimes overlapping and enabled us to satisfactorily map the phase diagram.  The method for identifying regimes is illustrated in the Appendix.

Rather than assuming that the regimes fill all of phase space, we will be careful to split the entrance to and exit from a particular regime. For example, the exit from the V-regime will be labelled as $r_{min}^{V\to}$ and the entrance to the VI-regime as $r_{min}^{\to VI}$, rather than assuming a-priori these coincide at $r_{min}^{V\to VI}$.

\subsection{Analysis of the Regimes}\label{S:analysis}

To understand the dynamics in the different regimes and determine when the assumptions behind the similarity solutions of \S\ref{S:regimes} are valid, local to the breakup point we analyse the assumptions made about (a) the relative magnitudes of viscosity and inertia and (b) the slenderness of the thread.  In practice, this means calculating (a) the local Reynolds number $Re_{local}$ and (b) a measure for the slenderness of the free surface profile.  

In both cases, one needs to define an appropriate axial length $L_z$ and velocity $U_z$ scale near the pinch point.  The obvious choice for $U_z$ is the maximum axial velocity in the thread $U_z=w_{max}$, which will occur near, but not at, the position where $r=r_{min}$.  Therefore, a characteristic axial scale can be defined as the vertical distance between the positions of minimum radius and maximum velocity, so that $L_z = |z(r=r_{min}) - z(w=w_{max})|$.  This defines (a) the local Reynolds number as $Re_{local}=Oh^{-2} U_{z} L_z$ and (b) the slenderness as the ratio of the characteristic radial length scale $L_r=r_{min}$ with $L_z$, so that the geometry is slender when $L_r/L_z\ll1$. Importantly, $w_{max}=\max(w)$ is used rather than $\max |w|$ (see Figure~\ref{F:Oh0p16}(c)) to avoid discontinuities in $L_z$, and hence $Re_{local}$, which occur when the position of $\max |w|$ jumps from being above the pinch point to below it (as in the V$\to$VI transition).

Other possible definitions of $U_z$ and $L_z$ exist, see \cite{castrejon15}, so it is not a-priori clear we have made the correct choices.  However, computations will show that our definition allows us reproduce the expected scalings for $Re_{local}$ in both the I-regime (\ref{inertial}) and the V-regime (\ref{papa}), which serves to validate our approach.
%
%In \cite{castrejon15}, the local Reynolds number is defined using $L_z=z|_{1.2r_{min}} - z|_{r_{min}}$ and $U_z=w|_{1.2r_{min}}$, so that the length scale is defined by the distance it takes for the thread's radius to increase by 20\% and the velocity is sampled at this point too.  Our computations confirm that although there are quantitative differences between the predictions of the two local Reynolds numbers, the qualitative trends appear to be the same.  

\section{Overview}

In \S\ref{S:V-regime}--\S\ref{S:I-regime}, respectively, the dynamics of breakup for large ($Oh>1$), intermediate ($0.1<Oh<1$) and small ($Oh<0.1$) Ohnesorge numbers are analysed separately and used to establish the limits of applicability of the different similarity solutions described in \S\ref{S:regimes}.  This will then be used to define the regimes of breakup on a phase diagram.  

In each section, the initial focus will be on a breakup event which highlights the features of the prominent regime, namely the V-regime ($Oh=10$), VI-regime ($Oh=0.16$) and I-regime ($Oh=10^{-3}$), with a movie of each breakup in the Supplementary Material. In each case, snapshots of the free surface shape, axial velocity and pressure distributions in the breakup region will be presented.  Evidence that the breakup is in a given regime will be provided by comparing the computational results with the similarity solution's predictions for the evolution of the free-surface shape, minimum bridge radius $r_{min}$, maximum axial velocity $w_{max}$ and scaling of the local Reynolds number $Re_{local}$.

Having confirmed the existence of a regime, we then identify its boundaries on the phase diagram to determine the regime transitions.  As explained in \S\ref{S:quant_regimes}, these boundaries are specified by calculating the values of $\dot{r}_{min}$ and $\dot{l}_{min}$ as a function of $r_{min}$.  Having discovered the transitional behaviour, the local measures $Re_{local}$ and slenderness $L_r/L_z$ (introduced in \S\ref{S:analysis}) are studied in an attempt to rationalise these findings.

The analysis in \S\ref{S:V-regime}--\S\ref{S:I-regime} will not only allow us to construct a phase diagram (c.f.\ Figure~\ref{F:allOh_transitions}) for the process but will also open up new questions about the breakup process; to avoid distracting from the main focus in \S\ref{S:V-regime}--\S\ref{S:I-regime}, these aspects will be considered in more detail in \S\ref{S:discussion}.

\section{Viscous-Dominated Flow (Large $Oh$)}\label{S:V-regime}

\begin{figure}
     \centering
\subfigure[Free surface evolution.]{\includegraphics[scale=0.4]{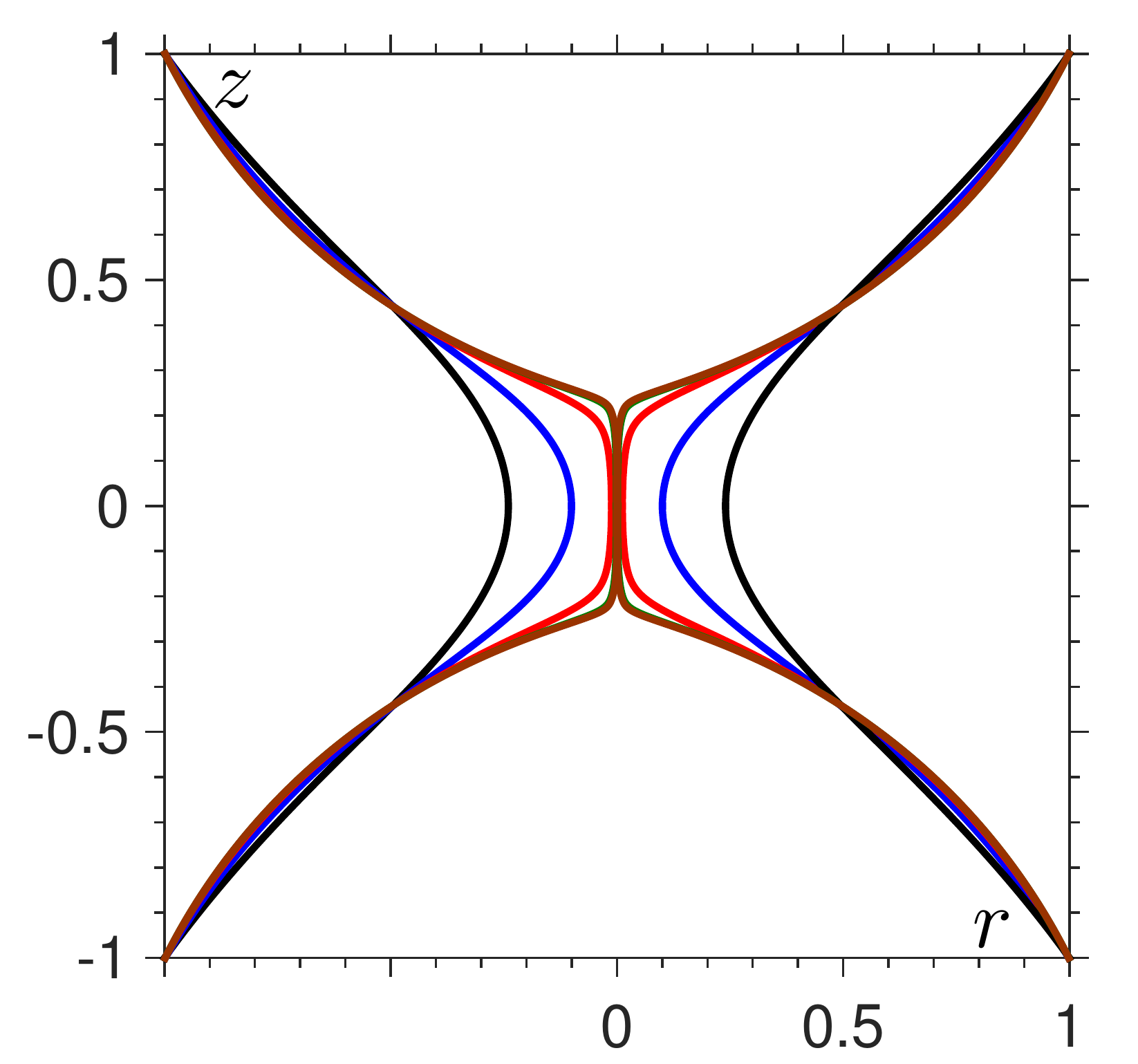}}\\
\subfigure[Close-up of breakup.]{\includegraphics[scale=0.3]{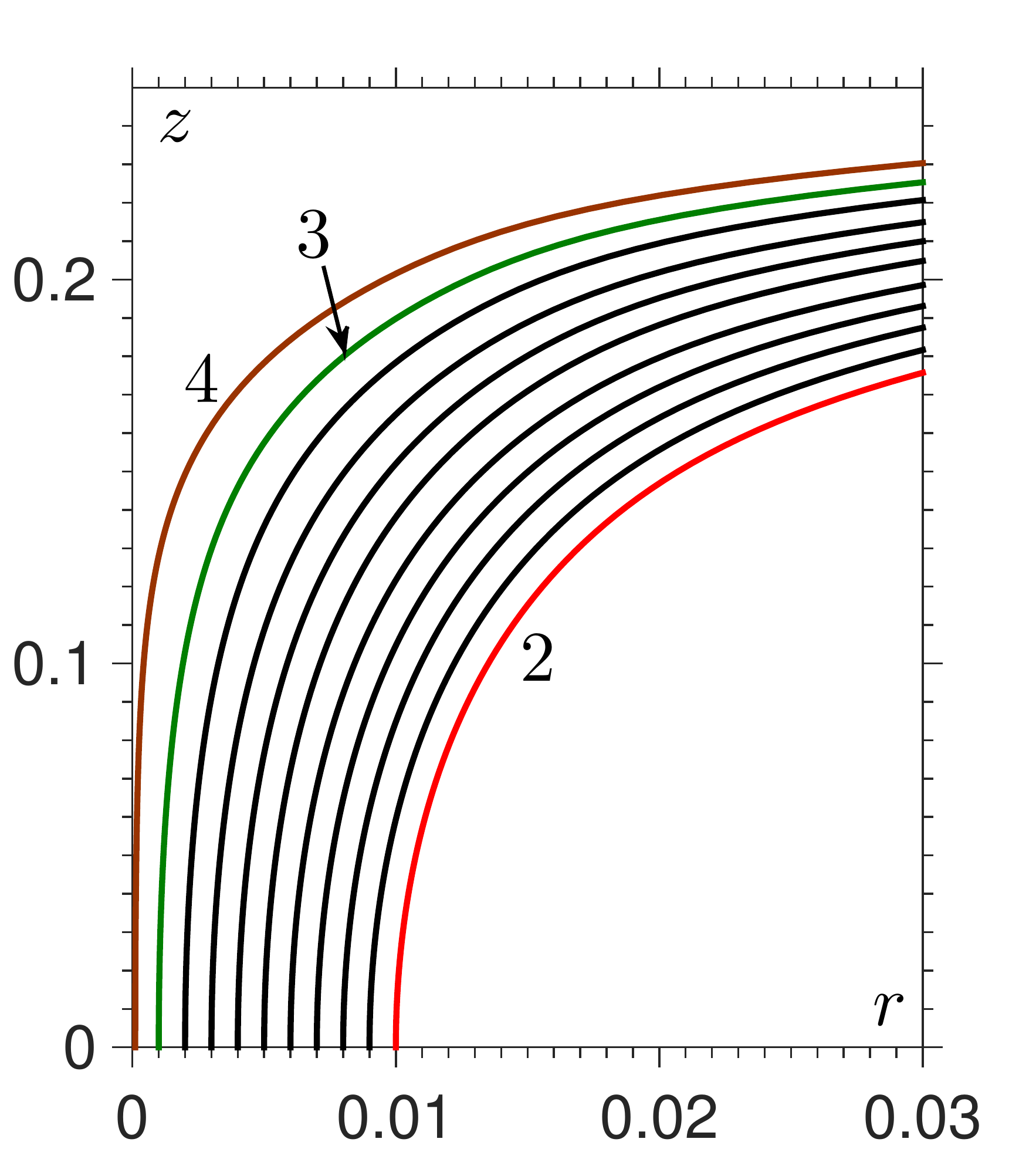}} 
\subfigure[Axial velocity.]{\includegraphics[scale=0.3]{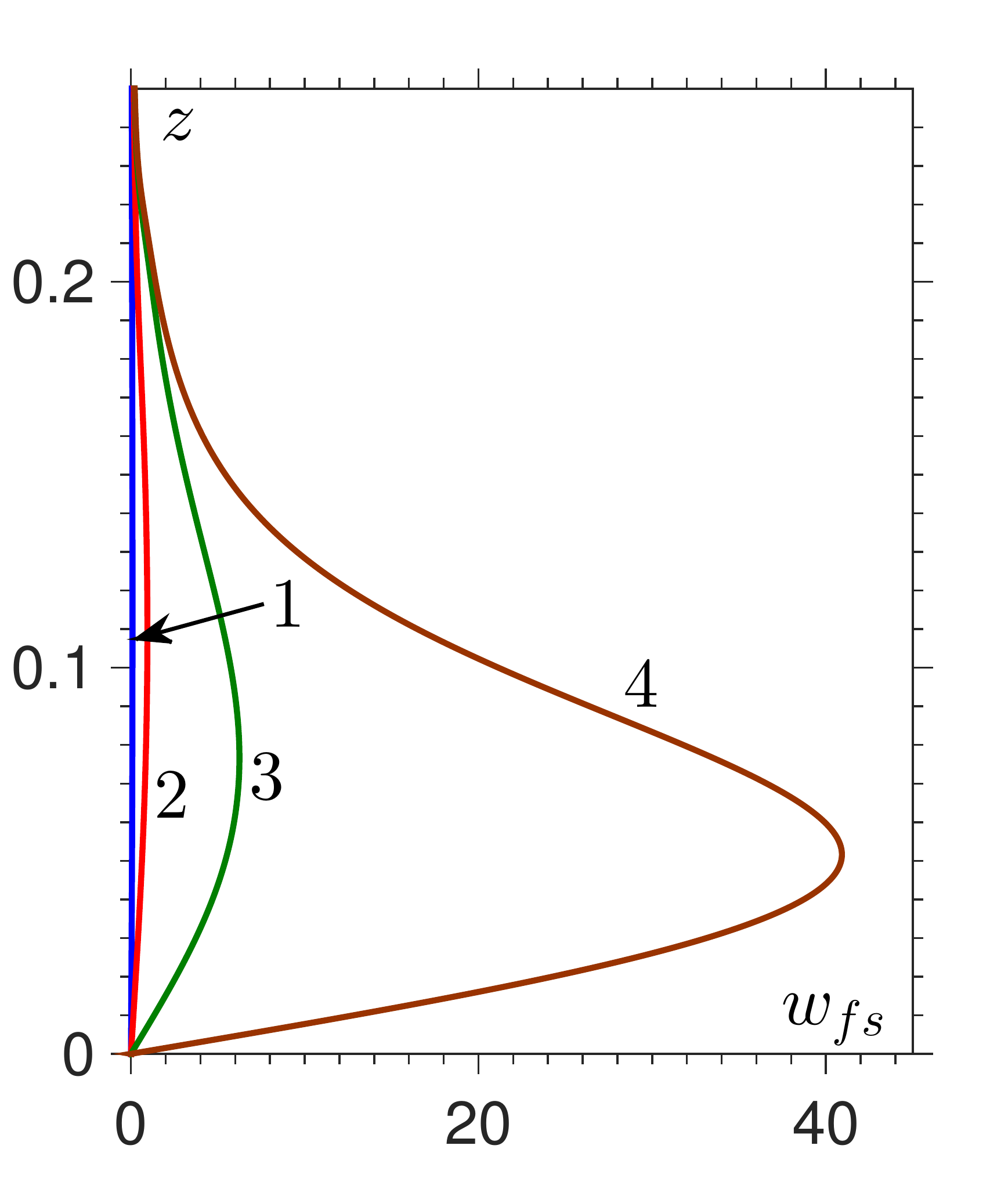}}
\subfigure[Pressure.]{\includegraphics[scale=0.3]{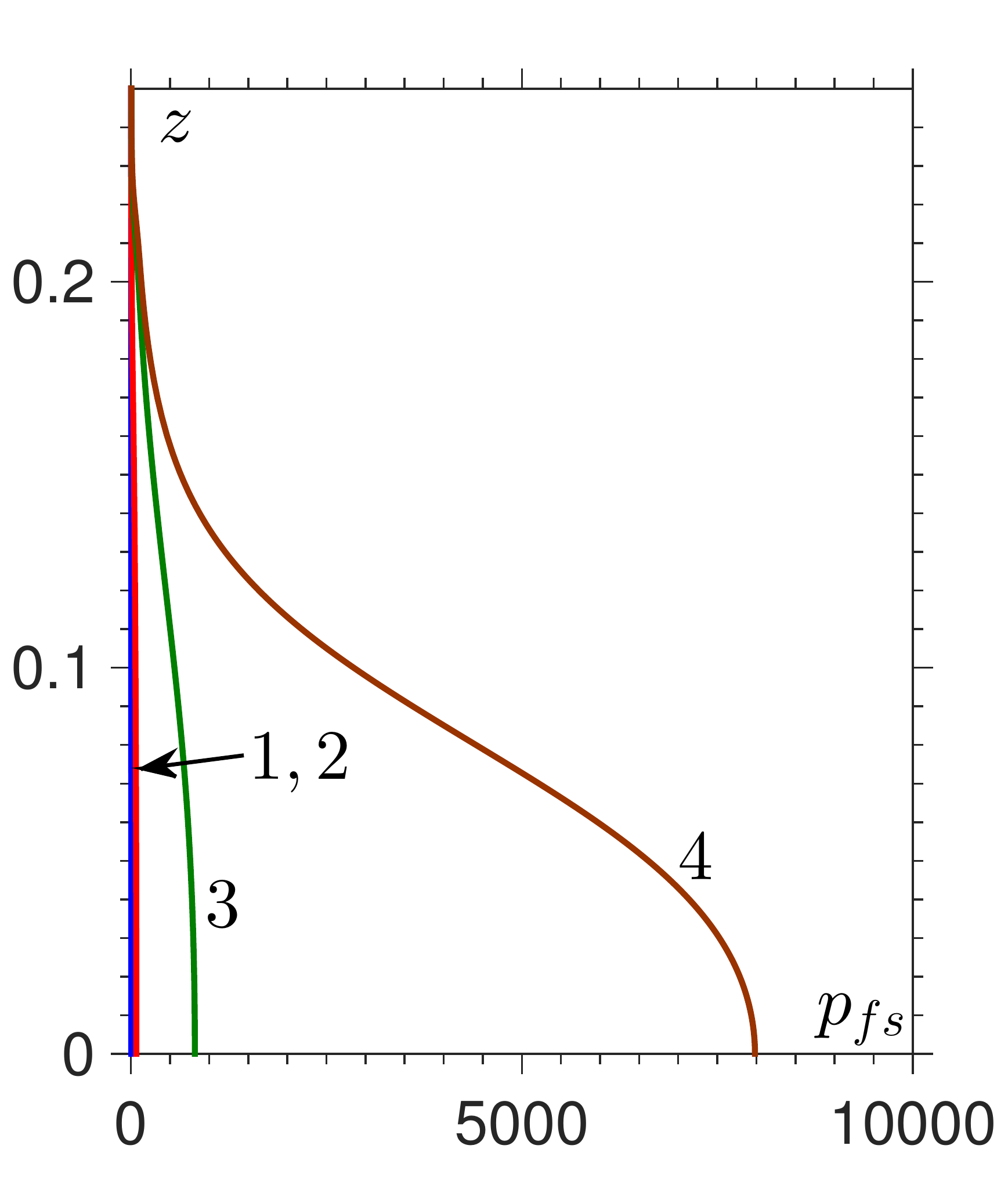}}
\caption{For $Oh=10$, characteristic of the V-regime, plots show (a) the entire free-surface, (b) a close-up of the breakup region, (c) the axial velocity at the free-surface $w_{fs}$ and (d) the pressure at the free-surface $p_{fs}$ at 1: $r_{min}=10^{-1}$ (blue), 2: $r_{min}=10^{-2}$ (red), 3: $r_{min}=10^{-3}$ (green) and 4: $r_{min}=10^{-4}$ (brown). }
\label{F:Oh10}
\end{figure}

For $Oh=10$ free surface profiles in Figure~\ref{F:Oh10}(a,b) show the development of a thin thread of liquid whose smallest minimum radius remains at $z=0$ throughout, so that no satellite drops form.  The axial velocity (Figure~\ref{F:Oh10}(c)) at the free surface $w_{fs}$, which is approximately $r$-independent, shows a rapid drainage from the thread driven by a pressure gradient from $z=0$ (Figure~\ref{F:Oh10}(d)).  

\begin{figure}
     \centering
\subfigure[Free surface profile in the VI Regime]{\includegraphics[scale=0.26]{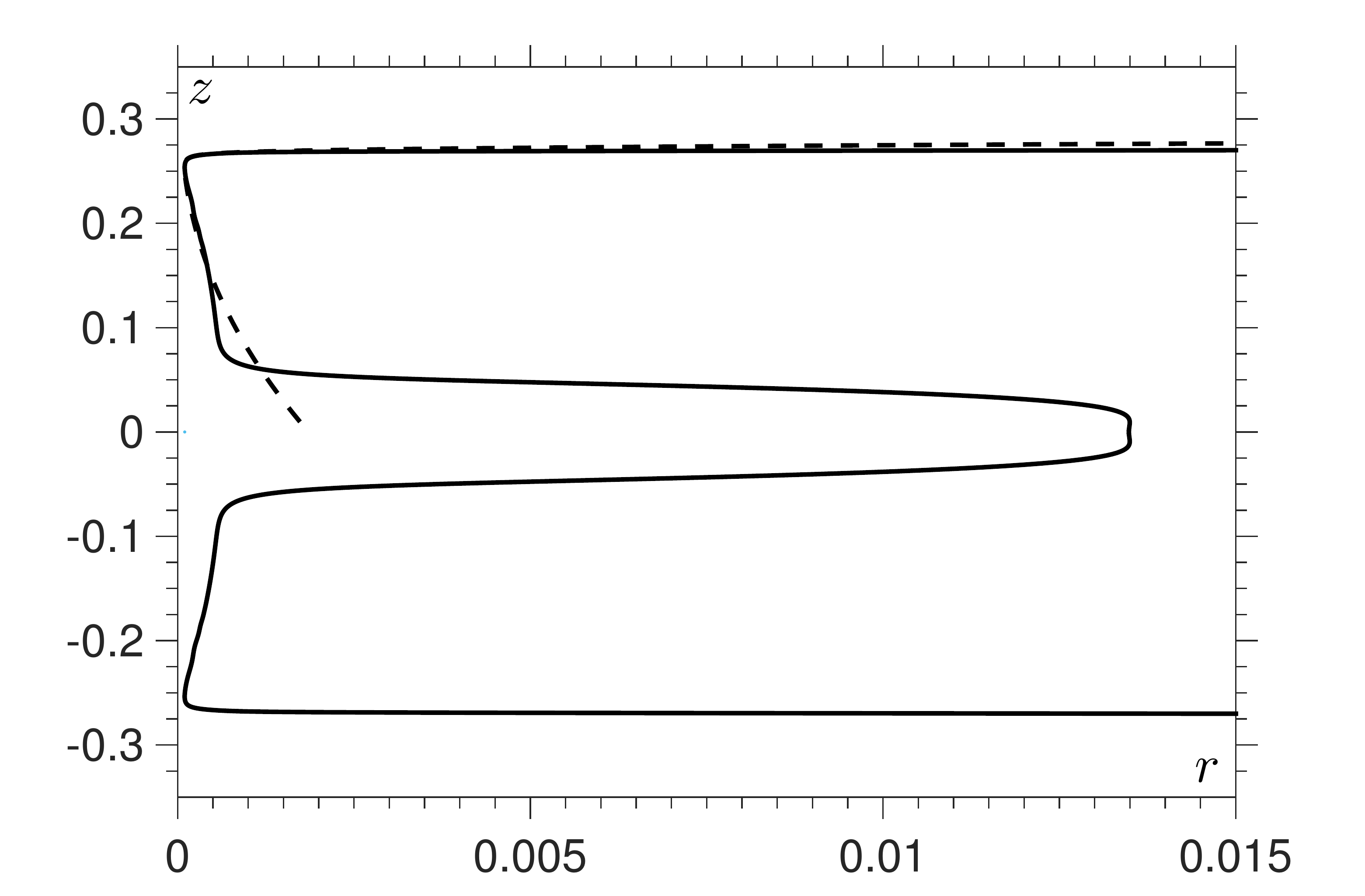}}
\subfigure[Free surface profile in the V Regime]{\includegraphics[scale=0.26]{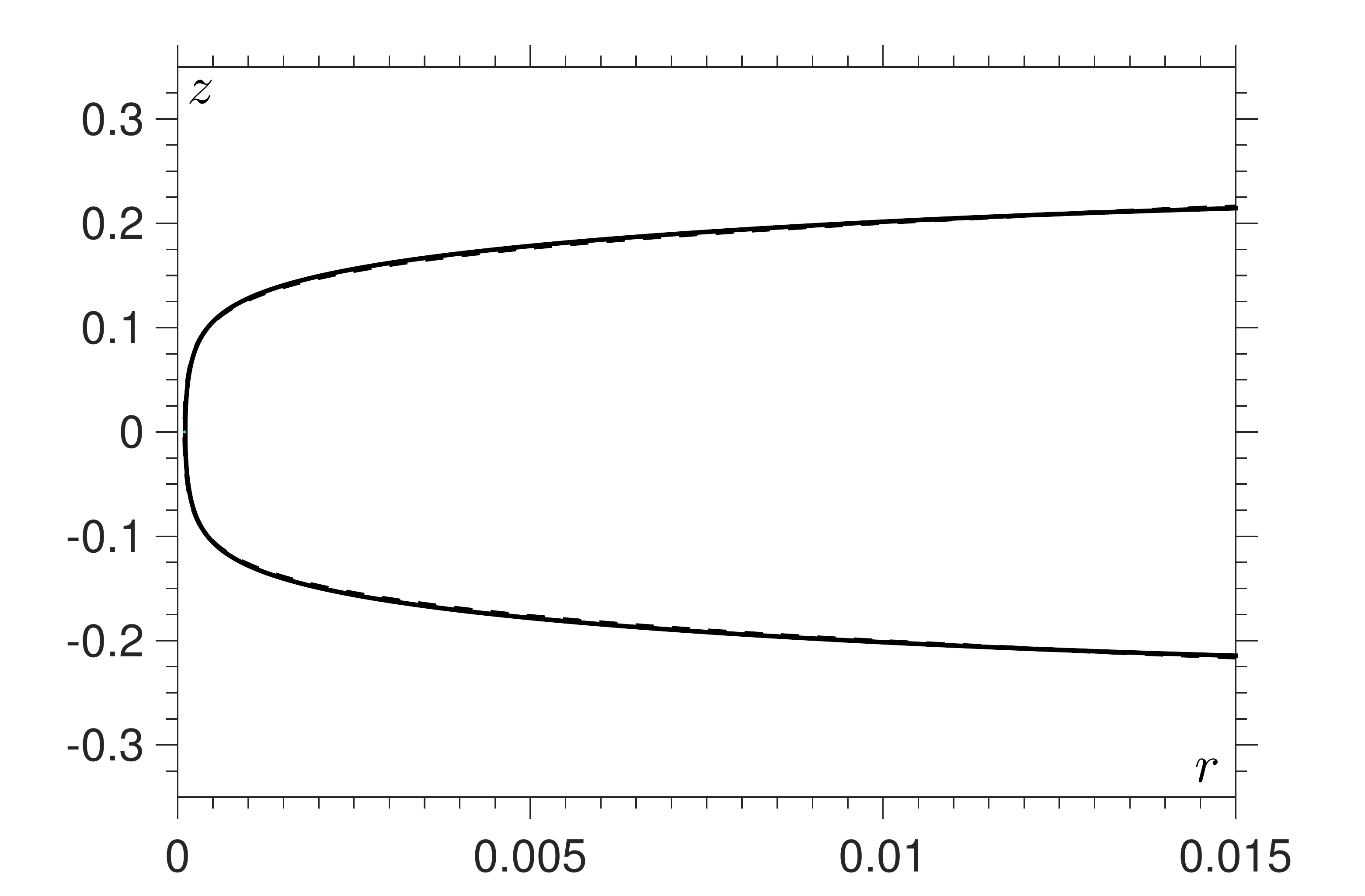}}
\caption{A comparison of the computed free surface profiles with the similarity solutions for the VI- and V-regimes.  Computed profiles are at the minimum radius $r_{min}=10^{-4}$ for (a) $Oh=0.16$ and (b) $Oh=10$ typical of breakup in the VI and V-regimes, respectively.  These profiles compare well to the similarity solutions (dashed lines) in (a) with ($\xi_{VI},\phi_{VI}$) provided in \cite{eggers93} giving $(r,z) = (\tau\phi_{VI}, z_0+Oh\tau^{1/2}\xi_{VI})$ with no free parameters and in (b) with ($\xi_{V},\phi_{V}$) from \cite{papageorgiou95,eggers97} so that $(r,z) = (\tau\phi_{V}, Oh^{2-2\beta}\tau^{\beta}\ell_z\xi_{V})$  with $\ell_{z}=1.85\times10^{-3}$.  In (a), the height of the pinch point $z_0=0.268$ for the VI-regime is calculated using the expression in \cite{eggers93} for the drift of this point $z(r=r_{min})=z_0-1.6~Oh~\tau^{-1/2}$ (shown in Figure~\ref{F:zmin} as the dashed line).}
\label{F:fs_vs_similarity}
\end{figure}

\begin{figure}
     \centering
\subfigure[Evolution of the maximum axial velocity.]{\includegraphics[scale=0.26]{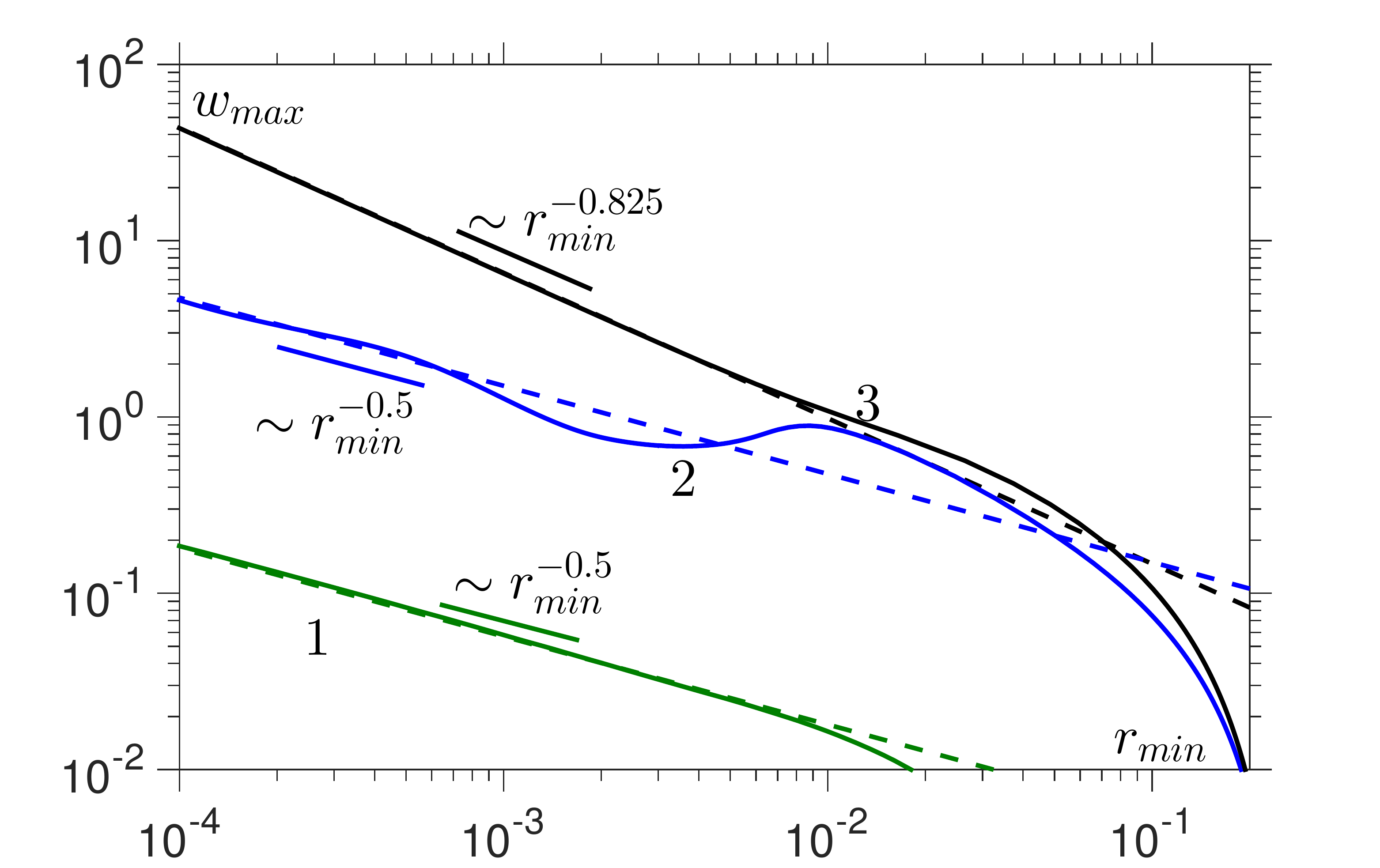}\label{F:wmax}} 
\subfigure[Minimum bridge radius against time from breakup.]{\includegraphics[scale=0.26]{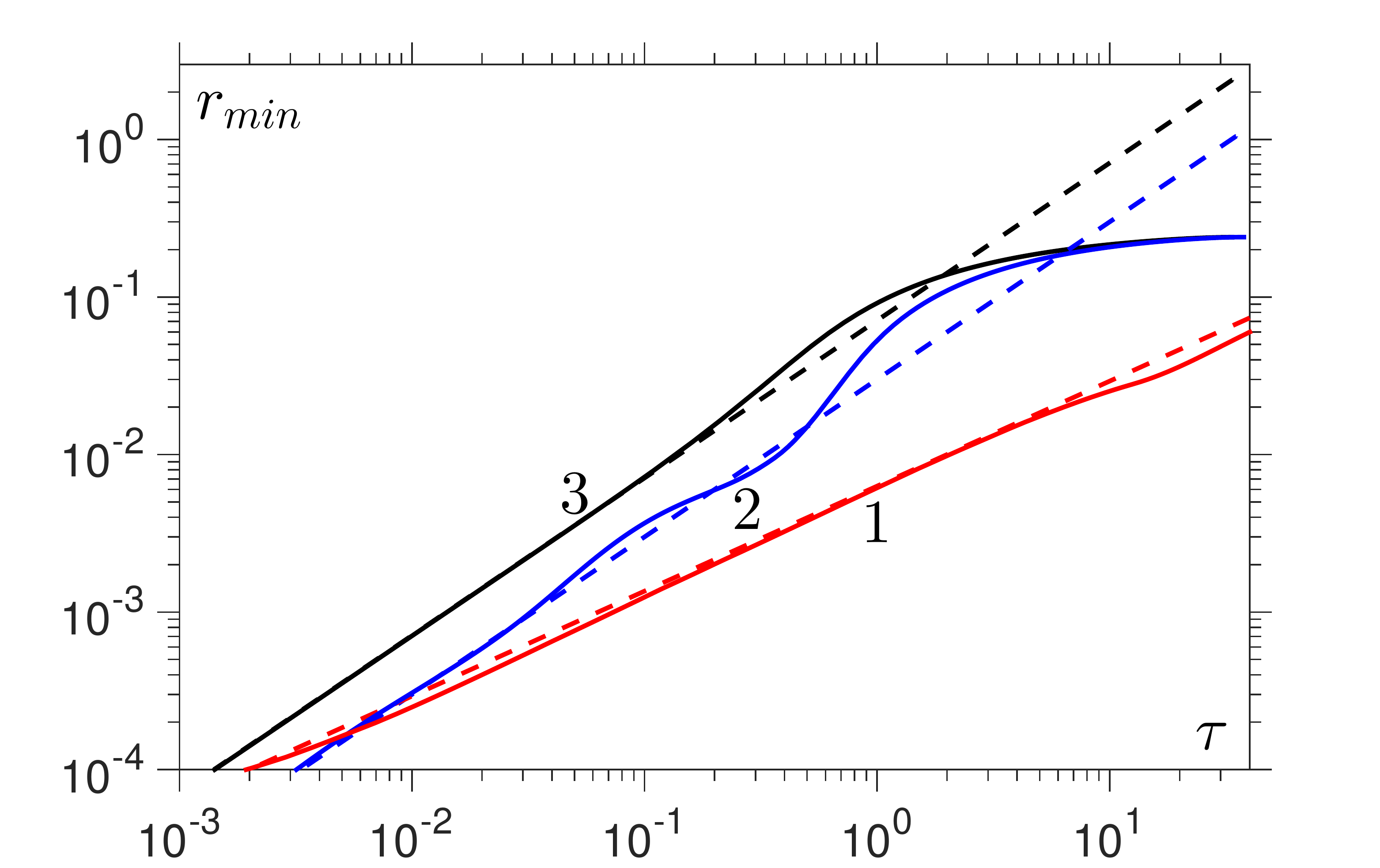}\label{F:rvst}} 
\caption{Curves are for 1: $Oh=10^{-3}$, 2: $Oh=0.16$ and 3: $Oh=10$.  (a) Evolution of $w_{max}$ as minimum radius $r_{min}$ decreases showing excellent agreement with the similarity solutions (dashed lines) from the I-regime, that $w_{max}\sim r_{min}^{-0.5}$, the VI-regime, that $w_{max}=0.3~Oh~r_{min}^{-0.5}$, and V-regime, that $w_{max}\sim r_{min}^{-0.825}$. (b) The minimum bridge radius $r_{min}$ against time from breakup $\tau$.  Computations converge to the similarity solutions for the I, VI and V-regimes, respectively, with the lower dashed line the similarity solution for the I-regime (\ref{inertial} with $A_I=0.63$), the middle one for the VI-regime (\ref{eggers}) and the highest one for the V-regime (\ref{papa}).}  %wmax=1.8e-3 r^-0.5 in I-regime
\end{figure}

Evidence that this breakup is in the V-regime are seen in Figure~\ref{F:fs_vs_similarity}(b), where the free surface profile obtained at $r_{min}=10^{-4}$ is seen to be almost indistinguishable from the similarity solution for the V-regime, and Figure~\ref{F:wmax}, in which the computed axial velocity (curve 3) follows the predicted (singular) form $w_{max}\sim r_{min}^{-0.825}$ as $r_{min}\to 0$.
%
% To plot the VI solution we use (with $\tau=3.19\times10^{-3}$ and $z_0=0.268$) 
% r = tau \phi_{sim}
% z = \sqrt{tau} Oh \xi_{sim} + z_0
%  plot(tau_0p0001*ss(181:430,2),0.268+0.16*sqrt(tau_0p0001)*ss(181:430,1))
%Velocity would be plot(0.16*ss(180:430,3)/sqrt(tau_0p0001),0.265+0.16*sqrt(tau_0p0001)*ss(180:430,1),'.')
%
% To plot the V solution we use:with $\tau=1.419\times10^{-3}$)
% r = tau \phi_{sim}
% z = \sqrt{tau} Oh \xi_{sim} + z_0
% plot(taus_0p0001*sss(:,2),0.00185*(10^(2-2*0.175))*(taus_0p0001^0.175)*sss(:,1),'.')
%
%

Assuming that the breakup remains in the V-regime, the data can be extrapolated to obtain the breakup time $t_b$ and, as a result, plot $r_{min}$ against $\tau$ in Figure~\ref{F:rvst} (curve 3) where there is excellent agreement with the similarity solution of (\ref{papa}) for $r_{min}<10^{-2}$.

\subsection{Identification of the V-Regime}

\begin{figure}
\subfigure[Evolution of the breakup speed $\dot{r}_{min}$ for 1: $Oh=1$, 2: $Oh=1.5$, 3: $Oh=2.5$, 4: $Oh=5$ and 5: $Oh=10$.]{\includegraphics[scale=0.26]{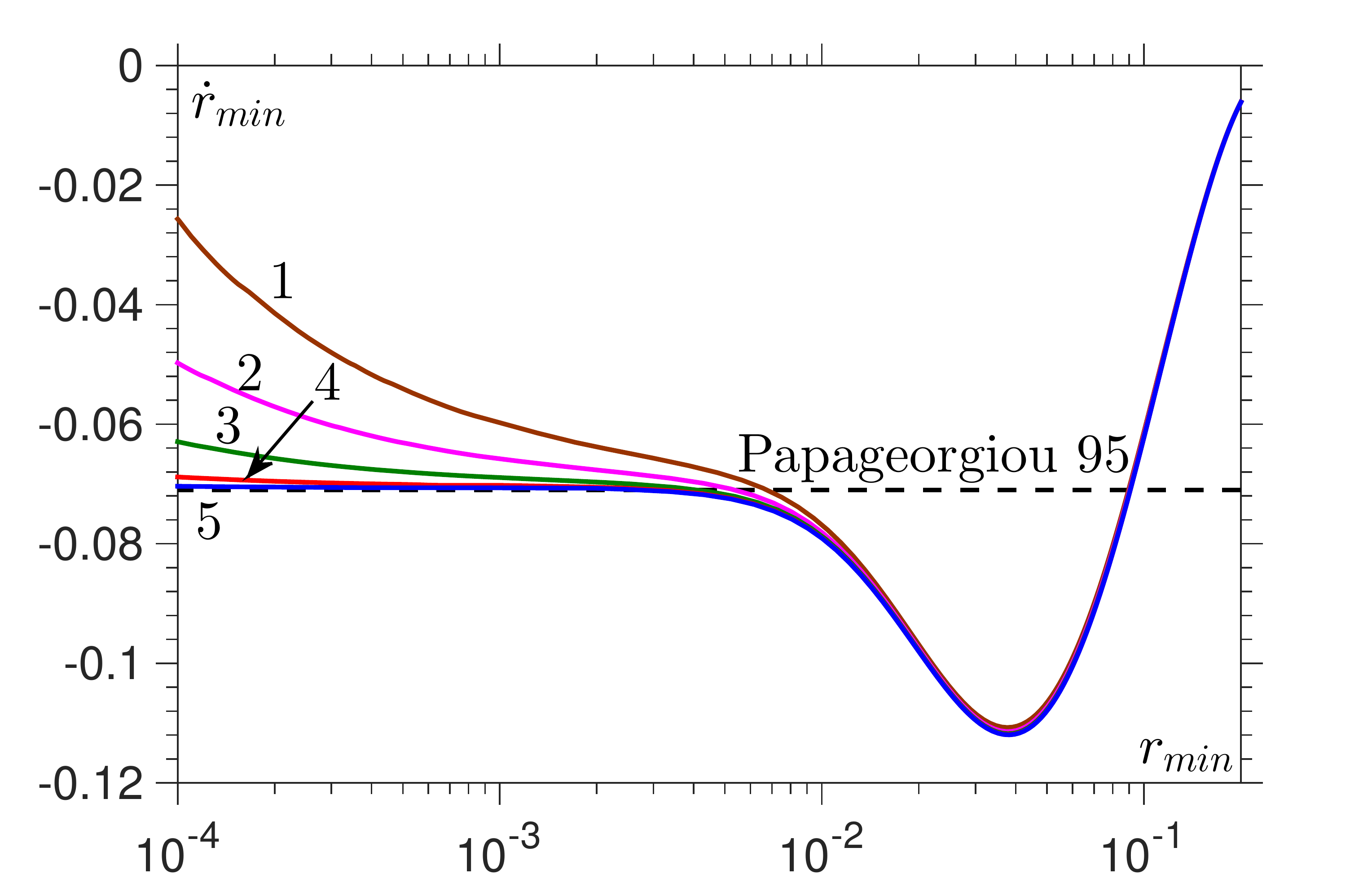}\label{F:highOh_umin}} 
\subfigure[Variation of the local Reynolds number for 1: $Oh=0.25$, 2: $Oh=0.5$, 3: $Oh=1$, 4: $Oh=2.5$ and 5: $Oh=10$.]{\includegraphics[scale=0.26]{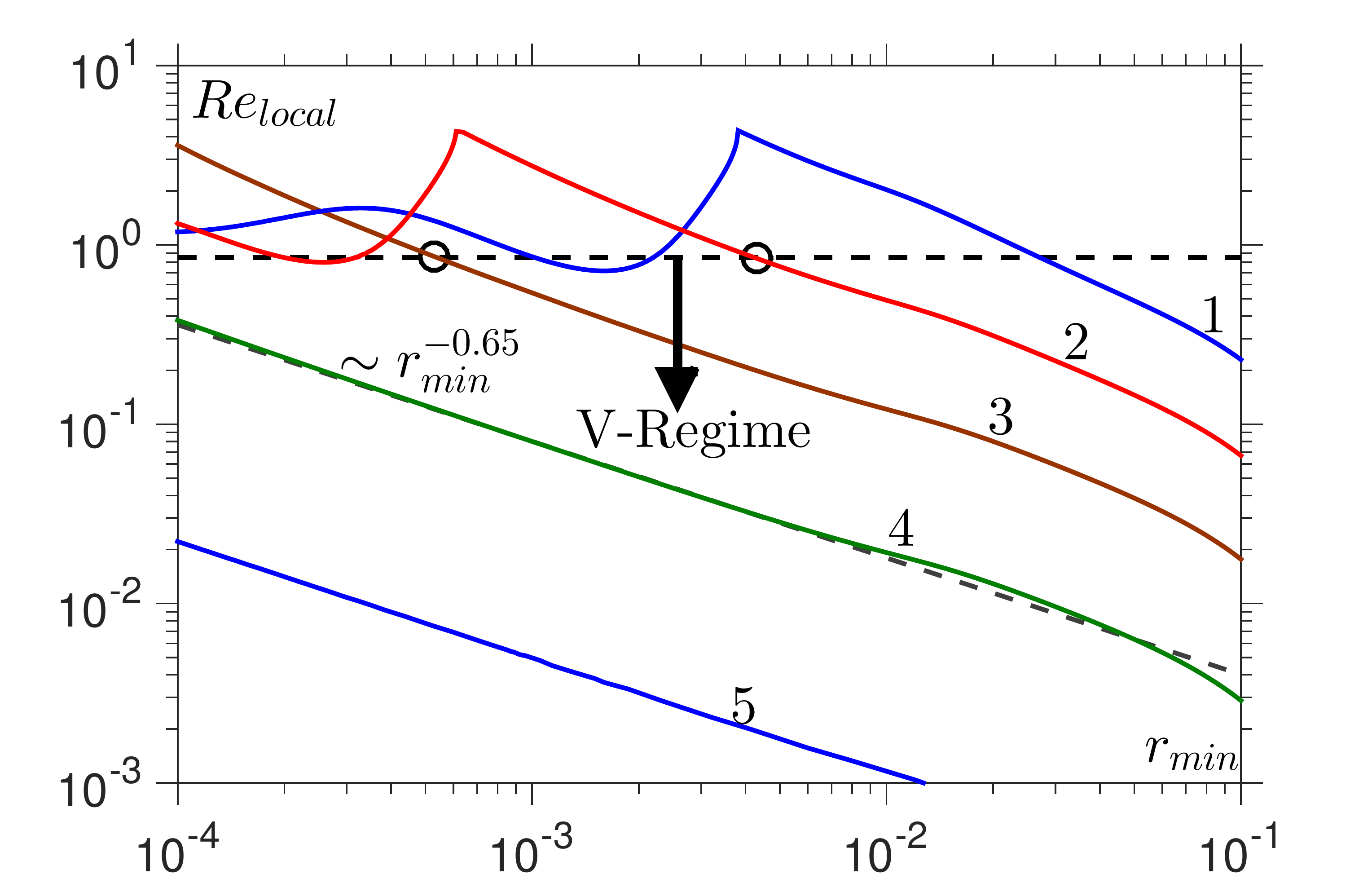}\label{F:Re_local_Ohg0p15}}
\caption{(a) The similarity solution in the V-regime ($\dot{r}_{min}=-0.071$) is shown as a dashed line.  At higher $Oh$ (e.g. curve 5) the speed remains at $-0.071$ whilst at lower values (e.g. curve 1), there is a transition towards the VI-regime where $\dot{r}_{min}=-0.030$. (b) Curves show in the V-regime $Re_{local}$ gradually increases until a critical value is achieved at which point the V-regime is exited.  The increase of $Re_{local}$ in the V-regime is shown to follow the predicted scaling from (\ref{papa}) of $Re_{local}\sim r_{min}^{-0.65}$ (lower dashed line).The transition out of the V-regime ($r_{min}=r_{min}^{V\to}$ marked as circles) is found to occur when $Re_{local}\approx0.85$ (horizontal dashed line). }
\end{figure}

When $Oh\gg1$ the breakup is expected to start in the V-regime before transitioning into the VI-regime.  Figure~\ref{F:highOh_umin}, shows that when $Oh=10$ (curve 5), the breakup speed $\dot{r}_{min}$ reaches the value predicted in the V-regime, of $-0.071$, at around $r_{min}=10^{-2}$ and remains there until $r_{min}=10^{-4}$, i.e.\ until the end of the computation. For higher Oh, curves are graphically indistinguishable from the case of $Oh=10$, so this is the Stokes flow solution.  Therefore, for $r_{min}\ge10^{-4}$ the VI-regime is not observed at large $Oh$.  

\begin{figure}
     \centering
\includegraphics[scale=0.26]{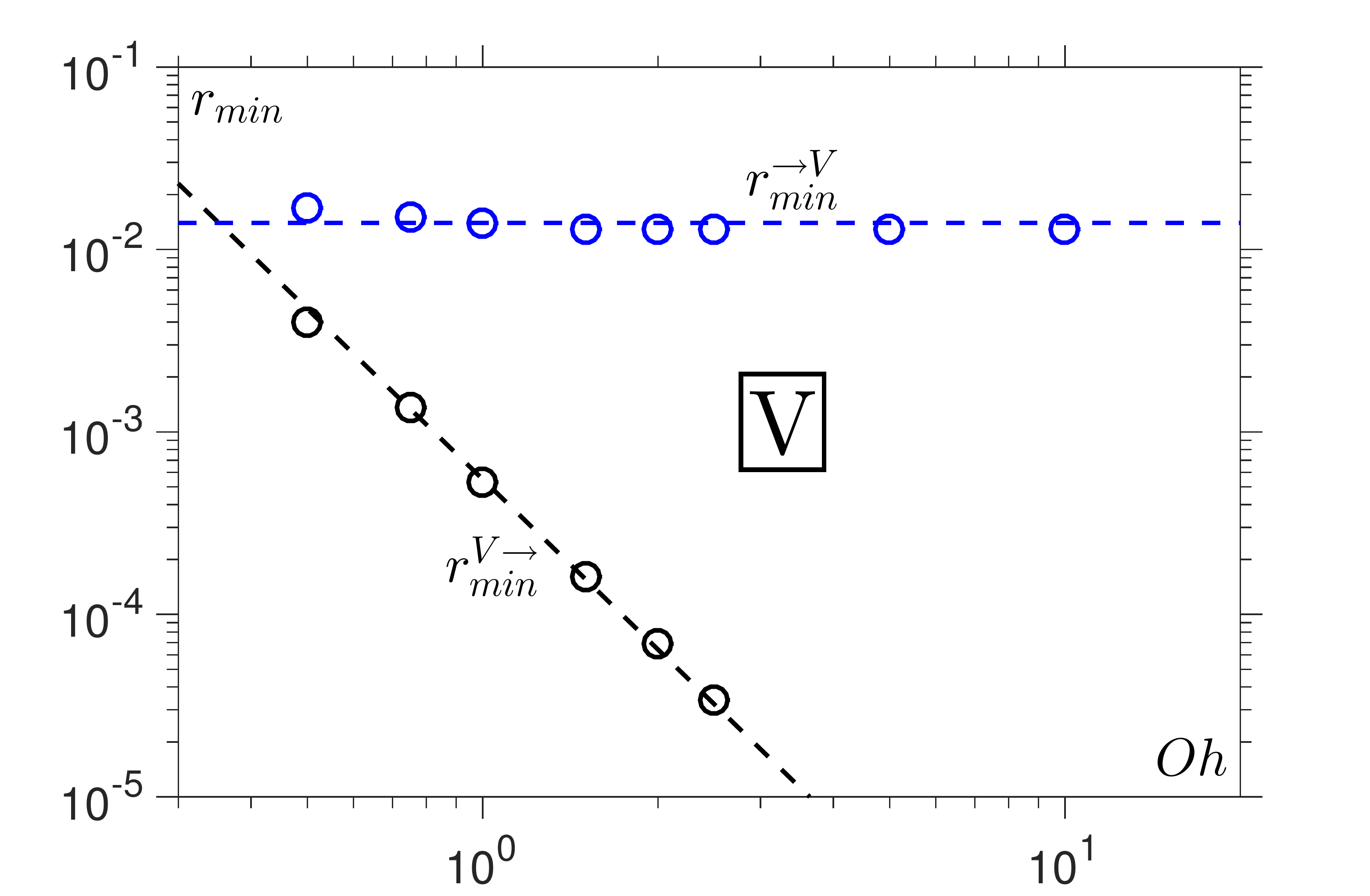}
\caption{Phase diagram showing the existence of a V-regime where there is a balance between viscous and capillary forces.  The transition $r^{\to V}_{min}$ is where the breakup enters this regime and $r^{V \to}_{min}$ is where it leaves.  Computational results (circles) show that the entrance is at a constant $Oh$-independent $r^{\to V}_{min}=0.014$ (horizontal dashed line) whilst the exit follows $r^{V\to}_{min}=5.5\times10^{-4}~Oh^{-3.1}$ (lower dashed line).}
\label{F:highOh_transition}
\end{figure}

Figure~\ref{F:highOh_transition}, shows the region of $Oh$-$r_{min}$ phase space where V-regime dynamics are encountered, with the circles marking the computationally determined boundaries of this regime.  The procedure for identifying this regime is shown in the Appendix and, as discussed in \S\ref{S:quant_regimes}, it is based on the speed of breakup $\dot{r}_{min}$ remaining between $-0.071\pm0.015$.  Next, the changes in behaviour triggering the entrance and exit from this V-regime are considered.

\subsubsection{Entrance into the V-Regime}

\begin{figure}
     \centering
\includegraphics[scale=0.26]{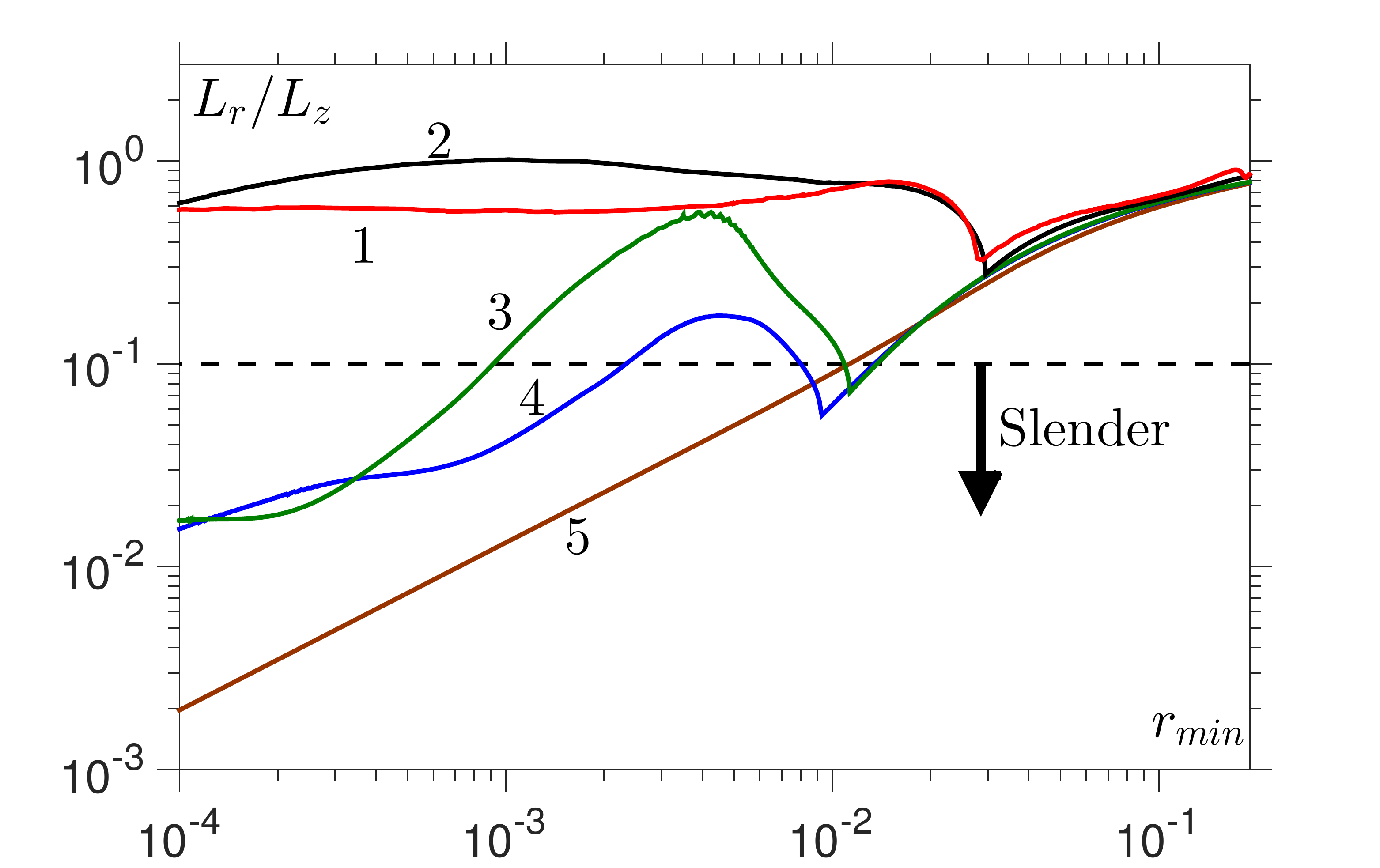} 
\caption{Evolution of the slenderness of the free-surface in the breakup region for 1: $Oh=10^{-3}$,  2: $Oh=10^{-2}$,  3: $Oh=0.14$,  4: $Oh=0.16$,  5: $Oh=10$. This quantity, which must be small for the thread to be slender (as is required by the V- and VI-regimes) is defined using $L_r=r_{min}$ and $L_z=|z(r=r_{min})-z(w=w_{max})|$.}
\label{F:slenderness}
\end{figure}

The V-regime is entered ($r^{\to V}_{min}$) at a constant $Oh$-independent value $r_{min}\approx10^{-2}$ which suggests that this transition may occur when the thread can be considered slender, so that the assumptions behind the similarity solution (\ref{papa}) become valid.  To determine whether the geometry near the breakup point is `slender' or not, we plot $L_r/L_z$ in Figure~\ref{F:slenderness}.  For $Oh=10$ (curve 5), and generally for larger $Oh$, initially $L_r/L_z \sim 1$ (no scale separation) but as $r_{min}$ shrinks the breakup region becomes slender. Defining the region to be slender when $L_r/L_z<0.1$, it is found that for all $Oh>1$ slenderness occurs at $r_{min}\approx10^{-2}$ in agreement with our result for $r^{\to V}_{min}$.  Therefore, the transition into the V-regime appears to be dictated by the geometry of the thread.  
%Crossover now based on \pm 0.015
%Oh = [0.5 0.75 1 1.5 2 2.5 5 10]
%r_in= [0.017 0.015 0.014 0.013 0.013 0.013 0.013 0.013];
%r_out = [4.3e-3 1.36e-3 5.3e-4 1.6e-4 6.9e-5 3.4e-5 -10000 -1000]; 
%No effect of shorter time step

\subsubsection{Exit from the V-Regime}

The V-regime is left when the speed of breakup $\dot{r}_{min}$ diverges from $-0.071\pm0.015$ (e.g.\ curve 1 in Figure~\ref{F:highOh_umin}).  Figure~\ref{F:highOh_transition} shows that this transitions follows the proposed scaling for V$\to$VI that $r_{min}=A~Oh^{-3.1}$, with computations finding $A=5.5\times10^{-4}$.   

The transition out of the V-regime ($r_{min}^{V\to}$) occurs when inertial effects become non-negligible, so that (\ref{papa}) is no longer valid.  Figure~\ref{F:Re_local_Ohg0p15} shows the increase in $Re_{local}$ follows remarkably well the scaling predicted by the similarity solution for the V-regime that $Re_{local}\sim r_{min}^{-0.65}$.  This also serves to validate our definition of $Re_{local}$.  Circles placed at $r_{min}^{V\to}$ on curves in Figure~\ref{F:Re_local_Ohg0p15} show that the transition out of the V-regime occurs when $Re_{local}\approx 0.85$.  One may expect that at this point, where $Re_{local}\sim1$, the VI-regime will be encountered, and this will be the focus of the next section.
%
%%%rho.U.L/sigma = Re.U_l.L_l = (1/Oh^2) U_l.L_l
%Crossover now based on \pm 0.015
%Oh = [0.5 1]
%r_out = [4.3e-3 5.3e-4]; 
%Re_local_there=[0.843 0.865]

\section{Viscous-Inertial Balance (Intermediate $Oh$)}\label{S:VI-regime}

\begin{figure}
     \centering
\subfigure[Free-surface evolution.]{\includegraphics[scale=0.4]{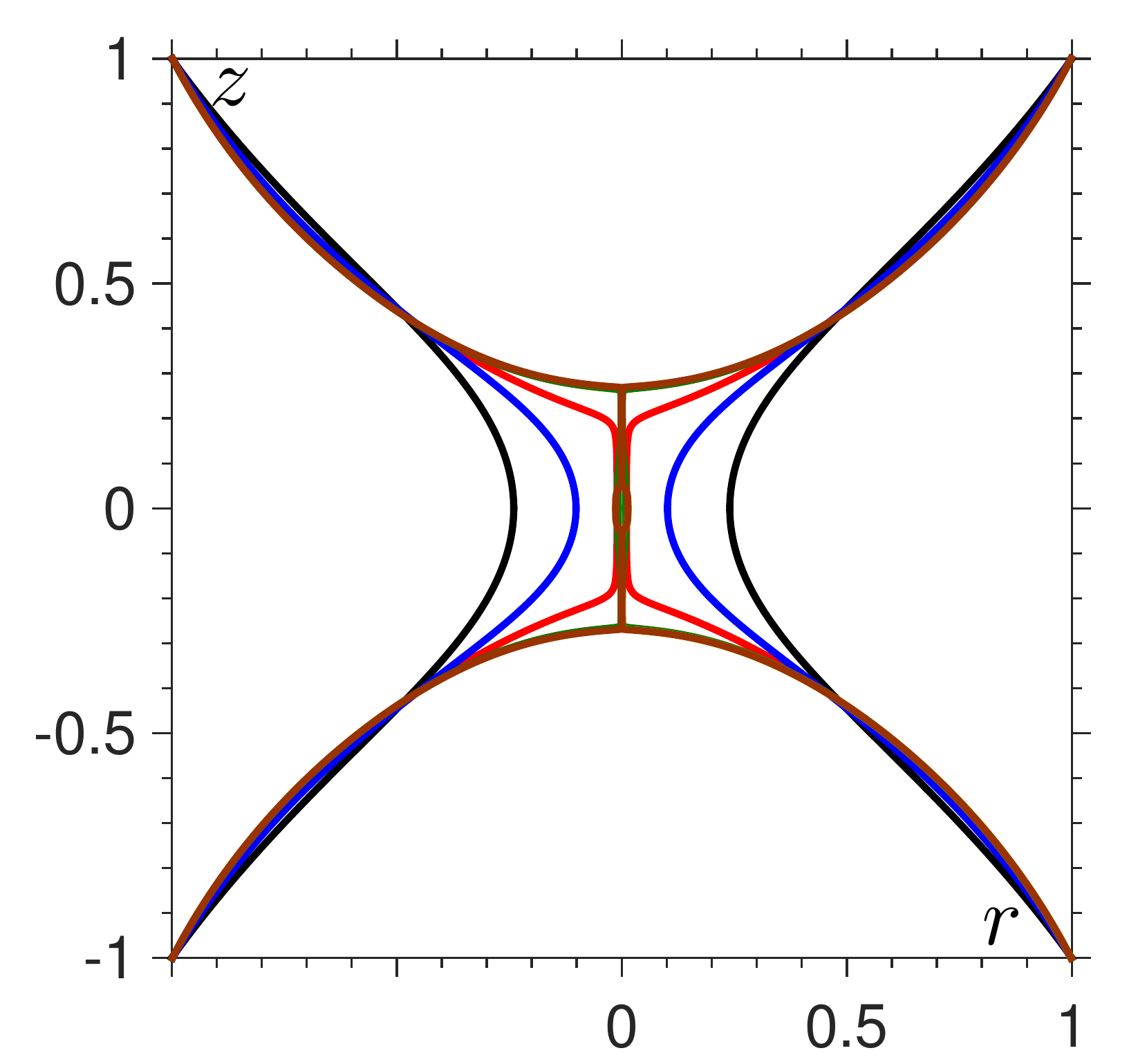}}\\
\subfigure[Close-up of breakup.]{\includegraphics[scale=0.3]{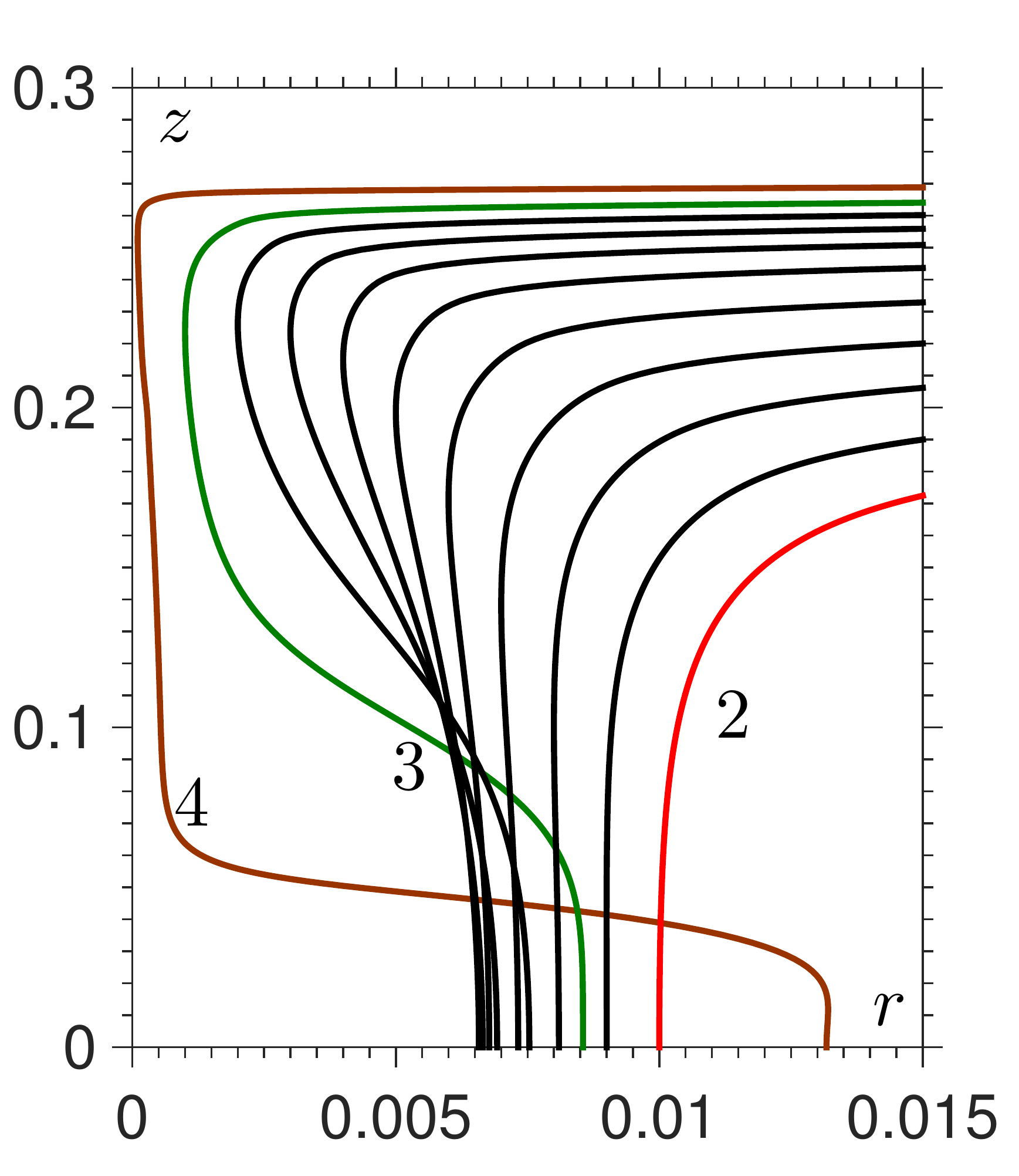}}
\subfigure[Axial velocity.]{\includegraphics[scale=0.3]{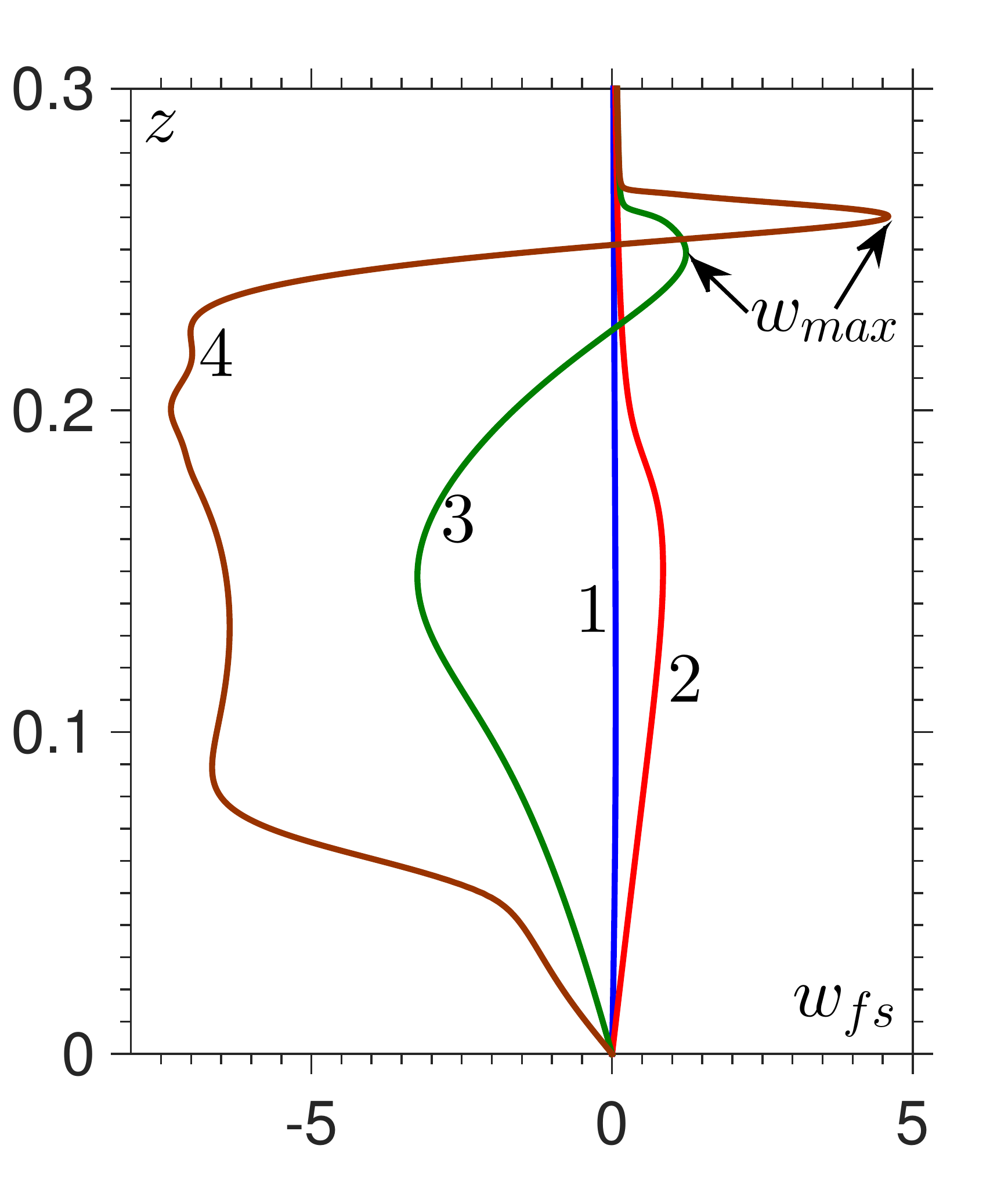}} 
\subfigure[Pressure.]{\includegraphics[scale=0.3]{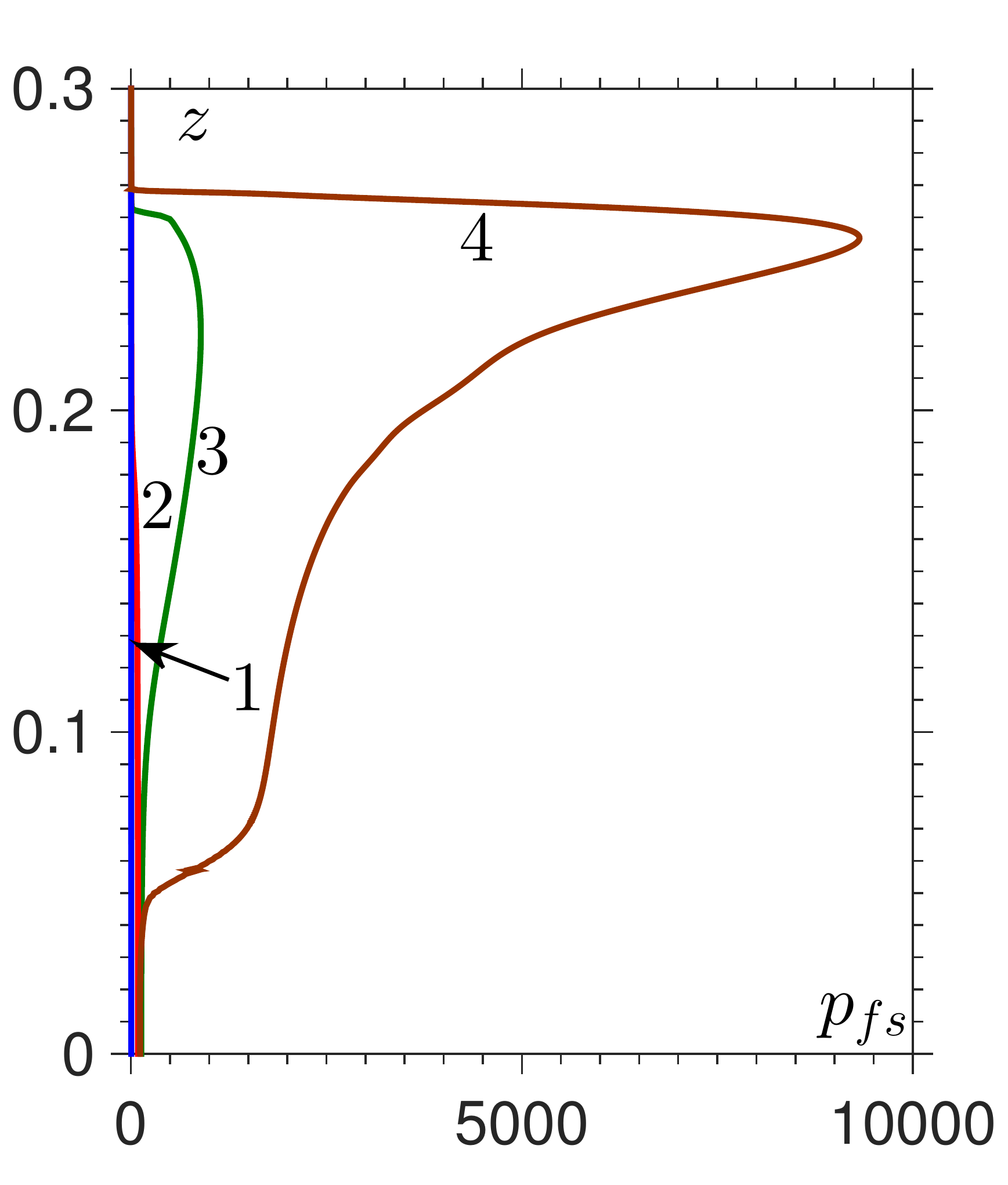}}
\caption{For $Oh=0.16$ plots show (a) the entire free-surface, (b) a close-up of the breakup region, (c) the axial velocity at the free-surface $w_{fs}$ (with $w_{max}$ when $r_{min}=10^{-3},~10^{-4}$ shown) and (d) the pressure at the free-surface $p_{fs}$ at 1: $r_{min}=10^{-1}$ (blue), 2: $r_{min}=10^{-2}$ (red), 3: $r_{min}=10^{-3}$ (green) and 4: $r_{min}=10^{-4}$ (brown).}
\label{F:Oh0p16}
\end{figure}

At $Oh=0.16$ in Figure~\ref{F:Oh0p16}(a,b) one can see the appearance of a satellite drop centred at $z=0$. The close up images show how this is driven by symmetry breaking about the pinch point after $r_{min}\approx10^{-2}$ (curve 2), as observed experimentally in \cite{rothert01}.  Consequently, two pinch points occur at a finite distance either side of the symmetry plane ($z=0$) with a large pressure acting to push fluid out of the thinnest region (Figure~\ref{F:Oh0p16}(d)).  In contrast to profiles in the I-regime (c.f.\ \S\ref{S:I-regime}), the geometry near the pinch point remains slender, see curve 4 in Figure~\ref{F:Oh0p16}(b) and the slenderness data in Figure~\ref{F:slenderness} (curve 4).  

Features that indicate the breakup is occurring in the VI-regime are as follows.  Figure~\ref{F:fs_vs_similarity}(a) shows that the free surface profile at $r_{min}=10^{-4}$ compares well with the similarity solution (dashed line) derived in \cite{eggers93} with no free parameters.    Below the pinch point ($z<0.25$) in Figure~\ref{F:fs_vs_similarity}, deviations of the computed free surface shape from the similarity solution occur as there is no `far field' in our setup (as the plane of symmetry is at $z=0$).  This also affects the velocity profiles (see Figure~\ref{F:Oh0p16}(c)) which agree with the similarity profiles in \cite{eggers93} above the pinch point but do not below; a point will be discussed further in \S\ref{S:bumps}. However, curve 2 in Figure~\ref{F:wmax} shows that the maximum axial velocity follows $w_{max}=0.3~Oh~r_{min}^{-0.5}$ from (\ref{eggers}), again with no adjustable parameters.    %

\begin{figure}
     \centering
\subfigure[Evolution of the speed of breakup when $Oh>Oh_c$.]{\includegraphics[scale=0.26]{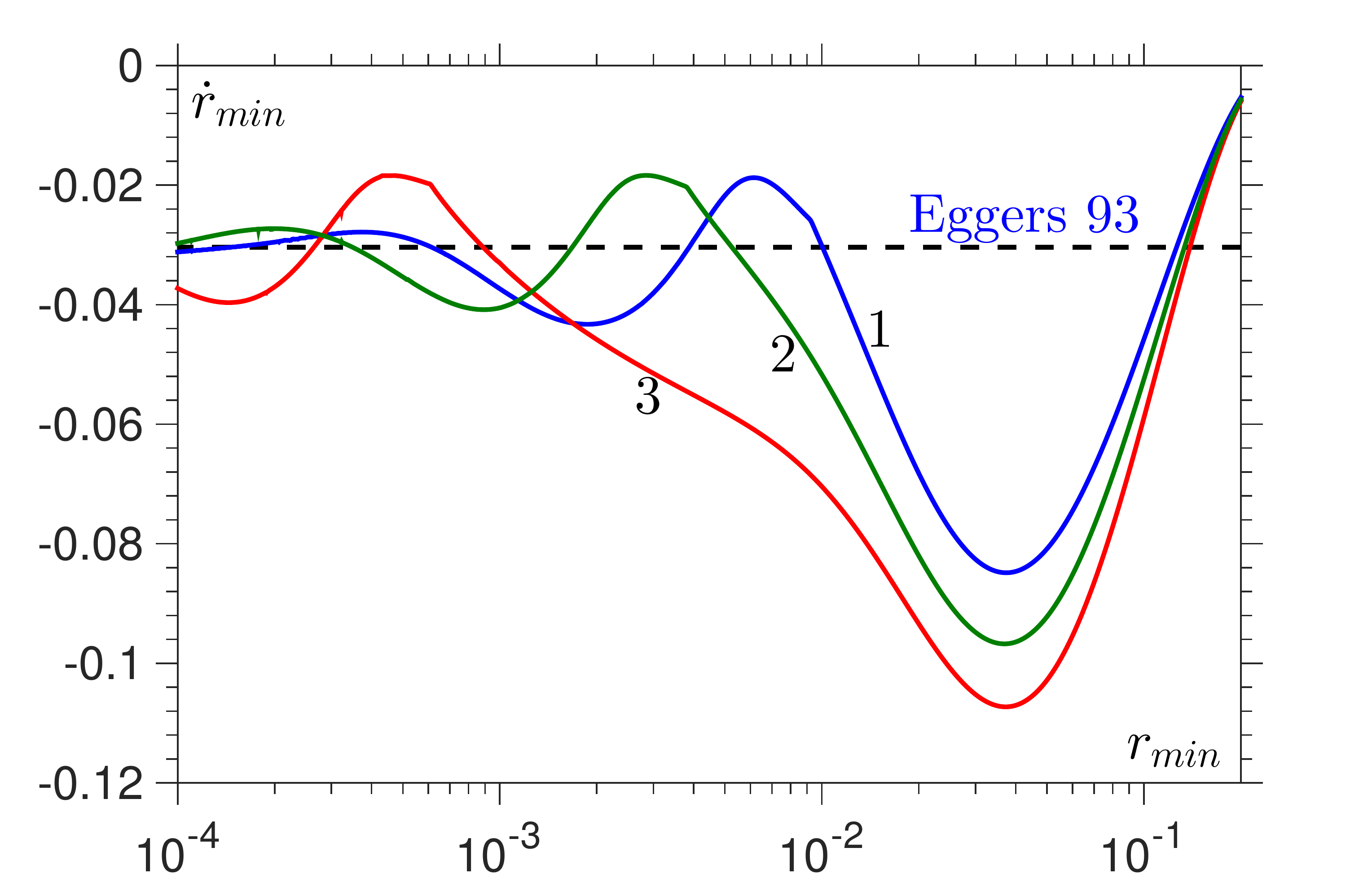}\label{F:Ohg0p15_rdotmin}} 
\subfigure[Comparison of convergence towards the breakup speeds of the V and VI regimes.]{\includegraphics[scale=0.26]{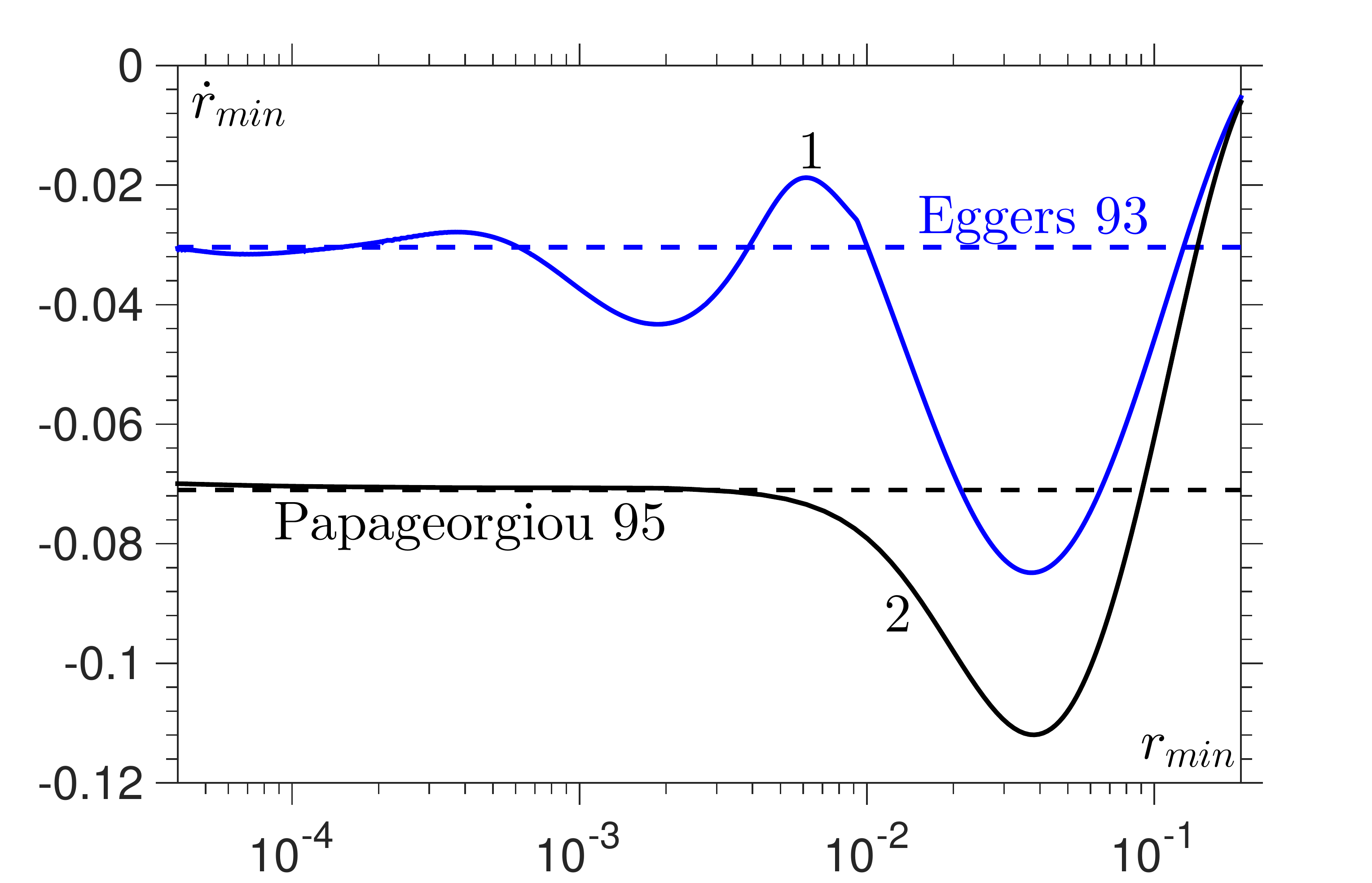}\label{F:VvsVI}}
\caption{(a) Curves are for 1: $Oh=0.16$, 2: $Oh=0.25$, 3: $Oh=0.5$.  At $Oh>Oh_c$ curves converge towards the similarity solution in the VI-regime given by (\ref{eggers}), i.e. that $\dot{r}_{min}=-0.030$. (b) The different dynamics observed for the breakup speed in the VI- and V-regimes, with 1: $Oh=0.16$, 2: $Oh=10$.  For these values of $Oh$, the computed solutions converge 1: oscillatorily and 2: monotonically towards similarity solutions in the VI- and V-regimes, given by (\ref{eggers})  and  (\ref{papa}), respectively, which are the lower and upper dashed lines.}
\end{figure}

In Figure~\ref{F:Ohg0p15_rdotmin}, the breakup speed is shown for $Oh>0.15$, values whose significance will become apparent. In this range, all curves tend towards the value of $-0.030$, indicating that breakup enters the VI-regime.
%
%%%rho.U.L/sigma = Re.U_l.L_l = (1/Oh^2) U_l.L_l
\begin{figure}
     \centering
\includegraphics[scale=0.26]{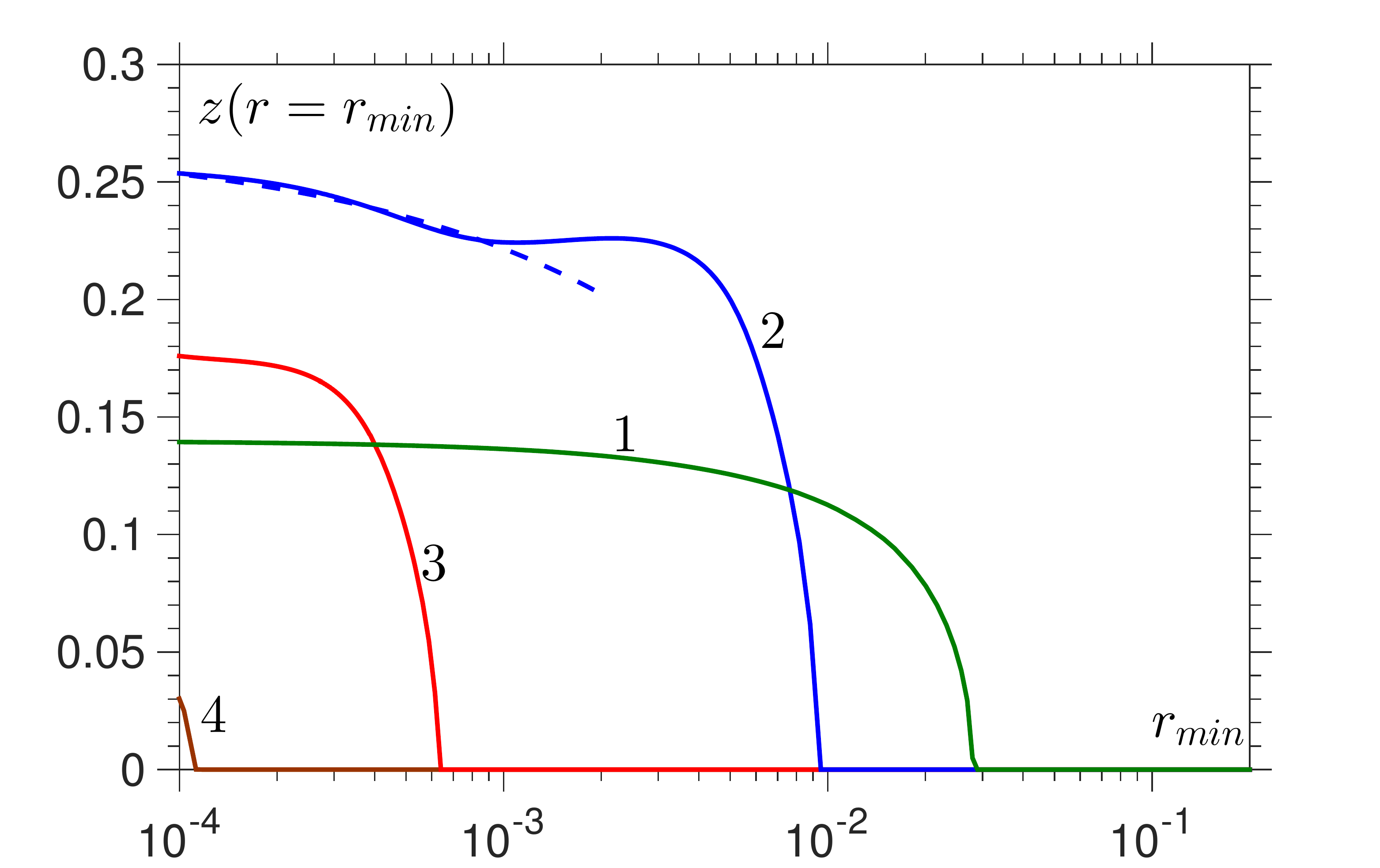}
\caption{Evolution of the vertical position of the pinch point, i.e.\ minimum thread radius $z(r=r_{min})$, indicating the formation of satellite drops once  the pinch point moves away from the centre line $z(r=r_{min}) >0$.  Curves are for 1: $Oh=10^{-3}$, 2: $Oh=0.16$, 3: $Oh=0.5$ and 4: $Oh=1$.  The dashed line shows the expression from \cite{eggers93} at $Oh=0.16$ for the drift of the pinch point $z(r=r_{min})=z_0-1.6~Oh~\tau^{-1/2}$, with $z_0=0.268$.}
\label{F:zmin}
\end{figure}

What is immediately striking about the curves in Figure~\ref{F:Ohg0p15_rdotmin} is that they appear to oscillate around the value of $-0.030$, converging towards it as $r_{min}\to0$.  To highlight this behaviour, Figure~\ref{F:VvsVI} compares curves for $Oh=10$, which remains in the V-regime, and $Oh=0.16$, where the VI-regime is most prominent. Whilst the value of $-0.071$ is approached monotonically from below by curve 2, the approach of curve 1 towards $-0.030$ is entirely different.  The motion in the VI-regime is clearly oscillatory in the radial direction and on closer inspection of Figure~\ref{F:wmax} and Figure~\ref{F:zmin} oscillations in axial quantities $w_{max}$ (and hence $Re_{local}$ in Figure~\ref{F:Re_local_Ohg0p15}) and $z(r=r_{min})$ can also be seen.  Having identified this behaviour, it can also be recognised in the plot of $r_{min}$ against $\tau$ in curve 2 of Figure~\ref{F:rvst}, with the bridge radius oscillating around and converging towards the dashed line predicted by the similarity solution (\ref{eggers}).  As far as we are aware, this is the first time this oscillatory behaviour has been observed and it will be discussed further in \S\ref{S:oscillations}. 

\begin{figure}
     \centering
\subfigure[Evolution of the breakup speed around the critical point $Oh_c=0.15$  and appearance of the low-$Oh$ V-regime.]{\includegraphics[scale=0.26]{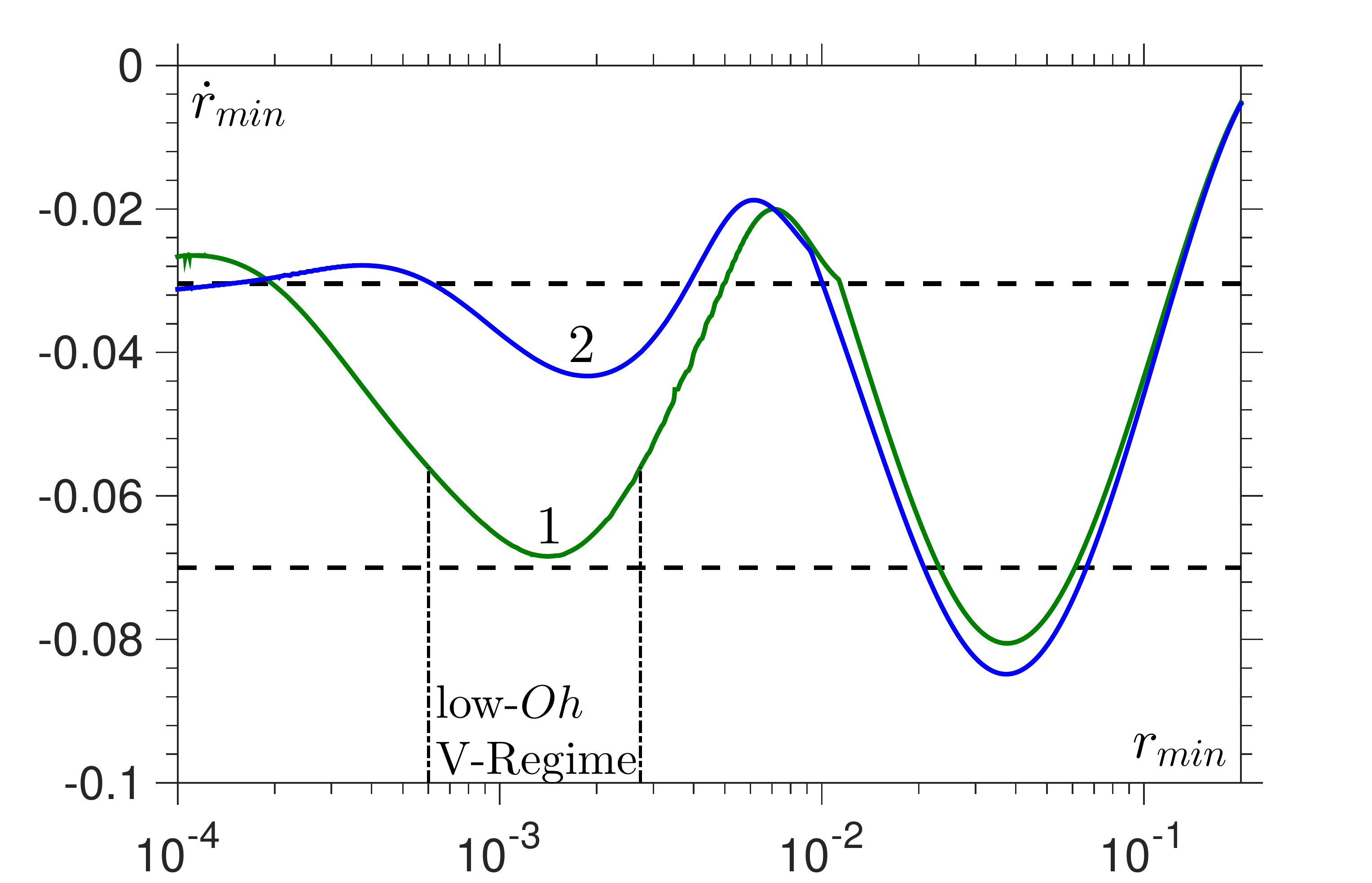}\label{F:Ohl0p15_rdotmin}}
\subfigure[Variations in the local Reynolds number near $Oh_c$.]{\includegraphics[scale=0.26]{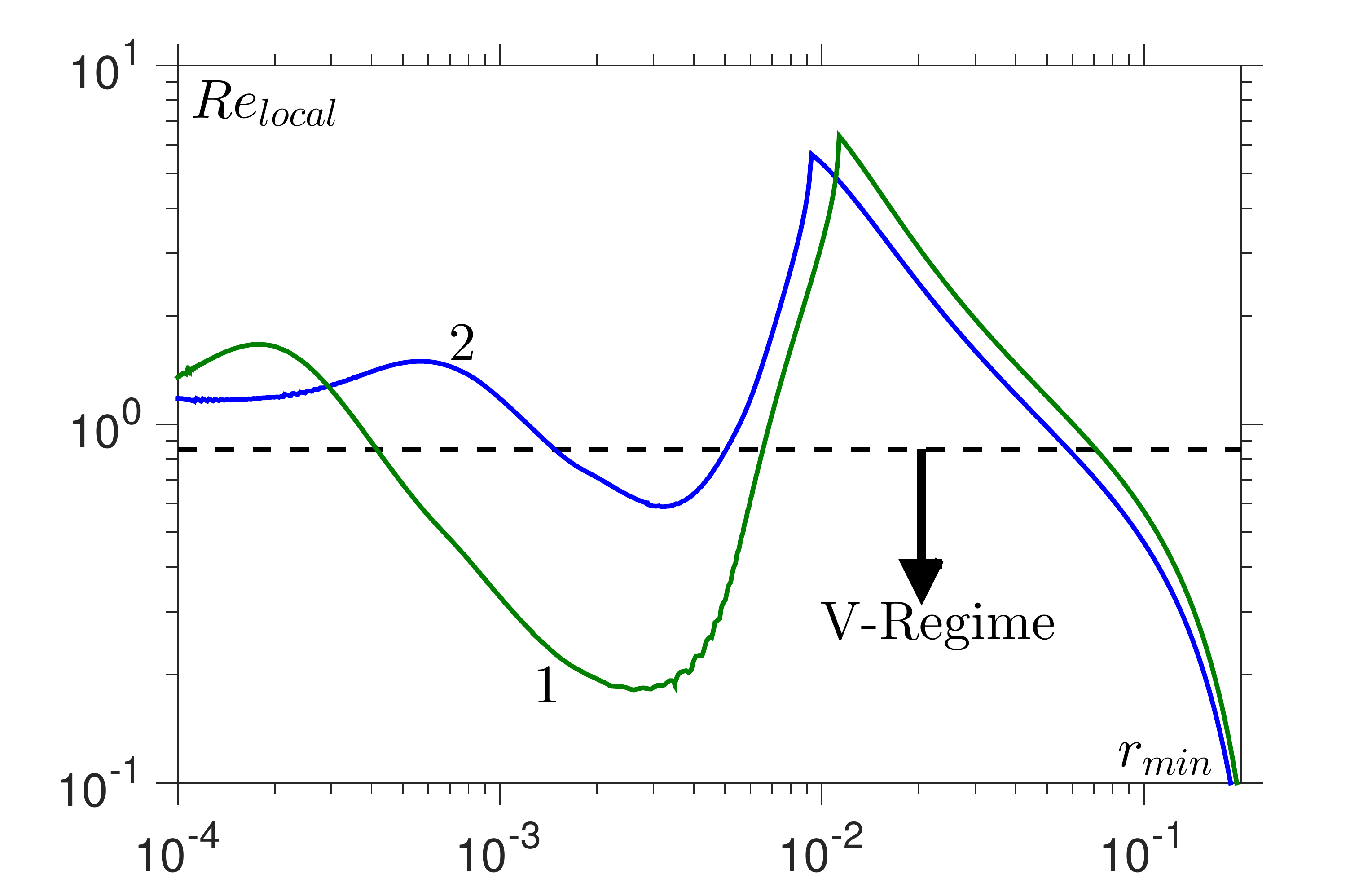}\label{F:Re_local_Ohl0p15}} 
\caption{ Curves are for 1: $Oh=0.14$ and 2: $Oh=0.16$.  (a) Changes in the evolution of the breakup speed $\dot{r}_{min}$, against $r_{min}$, around the critical point $Oh=0.15$.  Curve 1 shows the appearance of the low-$Oh$ V-regime around $r_{min}\approx 10^{-3}$ where the breakup speed dips to $\dot{r}_{min}\approx -0.071$ in contrast to curve 2 where the speed immediately tends towards the value from the VI-regime of $-0.030$. (b) The local Reynolds number $Re_{local}$ for the same values of $Oh$ shows that the low-$Oh$ V-regime coincides with a drop in $Re_{local}$.  The dashed line of $Re_{local}=0.85$ is the value below which V-regime dynamics are to be expected.}
\end{figure}

Figure~\ref{F:Ohl0p15_rdotmin} focuses on the narrow range $Oh=0.14-0.16$ to highlight a sharp transition in the dynamics of breakup which occurs at a critical $Oh_c\approx0.15$.  For $Oh=0.16>Oh_c$ (curve 2), the transition into the VI-regime occurs at $r_{min}\approx10^{-2}$ whilst for $Oh=0.14<Oh_c$ (curve 1) the transition into this regime is delayed until $r_{min}\approx4\times10^{-4}$.  Remarkably, we will show that this is due to the existence of a `low-$Oh$ V-regime', first discovered in \cite{castrejon15}, which precedes entry into the VI-regime for $Oh<Oh_c$. In Figure~\ref{F:Ohl0p15_rdotmin} for $Oh=0.14$ (curve 1), it can be seen that this recently discovered regime is entered  when $r^{\to V}_{min}=3\times10^{-3}$ and exited again when $r^{V\to}_{min}=6\times10^{-4}$ during which period the speed of breakup is in the V-regime range of $-0.071\pm0.015$.

\begin{figure}
     \centering
\subfigure[Free surface evolution at $Oh=0.14$]{\includegraphics[scale=0.26]{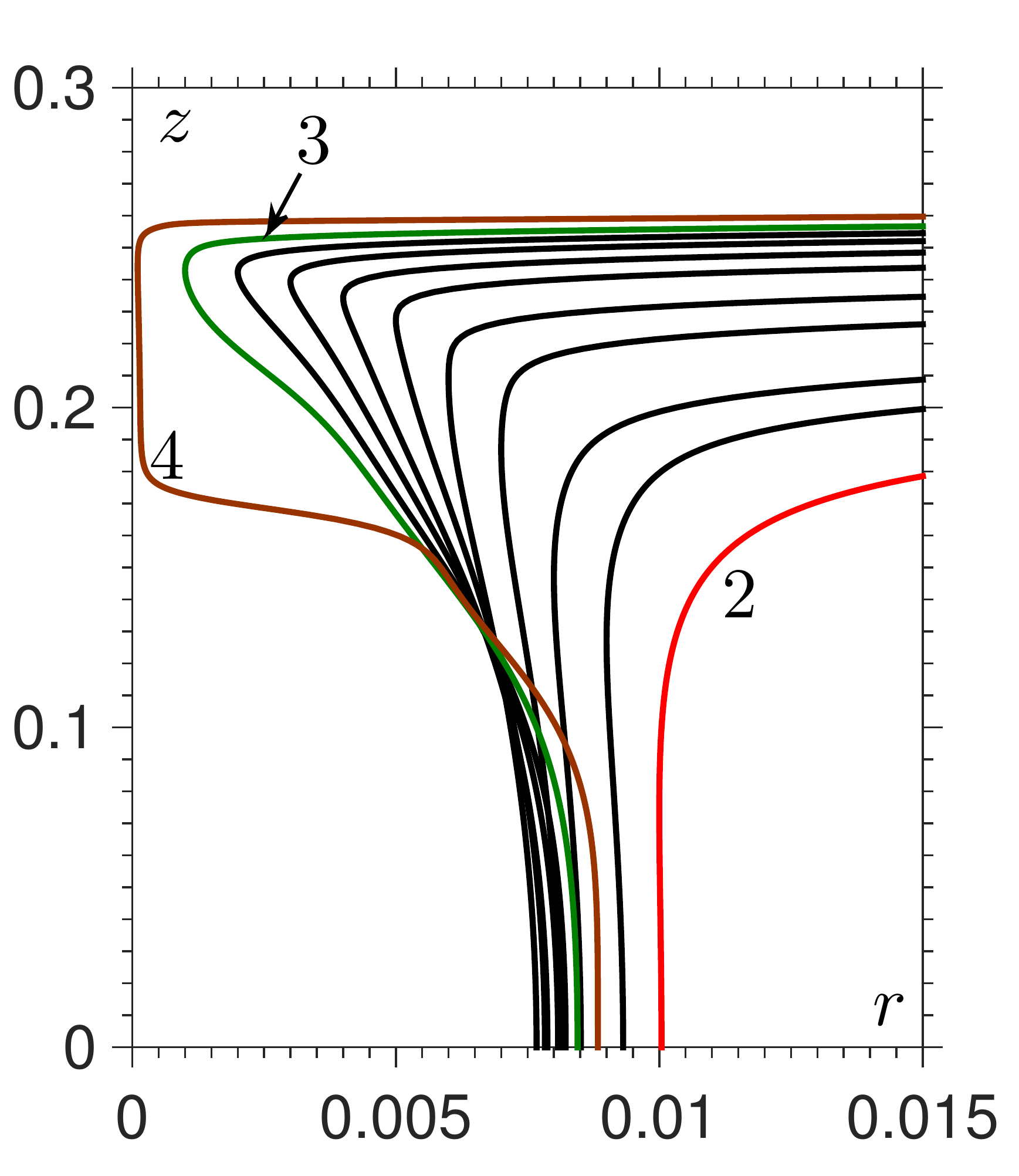} \label{F:Ohp14_fs}}
\subfigure[Free surface shapes above and below $Oh_c$]{\includegraphics[scale=0.26]{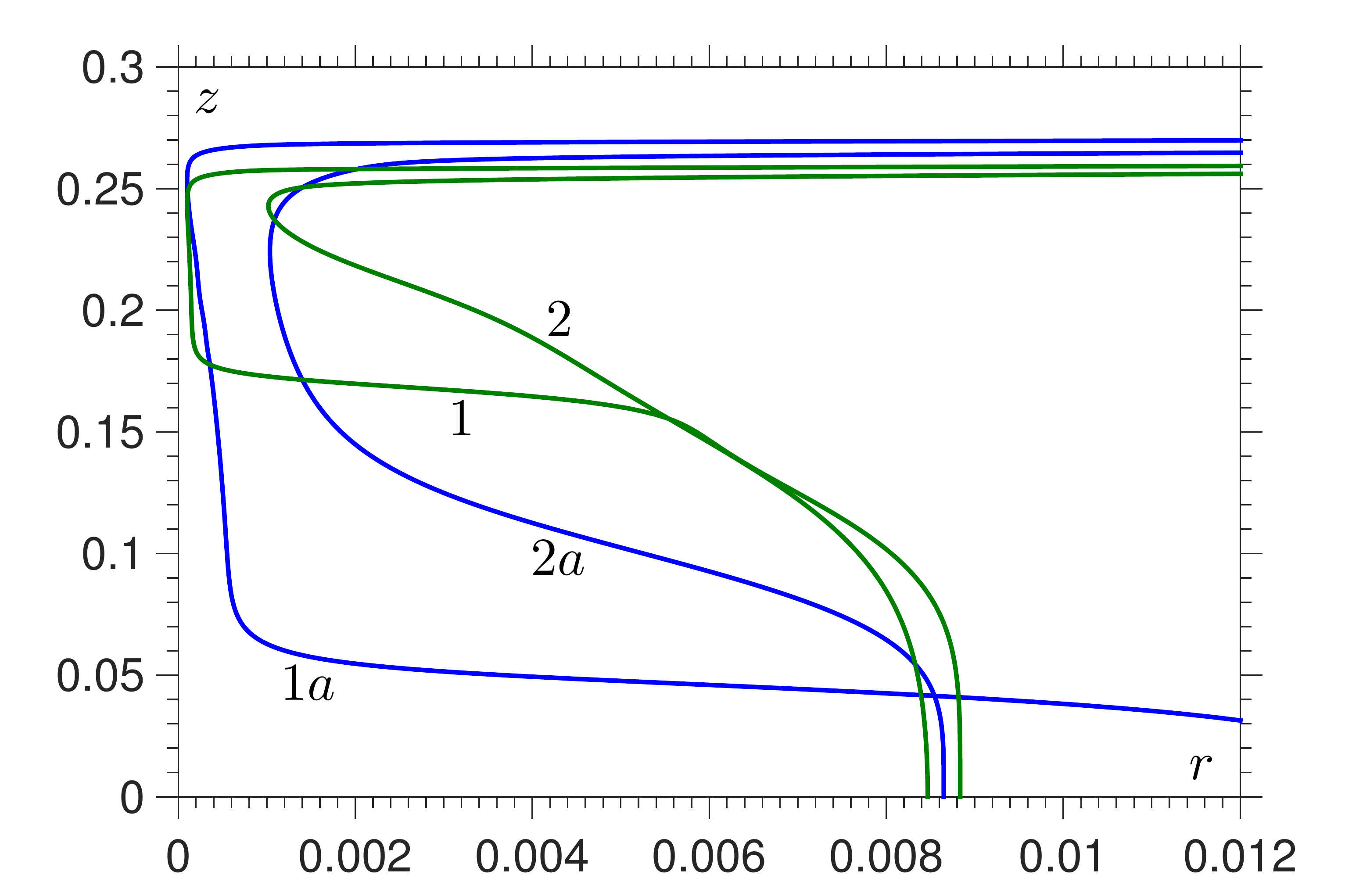} \label{F:evolution_Oh0p16_vs_Oh0p14}}
\caption{(a) Evolution of the free-surface for $Oh=0.14$ showing the development of a corner-like geometry (curve 3) while the flow is in the low-$Oh$ V-regime followed by the development of a thin thread once the VI-regime is entered (curve 4).  (b) Comparison of free-surface shapes below the critical point at $Oh=0.14$ (curves 1, 2) and above it $Oh=0.16$ (curves 1a, 2a) at 1,1a: $r_{min}=10^{-4}$ and 2,2a: $r_{min}=10^{-3}$ .}
\end{figure}

The transition in flow behaviour at $Oh_c$ is also seen from the free-surface shapes in the vicinity of the breakup region.  In Figure~\ref{F:evolution_Oh0p16_vs_Oh0p14} significant differences in the breakup geometry are shown for Ohnesorge numbers just below ($Oh=0.14$) and above ($Oh=0.16$) the critical value.  At $r_{min}=10^{-4}$, for $Oh=0.16$ (curve 1a) a long slender thread of length $\approx 0.22$ connects the hemispherical volume above ($z>0.27$) to a small satellite drop below ($z<0.05$) whilst at $Oh=0.14$ (curve 1) the thread is much shorter at $\approx 0.07$. Consequently, the satellite drops formed in each case differ considerably, with $Oh=0.16$ forming a much fatter drop connected to a longer thin thread.  
%In fact, going from $Oh=0.16$ down to $Oh=0.14$, the satellite drop volume increases by a remarkable $50\%$

The free-surface profile in the low-$Oh$ V-regime (curve 2) does not have the slender symmetric shape in the breakup region that is characteristic of the conventional V-regime, which suggests that this regime is encountered in a weaker sense than the high-$Oh$ V-regime which satisfies all expected characteristics (see \S\ref{S:V-regime}). Using more stringent criteria for how the regimes are defined, could result in this becoming a region of phase space where none of the similarity solutions are accurate; however, our definition is based on the breakup speed alone and this identifies it as a low-$Oh$ V-regime.

\subsection{Identification of Regimes}

\begin{figure}
     \centering
\includegraphics[scale=0.26]{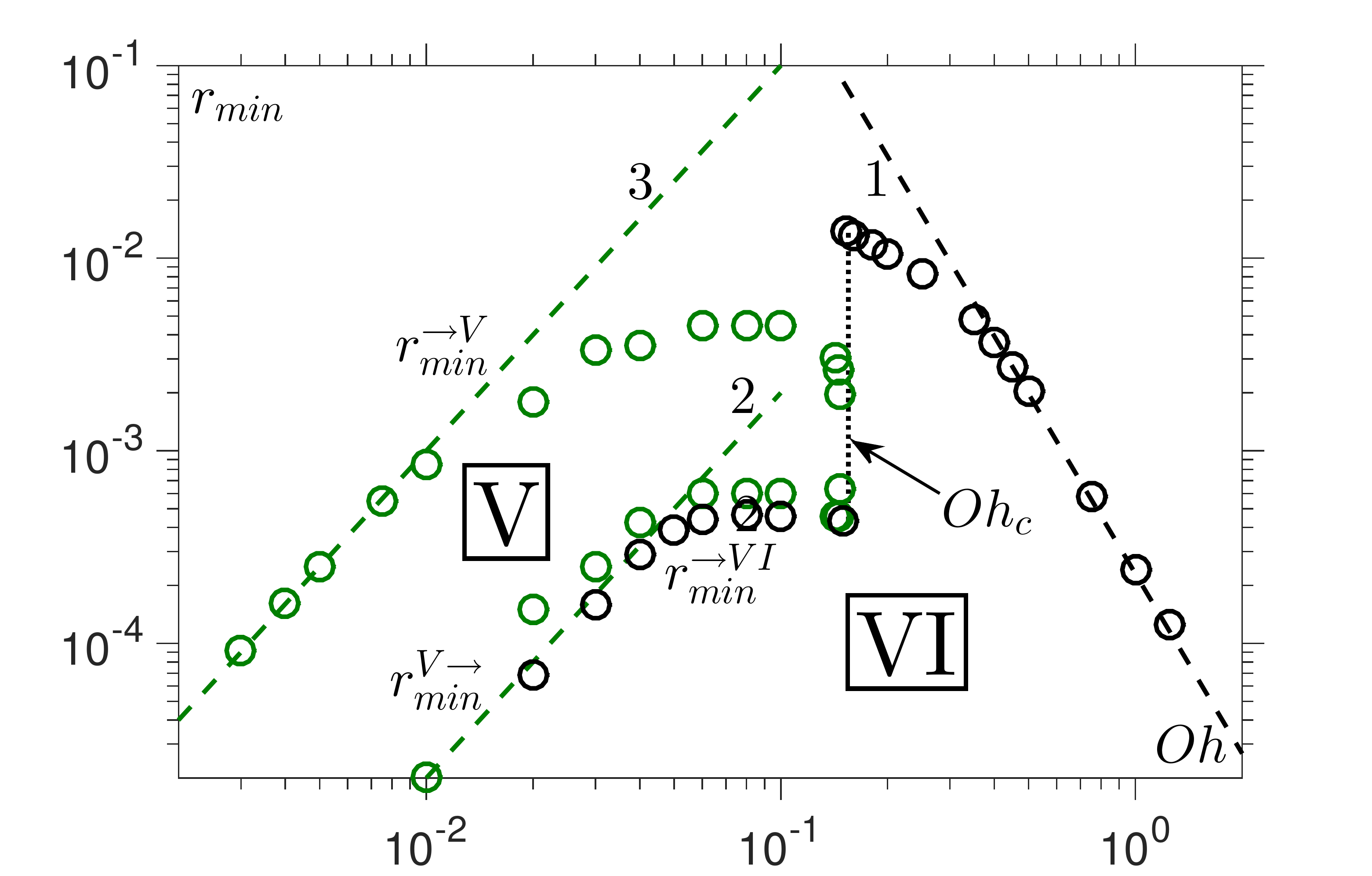}
\caption{Phase diagram showing the low-$Oh$ V-regime and the VI-regime.  The transition $r^{\to VI}_{min}$ follows the scaling $r_{min}=2.3\times10^{-4}~Oh^{-3.1}$ (curve 1) for $Oh>Oh_c=0.15$.  For $Oh<Oh_c$ the low-$Oh$ V-regime appears and is bounded by curves 2: $r_{min}=0.2~Oh^2$ and 3: $r_{min}=10~Oh^2$ as $r_{min}\to0$.}
\label{F:midOh_transition}
\end{figure}

In Figure~\ref{F:midOh_transition} the VI and low-$Oh$ V-regimes are shown on a phase diagram. The extreme change in behaviour at $Oh_c=0.15$ is clearly visible. 
%
%Crossover based on 0.015-0.045
%Oh=[0.02 0.03 0.04 0.05 0.06 0.08 0.1 0.15               0.15 0.16 0.18 0.2 0.25 0.35 0.4 0.45 0.5 0.75 1 1.25];		
%rc=[6.8e-5 1.58e-4 2.88e-4 3.85e-4 4.4e-4 4.64e-4 4.55e-4 4.3e-4             1.37e-2 1.37e-2 1.31e-2 1.18e-2 1.06e-2 8.34e-3 4.78e-3 3.64e-3 2.74e-3 2.02e-3 5.8e-4 2.4e-4 1.25e-4]; 
%
%Reynolds number at which VI-regime is entered:   %%%This doesn't tell us much really and therefore is omitted%%%
%Oh = 0.25, 0.5, 1 - 
%rc = 8.34e-3, 2.02e-3, 2.4e-4
%Re = 2.27, 1.51, 1.62
%
%
%Low-Oh V-Regime:
%Oh=[0.003 0.004 0.005 0.0075 0.01 0.02                 0.03 0.04 0.06 0.08 0.1 0.14]
% rin = [9.2e-5 1.6e-4 2.5e-4 5.5e-4 8.5e-4 1.8e-3      3.32e-3  3.5e-3  4.5e-3 4.5e-3 4.5e-3 <-???  2.73e-3] 
% rout=[1e-10 1e-10 1e-10 1e-10 2e-5 1.5e-4            2.5e-4 4.23e-4 6e-4 6e-4 6e-4 <-???  6e-4]
%
%New VI:
%Oh=[0.14]
% rin = [3.8e-4] 
%
%New Low-Oh V-Regime:   - 0.145 is not in low-oh anymore
%Oh=[0.02 0.03 0.04 0.06 0.08 0.1 0.12 0.14		]
% rin = [1.9e-3 3.9e-3 3.6e-3 4.2e-3 4.5e-3  4.4e-3 3.9e-3 2.73e-3	] 
% rout=[1.4e-4  2.4e-4 2.7e-4 4.1e-4 6.1e-4 6.7e-4 6.4e-4 5.6e-4	] ???
%
%rineg=[6.6e-5 1.4e-4 4.4e-4 4.8e-4 3.9e-4 4.5e-4
\subsubsection{Entrance into the VI-Regime for $Oh>Oh_c$}

For $Oh>Oh_c$, the transition into the VI-regime follows the expected scaling for V$\to$VI of $r_{min}=B~Oh^{-3.1}$, with $B=2.3\times10^{-4}$. Curves 1-3 in Figure~\ref{F:Re_local_Ohg0p15} show that as expected $Re_{local}$ increases until it reaches $\approx1$, a value characteristic of the VI-regime.  The sharp changes in the gradients of these curves, at the point where $Re_{local}$ is a maximum marks the beginning of the formation of satellite drops, i.e.\ the axial shift of the minimum bridge radius to $z(r=r_{min})>0$, as can be seen from Figure~\ref{F:zmin}. This signals the appearance of satellite drops and entry into the VI-regime. Although $L_r/L_z$ increases at this point (Figure~\ref{F:slenderness}), the thread remains slender.  Once in the VI-regime, we do not see any evidence of an exit from this regime.

\subsubsection{Entrance into the low-$Oh$ V-Regime for $Oh<Oh_c$}

When $Oh<Oh_c$, appearance of the VI-regime is delayed by the presence of a low-$Oh$ V-regime.  The entrance to this new regime follow a $r_{min}\sim Oh^2$ scaling as $r_{min}\to 0$ which is characteristic of the I$\to$VI transition although here, as shown in \S\ref{S:I-regime}, this is an I$\to$V transition.

Figure~\ref{F:Re_local_Ohl0p15} corroborates the argument for a low-$Oh$ V-regime by showing, counter-intuitively, that once $Oh<Oh_c$ lower values of $Re_{local}$ are recovered.  For example, for $Oh=0.16$ there is a dip to $Re_{local}=0.59$ whilst for $Oh=0.14$ it falls as low as $Re_{local}=0.18$.  For $Oh=0.14$, there is then a substantial period ($4\times10^{-4}<r_{min}<7\times10^{-3}$) during which $Re_{local}$ remains at a value characteristic of the V-regime ($Re_{local}<0.85$).  Identifying the V-regime from the breakup speed in Figure~\ref{F:Ohl0p15_rdotmin} results in a slightly smaller period ($6\times10^{-4}<r_{min}<3\times 10^{-3}$), a mismatch which appears to be caused by an initial lack of slenderness in the thread (Figure~\ref{F:slenderness}), with $L_r/L_z<0.1$ only once $r_{min}<9\times10^{-4}$.  This may explain why although $\dot{r}_{min}\approx -0.071$ in the new regime, there is no asymptotic approach to this value; rather a transient passing through speeds associated with a V-regime.

Entrance into the low-$Oh$ V-regime is triggered by the minimum radius moving away from $z=0$ as a satellite drop is formed.  This is accompanied by other features that are characteristic of an I-regime, namely a corner-like free-surface shape (curve 3 in Figure~\ref{F:Ohp14_fs}) and a decrease in the axial length scale $L_z$ due to the position of maximum velocity approaching the pinch point.  However, in the newly created breakup region the axial velocity $U_z$ is still relatively small, so that $U_z L_z$ (i.e. $Re_{local}$) is not large enough to trigger I-regime dynamics and instead a V-regime is encountered; a feature discovered in \cite{castrejon15}. What remains unclear is why this mechanism persists at smaller $Oh$ and prevents transitions from the I-regime directly into the VI-regime.

\subsubsection{Entrance into the VI-regime for $Oh<Oh_c$}
 
As can be seen from Figure~\ref{F:midOh_transition}, the transition out of the low-$Oh$ V-regime is quickly followed by a transition into the VI-regime for $Oh<Oh_c$.  This boundary appears to follow a $r_{min}\sim Oh^2$ scaling as $r_{min}\to0$, although for $Oh>4\times10^{-2}$ it is approximately constant. The $\sim Oh^2$ scaling was predicted for the I$\to$VI transition and observed above for the entrance into the low-$Oh$ V-regime.  However, why  this scaling is followed for V$\to$VI is unclear.

\section{Inertia-Dominated Flow (Small $Oh$)}\label{S:I-regime}

\begin{figure}
     \centering 
\subfigure[Free surface evolution.]{\includegraphics[scale=0.4]{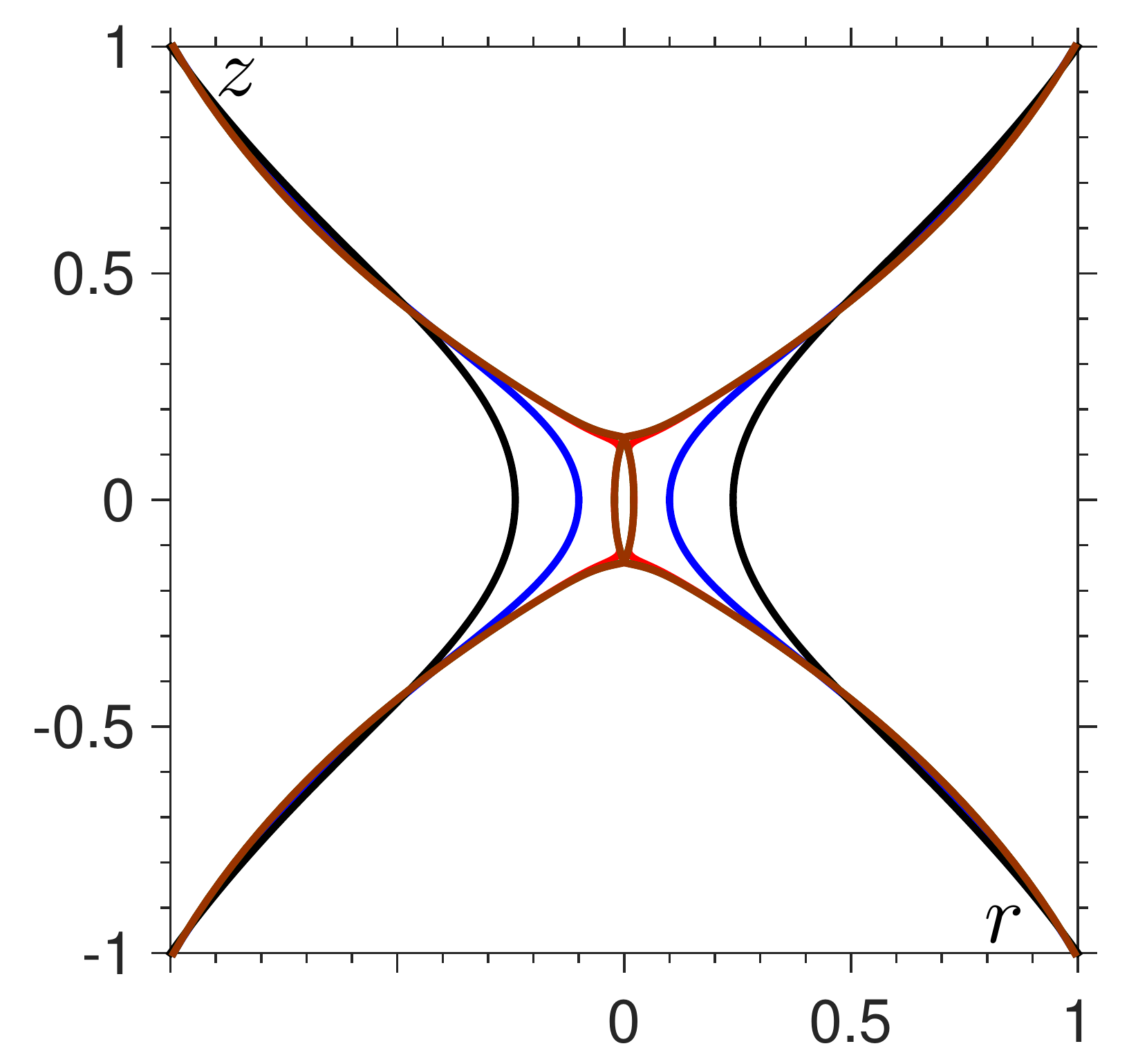}}\\   
\subfigure[Close-up of breakup.]{\includegraphics[scale=0.3]{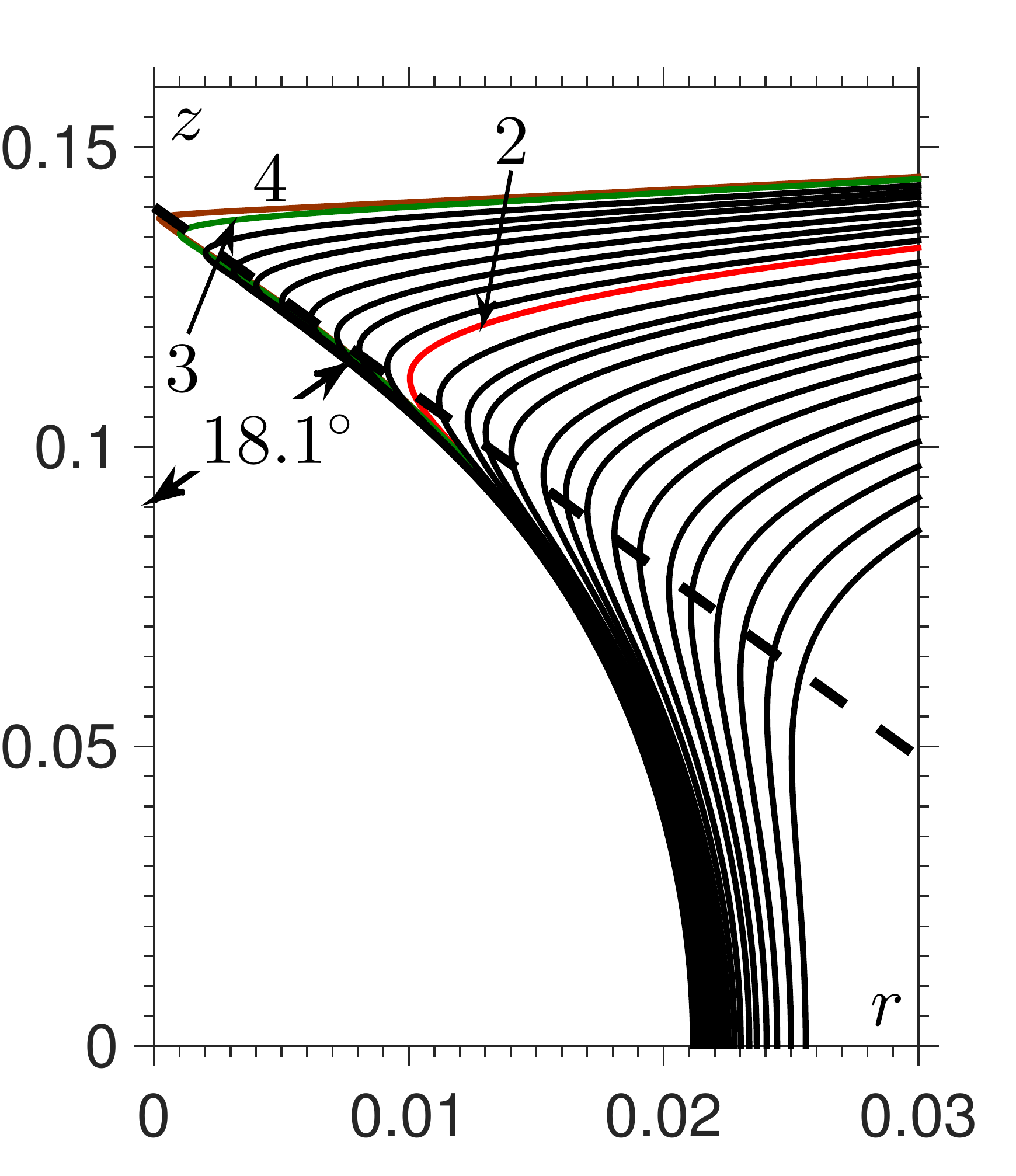}}
\subfigure[Axial velocity.]{\includegraphics[scale=0.3]{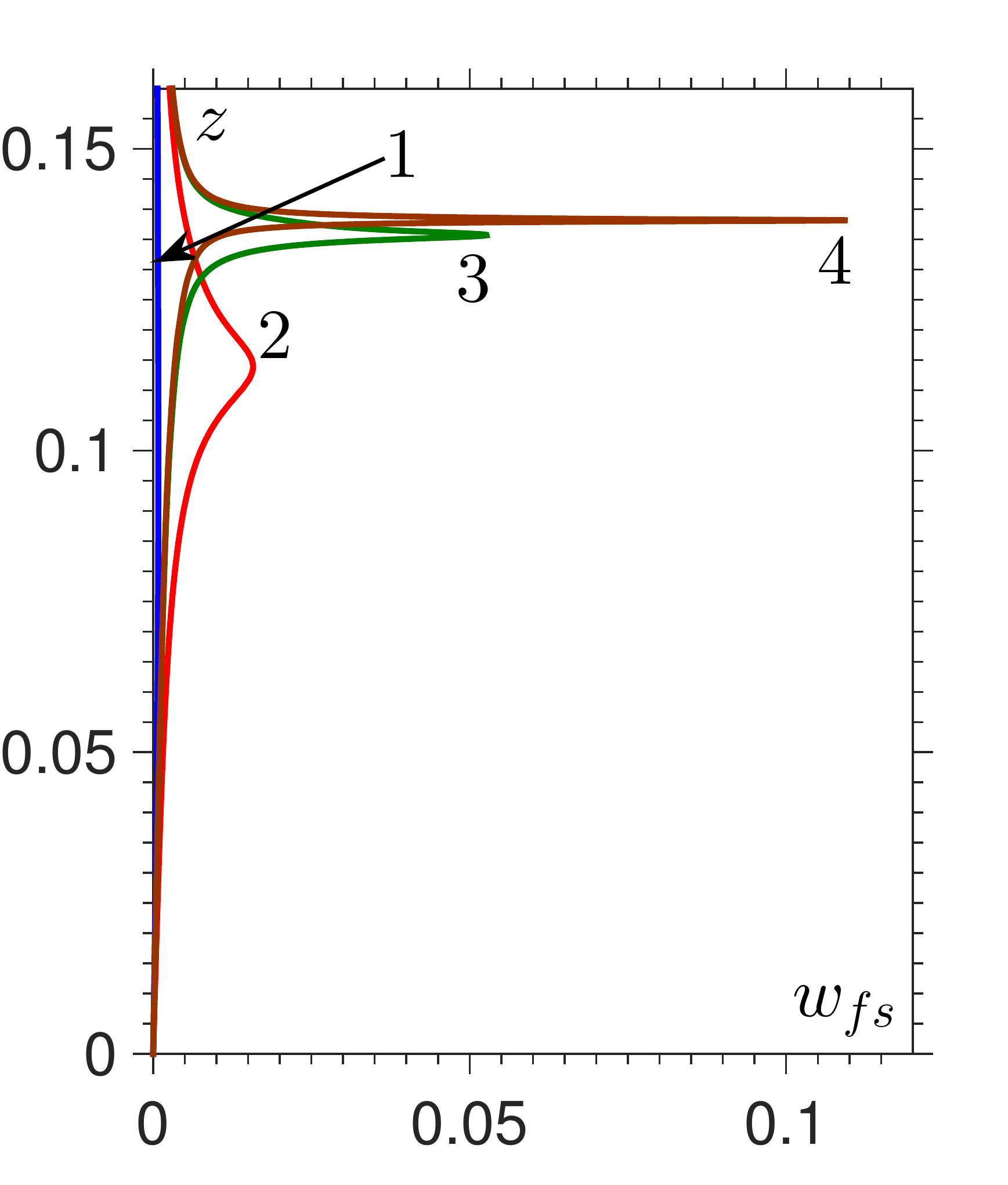}}
\subfigure[Pressure.]{\includegraphics[scale=0.3]{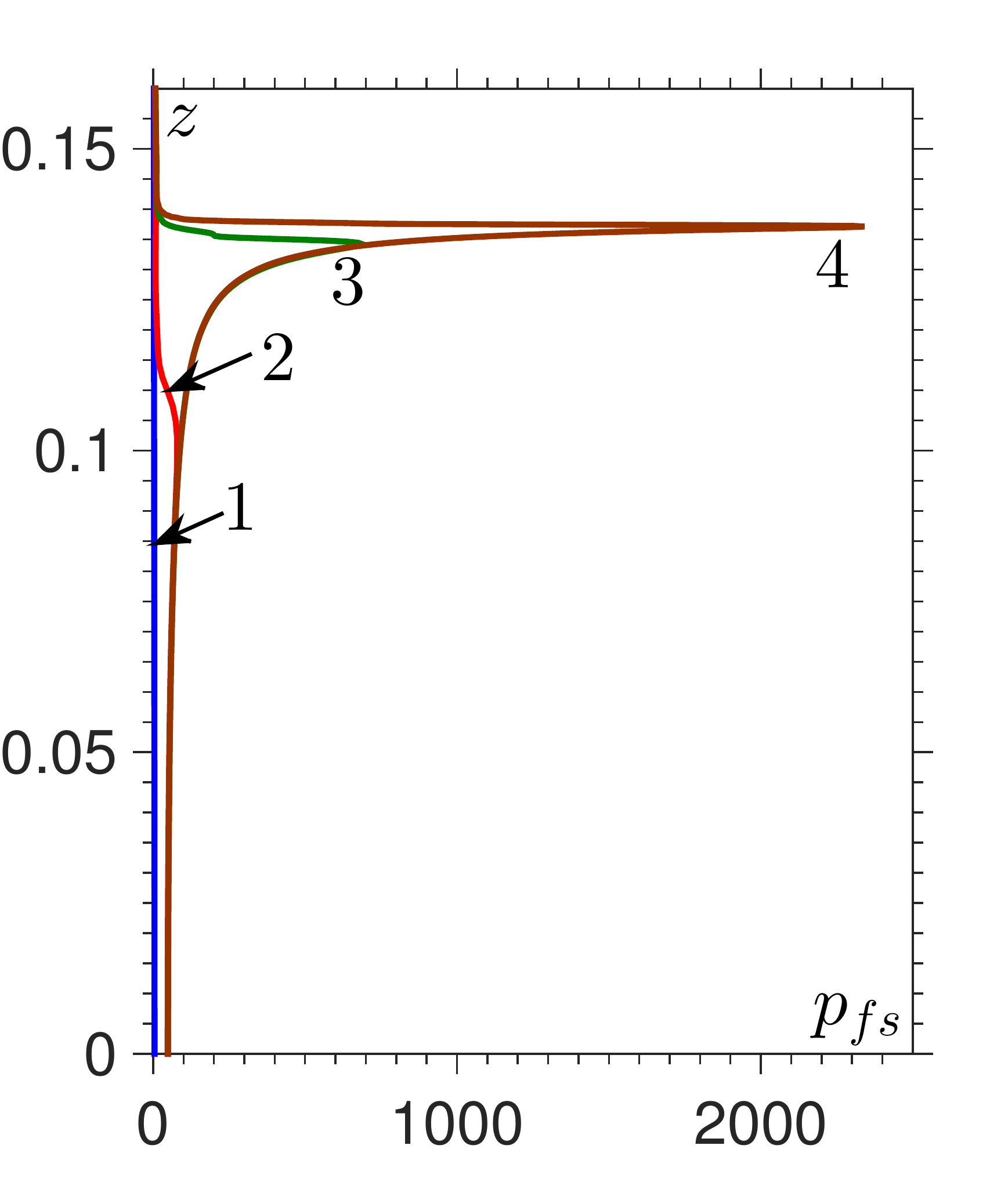}}
\caption{For $Oh=10^{-3}$ plots are for (a) the entire free-surface, (b) a close-up in the breakup region (with dashed line showing the angle $18.1^{\circ}$ from the z-axis predicted in \cite{day98}), (c) the axial velocity at the free-surface $w_{fs}$ and (d) the pressure at the free-surface $p_{fs}$ at 1: $r_{min}=10^{-1}$ (blue), 2: $r_{min}=10^{-2}$ (red), 3: $r_{min}=10^{-3}$ (green) and 4: $r_{min}=10^{-4}$ (brown).}
\label{F:Oh0p001}
\end{figure}

In Figure~\ref{F:Oh0p001} flow profiles for $Oh=10^{-3}$ are shown.  As seen in \S\ref{S:VI-regime}, a satellite drop is formed, but now the free-surface forms a corner at the pinch point (Figure~\ref{F:Oh0p001}(b)).  In a narrow region nearby a rapid increase in $w_{fs}$ and $p_{fs}$ (Figure~\ref{F:Oh0p001}(c,d)) can be seen.  Notably, $w$ is no longer $r$-independent.  Below the pinch point the free-surface forms an angle close to $18.1^\circ$ (predicted for the I-regime in \cite{day98}) with the $z$-axis (dashed line), but there is no agreement with the predicted angle above ($112.8^\circ$), which is found to be $78^\circ$.  Therefore, no `overturning' of the free-surface is observed, as discussed further in \S\ref{S:overturning}.  

The maximum axial velocity is shown in Figure~\ref{F:wmax} to scale as $\sim r_{min}^{-1/2}$ as predicted by the similarity solution (\ref{inertial}).  In contrast to the other two regimes, the radial velocity $-\dot{r}_{min}$ scales in the same way and is close to $w_{max}$ when multiplied by a factor of 5, showing that in the I-regime there is no separation of scales in the $r$ and $z$ directions. This is supported by observations that the thread is not slender as pinch-off is approached, as can be seen from curve 1 in Figure~\ref{F:slenderness}.

\begin{figure}
     \centering
\subfigure[Identification of the I-regime where $\dot{l}_{min}\approx-0.5$.]{\includegraphics[scale=0.26]{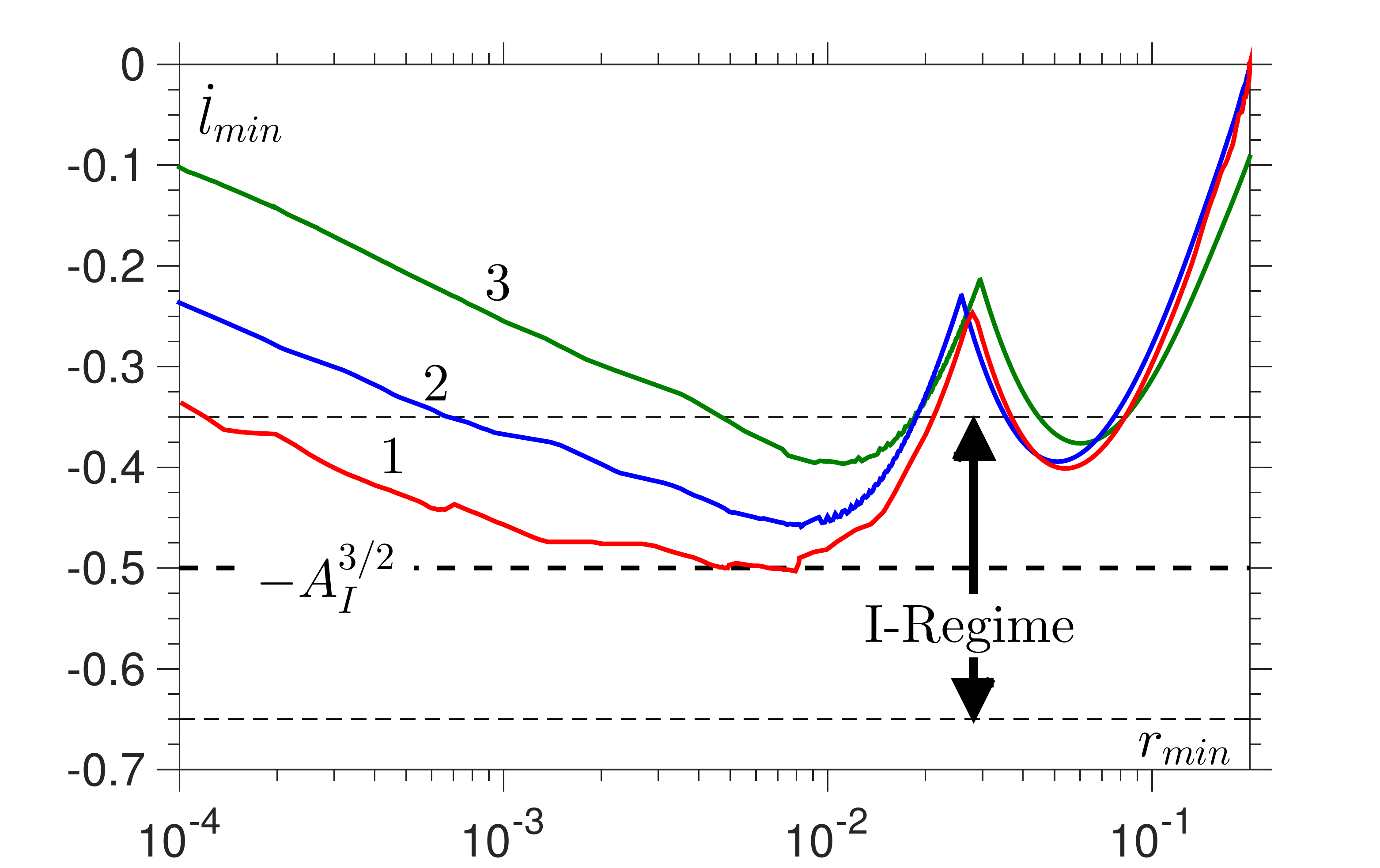}\label{F:Linviscid}}
\subfigure[Evolution of the local Reynolds number]{\includegraphics[scale=0.26]{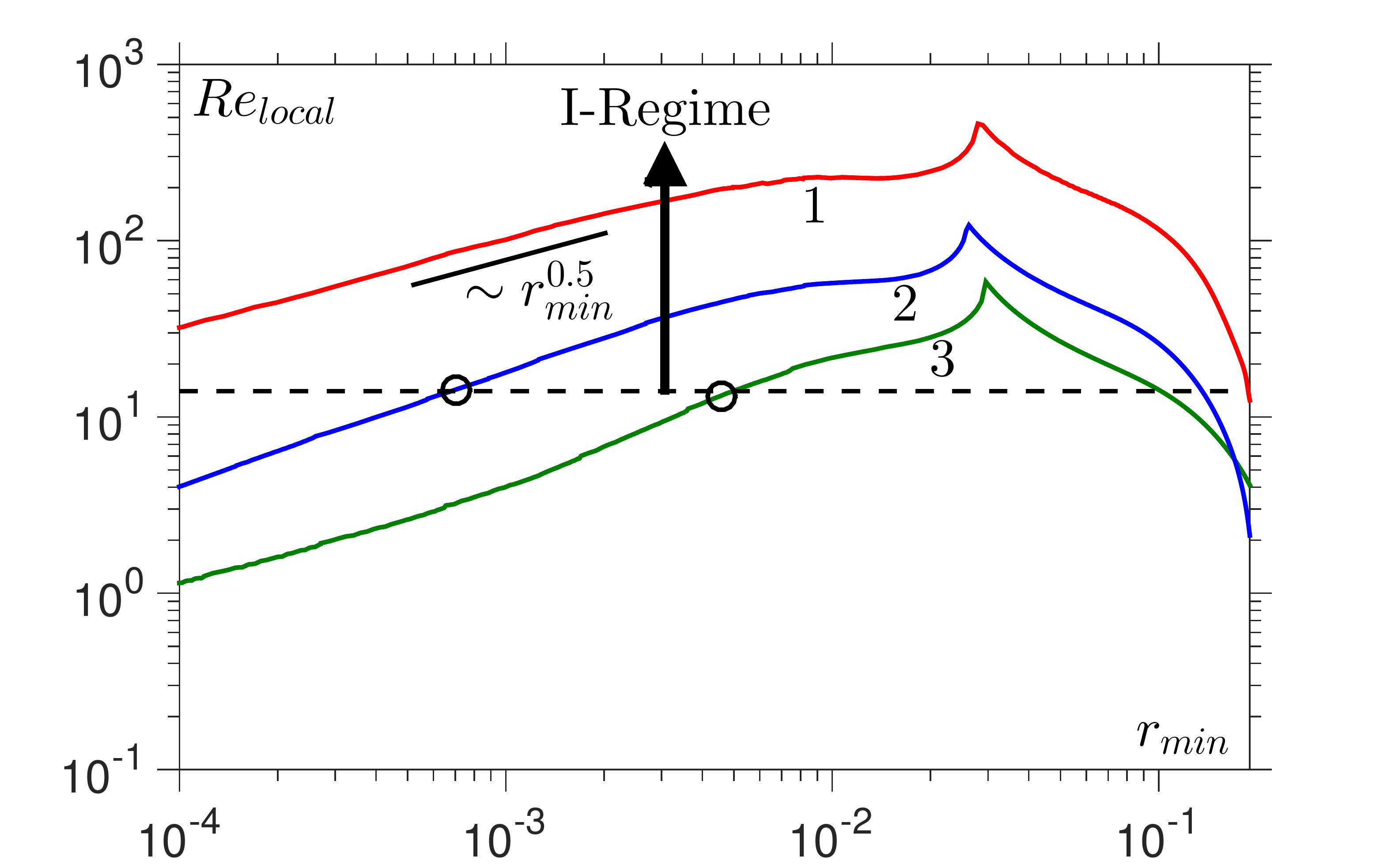}\label{F:Re_local_Ohl0p01}}
\caption{Curves are for 1: $Oh=10^{-3}$, 2: $Oh=4\times10^{-3}$ and 3: $Oh=10^{-2}$.  (a) The I-regime is defined when $\dot{l}_{min}=-0.5\pm0.15$ (dashed lines). (b) Transition points out of the I-regime at $r_{min}=r_{min}^{I\to}$ are marked as circles and show that this occurs when $Re_{local}\approx14$ (horizontal dashed line).  The decrease of $Re_{local}$ in the I-regime follows the predicted scaling from the similarity solution (\ref{inertial}) that $Re_{local}\sim r_{min}^{0.5}$.  }
\end{figure}

Recalling that regions where $\dot{l}_{min}$ is approximately constant results in a $r_{min}\sim\tau^{2/3}$ scaling indicative of the I-regime, curve 1 in Figure~\ref{F:Linviscid} shows the appearance of this regime for $Oh=10^{-3}$ around $10^{-3}<r_{min}<10^{-2}$.   Computations show that $\dot{l}_{min}=-A_I^{3/2}\approx -0.5$ so that $A_I=0.63$, which can be used to extract the breakup time and plot $r_{min}$ against $\tau$ in curve 1 of Figure~\ref{F:rvst}.  The value of $A_I$ is close to that given in \cite{eggers08} of $0.7$.  

\subsection{Identification of the I-regime}

\begin{figure}
     \centering
\includegraphics[scale=0.26]{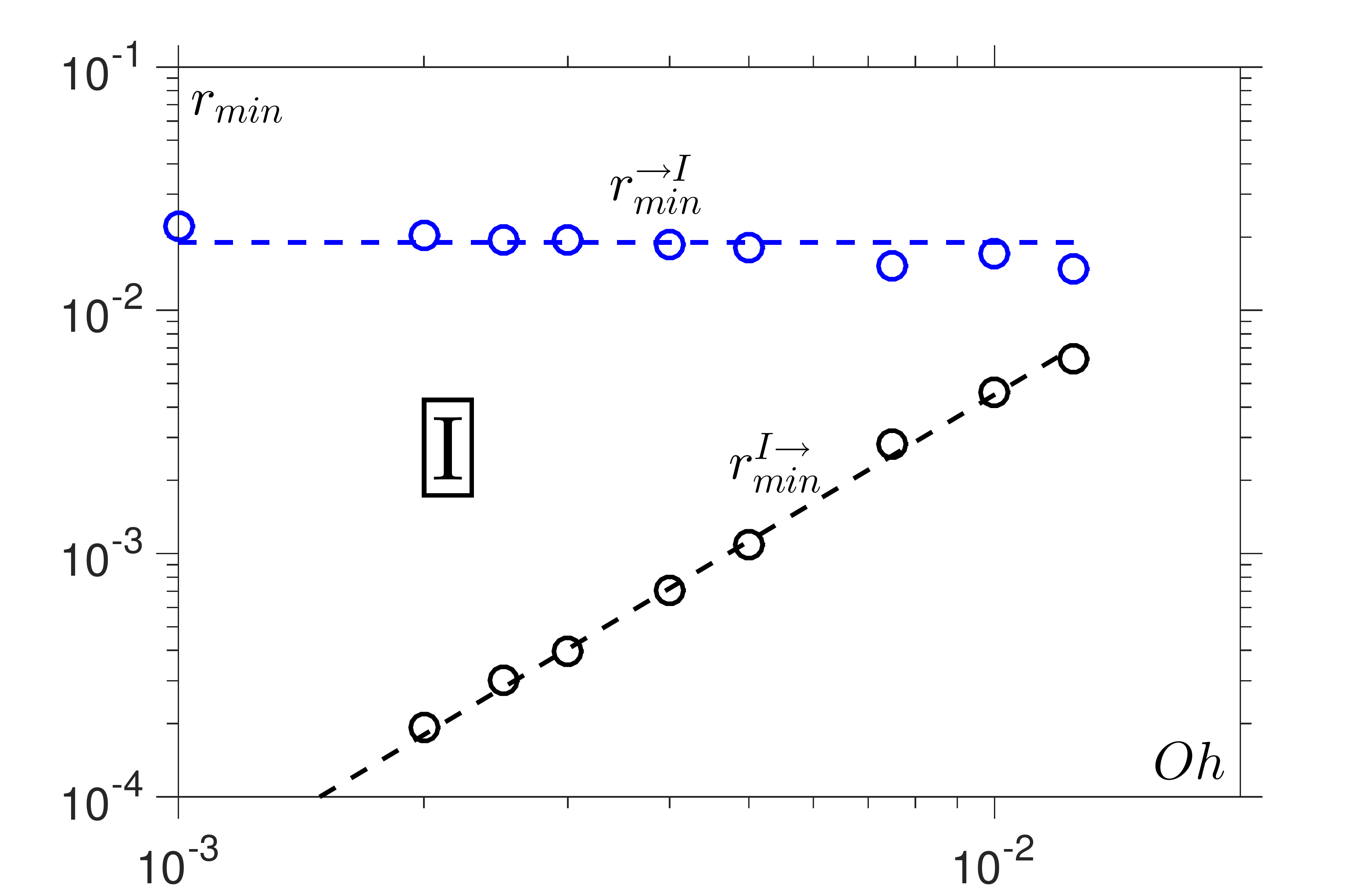}
\caption{Phase diagram showing the I-regime.  The transition into this regime $r^{\to I}_{min}$ is constant $\approx 0.019$ whilst the exit $r^{I \to }_{min}$ scales as $Oh^2$ with $r_{min}=45~Oh^2$.}
\label{F:lowOh_transition}
\end{figure}

Using $\dot{l}_{min}$ to define the I-regime, and starting from the point at which the satellite drop is formed, i.e.\ $r_{min}<3\times10^{-2}$ (Figure~\ref{F:zmin}), where the geometry of the I-regime start to take shape, the boundaries of the I-regime are shown in Figure~\ref{F:lowOh_transition}.  
%Oh = [0.0005, 0.00075, 0.001 0.002, 0.0025, 0.003 0.004 0.005, 0.0075 0.01 0.0125]
%r_in = [0.021, 0.021, 0.022, 0.0202, 0.0195,  0.0194, 0.0186, 0.018, 0.0153, 0.017 0.0149]
%r_out = [1e-10, 1e-10, 1e-10, 1.92e-4 3e-4, 3.96e-4, 7.05e-4, 1.08e-3, 2.8e-3, 4.62e-3 6.3e-3]

\subsubsection{Entrance into the I-Regime}

Similar to the V-regime, transitions into the I-regime $r^{\to I}_{min}$ are due to geometry, with a certain time required for the thread to form the corners which are characteristic of this regime.  As one can see from curves 1 and 2 in Figure~\ref{F:slenderness}, the breakup region is not slender and $L_r/L_z\approx 0.6$ remains approximately constant as $r_{min}\to0$.  

\subsubsection{Exit from the I-Regime}

The exit from the I-regime $r^{I \to }_{min}$ occurs when viscous effects become significant and the local Reynolds number drops below a critical value. Figure~\ref{F:Re_local_Ohl0p01} shows that the reductions in $Re_{local}$ follow the $\sim r_{min}^{0.5}$ scaling predicted for the I-regime (\ref{inertial}), which confirms our definition of $Re_{local}$ is a good one.  It is found that the I-regime is exited when $Re_{local}\approx 14$.  From \S\ref{S:VI-regime} we know that at this point the breakup will transition into the low-$Oh$ V-regime and Figure~\ref{F:lowOh_transition} shows that this entire boundary follows an $\sim Oh^2$ scaling.  
%
%
%%%rho.U.L/sigma = Re.U_l.L_l = (1/Oh^2) U_l.L_l 
%
%Old results with wmax at fs:
%Oh=[0.001 0.0025 0.005 0.01]
%r_out=[?? 2.8e-4 1.1e-3 4.6e-3]
%Re_local_there=[?? 3 2 2]
%
%New ones with proper maximum
%Oh=[0.004  0.01]
%r_out=[7.05e-4 4.6e-3]
%Re_local_there=[14.3 13.2]

\section{A Phase Diagram for Breakup}\label{S:phasediagram}

\begin{figure}
     \centering
\includegraphics[scale=0.26]{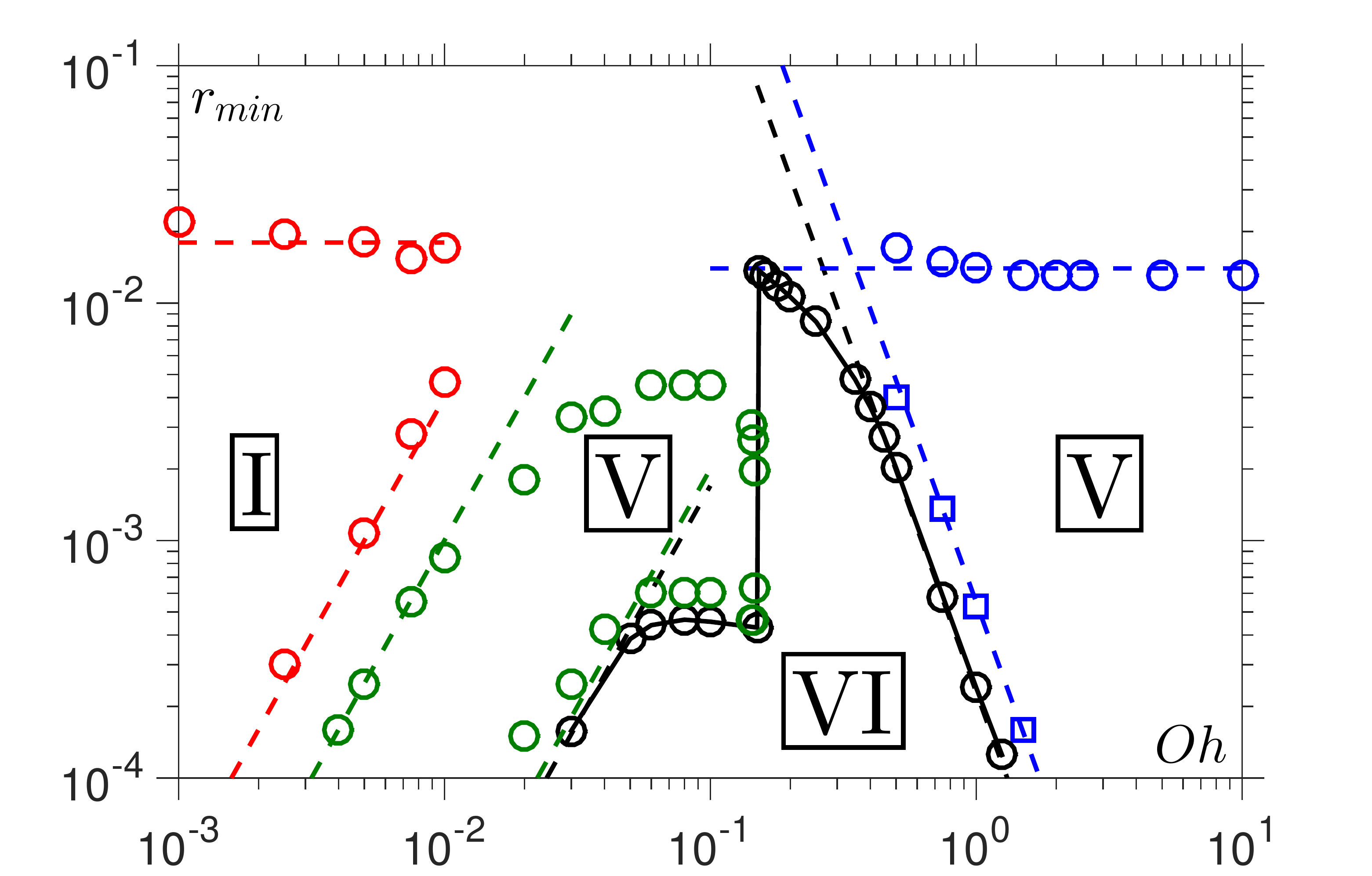}
\caption{Phase diagram for the breakup phenomenon which stitches together results from Figure~\ref{F:highOh_transition}, Figure~\ref{F:midOh_transition} and Figure~\ref{F:lowOh_transition}, where specific expressions for the transitions between regimes can be found.}
\label{F:allOh_transitions}
\end{figure}

Having computed transitions into and out of the three regimes across parameter space, the results can be stitched together to produce the phase diagram shown in Figure~\ref{F:allOh_transitions}. Notably, once a regime is reached, which is around $r_{min}\approx10^{-2}$ across $Oh$, the transitions between regimes are relatively sharp, so that there are not vast regions of parameter space where none of the scaling laws are applicable.  There are also no overlaps between the regimes, partially due to our choices of defining being `in' each regime.  Consequently, rather than worry about transitions into and out of the different regimes, it is now sensible to talk about transitions \emph{between} different regimes.  This is how Figure~\ref{F:allOh_transitions} was converted to produce the neat phase diagram Figure~\ref{F:actual} in \S\ref{S:multiple}, where (a) transitions between the regimes are provided and (b) only the asymptotic form of these transitions as $r_{min}\to0$ are kept (i.e.\ using only the dashed lines in Figure~\ref{F:allOh_transitions}). It is of course possible that there is further complexity in the phase diagram below $r_{min}=10^{-4}$. Here, analytic work is required as computational studies are inherently confined to a finite resolution below which the breakup behaviour can only be implied.

The most surprising results are the appearance of a low-$Oh$ V-regime which prevents I$\to$VI transitions and the dramatic change in behaviour around a critical value $Oh_c=0.15$.  For $Oh>Oh_c$ the scaling for the V$\to$VI transition follows that expected from theory ($r^{V\to VI}_{min}\sim Oh^{-3.1}$) whilst both the transitions from the I-regime into the low-$Oh$ V-regime and the exit into the VI-regime follow $r_{min}\sim Oh^2$ scaling predicted for the I$\to$VI transition, but only as $r_{min}\to0$.  Notably, it was shown that many of the transitions between regimes can be predicted from our measures $Re_{local}$ and $L_r/L_z$.   

An experimentally verifiable prediction for the existence of a sharp transition at $Oh_c=0.15$, is that going from $Oh=0.16$ down to $Oh=0.14$ decreases the length of the thin liquid thread which connects the satellite drop to the main volume by a factor of three, see Figure~\ref{F:evolution_Oh0p16_vs_Oh0p14}.
%
%ALl transitions are shown in:
%
%\begin{figure}
%   \centering
%\includegraphics[scale=0.3]{rvst_transitions-eps-converted-to.pdf} this doens't look right
%\caption{Plots of the bridge radius against time showing all of the transition types observed: 1: V$\to$VI transition for $Oh=0.5$, 2: V$\to$VI transition for $Oh=0.1$, and 3: I$\to$V transition for $Oh=7.5\times10^{-3}$.  The dashed curves labelled V, VI and I give the appropriate scalings of these regimes.}
%\label{F:rvst_transitions}
%\end{figure}
% $r^{\to V}_{min}=10Oh^2$ meets $r^{\to VI}_{min}=2.3\times10^{-4}Oh^{-3.1}$ 

\section{Discussion}\label{S:discussion}

Computations using a multiscale finite element method have allowed us to accurately construct a phase diagram for breakup and determine the transitions between different regimes. During the analysis, a number of features have been encountered which are worthy of further attention, and these are now considered.

\subsection{Unexpected Transitional Behaviour}

The possibility of multiple flow transitions was suggested in \cite{eggers05} and first observed in \cite{castrejon15}.  The low-$Oh$ V-regime can be seen in Figure~2 of \cite{castrejon15} at similar values of $Oh$ to those found here.  However, in contrast to \cite{castrejon15}, no evidence of a transient high-$Oh$ I-regime (their Figure~4 for the case of $Oh=1.81$), has been observed.  This could be because our method of defining the I-regime does not pick up transient regimes far from breakup; however, as can be seen from Figure~\ref{F:allOh_transitions}, our phase diagram is full for $r_{min}<10^{-2}$ so that any I-regime that has been missed will not appear in the later stages of breakup.  Moreover, for all cases where a high-$Oh$ I-regime could exist, $Re_{local}$ remained at values well below that seen for the I-regime, where $Re_{local}>14$, giving strong evidence for the absence of a high-$Oh$ I-regime.
%
%A possible method for picking up transient I-regimes far from breakup, which have the form $r_{min}=A_I (Oh\tau)^{2/3}+C$, would be to monitor $M_{min} =(\frac{-9\dot{r}_{min}^4}{8Oh^2~\ddot{r}_{min} })^{1/3}$ which satisfies $M_{min}=A_I$ in the I-regime, but the requirement of the second derivative $\ddot{r}_{min}$ makes this method difficult to implement.

To make unambiguous conclusions about the dominant forces as breakup is approached, one could directly extract the size of the inertial and viscous forces from the computed terms in the Navier-Stokes equations and compare their magnitudes in the breakup region.  This would be a more advanced method for calculating $Re_{local}$ that may provide additional insight into the transitions between different regimes.  However, we have found such measurements difficult to obtain in the breakup region due to the tiny elements found there and the singular nature the dynamics. Such issues are amplified when determining second derivatives of the velocity, which itself is only approximated quadratically across each element, in order to calculate the viscous forces.  This will be the focus of some of our future work in this area.

An intriguing possibility is that the low-$Oh$ V-regime is responsible for the transitional behaviour observed in \cite{rothert03}, where water-glycerol mixtures in the approximate range $0.06<Oh<3$ were shown to transition from V$\to$VI at a constant $r_{min}$ rather than following the predicted scaling $r_{min}\sim Oh^{-3.1}$.  Given the relatively low $Oh$ considered in these experiments, it seems possible that it was the low-$Oh$ V-regime that was being encountered rather than the usual one.  Furthermore, Figure~\ref{F:allOh_transitions} shows that in the range $0.04<Oh<0.14$ the transition from the low-$Oh$ V-regime into the VI-regime ($r_{min}^{V\to VI}$) is approximately constant.   Unfortunately, no direct comparison to \cite{rothert03} is possible due to the different flow configurations used, so at present we can only speculate about the possible observation of a low-$Oh$ V-regime in their experiments.  

\subsection{Oscillatory Convergence to the VI-Regime}\label{S:oscillations}

\begin{figure}
     \centering
\includegraphics[scale=0.26]{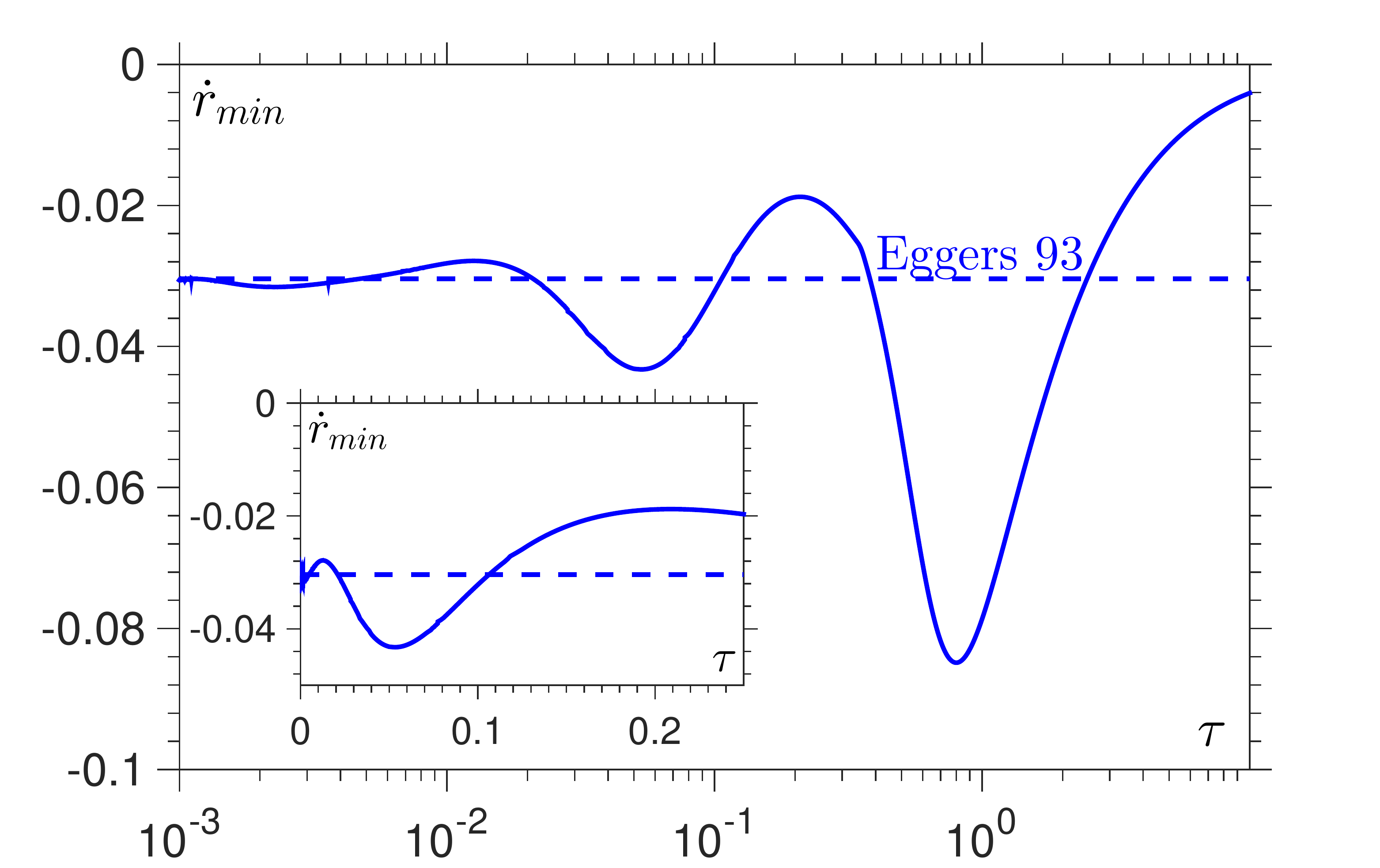}
\caption{Speed of breakup $\dot{r}_{min}$ against time from breakup $\tau$ for the case of $Oh=0.16$, where oscillations characteristic of the VI-regime can be observed most strongly.}
\label{F:VIvstau}
\end{figure}

Figure~\ref{F:VvsVI} clearly illustrated that convergence towards the V- and VI-regimes is fundamentally different: at $Oh=10$ the bridge speed increases monotonically towards the value predicted by the similarity solution in the V-regime of $-0.071$ whilst for $Oh=0.16$ the bridge speed converges in an oscillatory manner towards the speed predicted by the similarity solution in the VI-regime of $-0.030$ \footnote{Preliminary results for the case where the plates confining the liquid bridge move apart are similar except the VI-regime is entered earlier, i.e.\ at larger $r_{min}$.}. The results are reinforced by Figure~\ref{F:VIvstau} which shows the same oscillations when $\dot{r}_{min}$ is plotted against the time from breakup $\tau$. Notably, the wavelength of the oscillations is approximately constant in logarithmic time $\ln\tau$, a natural choice of variable used, e.g.\ in \cite{eggers97}, in the derivation of similarity solutions for breakup. Our results suggest that the similarity solutions in both the V and VI-regimes are stable to perturbations, but that in the former case the associated eigenvalues of the perturbation are negative and purely real whilst in the latter case they are complex with negative real part.  

For the V-regime our findings agree with those in \cite{eggers12}, where it is shown that out of an infinite hierarchy of similarity solutions, only (\ref{papa}) is stable with purely real eigenvalues. Our findings for the VI-regime are entirely new and have not been predicted computationally, analytically or experimentally.  The likely reasons for this are as follows:
\begin{itemize}

\item Computations: Very few simulations have captured the small-scales resolved in our work for the axisymmetric Navier-Stokes system.  However, in cases where they have, e.g.\ in \cite{castrejon15}, the focus has been on $r_{min}$ against $\tau$, whereas the oscillations are most clear when plotting $\dot{r}_{min}$ against $\tau$ or $r_{min}$.  What is surprising is that this behaviour has not been observed in slender jet codes which recover the similarity solution in \cite{eggers93}.  Our own preliminary simulations indicate that the same oscillations can be observed in this setting too.

\item Analysis: The current situation is summarised in p.39 of \cite{eggers08}: thus far no one has analytically calculated the eigenvalues of the Navier-Stokes system for breakup.  Consequently, it is unknown (a) whether the similarity solution is stable and (b) if the associated eigenvalues are purely real or complex.  The framework for the stability analysis is provided in Appendix B of \cite{brenner96}, and progress on similar flows has been achieved in \cite{bernoff98}, but extending these results to the Navier-Stokes pinch-off remains an open problem.

\item Experiments: As with computations, so far the focus has been on $r_{min}$ against $\tau$, where the oscillations can be missed, so that converting some of this data into $\dot{r}_{min}$ against $\tau$ or $r_{min}$ would be useful.  However, here one faces the additional complication of noisy data which could make the accurate extraction of derivatives a tricky procedure.  Despite this, the oscillations observed are quite significant, at $Oh=0.16$, between $r_{min}=10^{-3}$--$10^{-2}$ the bridge speed doubles, from $-0.02$ up to $-0.04$ (Figure~\ref{F:VvsVI}).  If the length scale is $R=1$~mm this occurs when the dimensional bridge radius is around $1$--$10\mu$m, i.e.\ near the limits of optical resolution.  Therefore, there is hope that experimental data can identify these changes.  
\end{itemize}
%What do they actually look like?
%Why does a V-regime have to precede a VI one?

\subsubsection{Bumps on the VI Similarity Solution}\label{S:bumps}

Interestingly, continuing some of our simulations in the VI-regime beyond our self-imposed minimum radius of $r_{min}=10^{-4}$ resulted in the growth of a number of interesting features which, as far as we can see, appear unrelated to the aforementioned oscillations. These should be treated with caution, as the accuracy of the numerical scheme is close to its limitations at such small radii, but the features appear to be robust. 

\begin{figure}
     \centering
\subfigure[Bumps on the free surface profiles]{\includegraphics[scale=0.3]{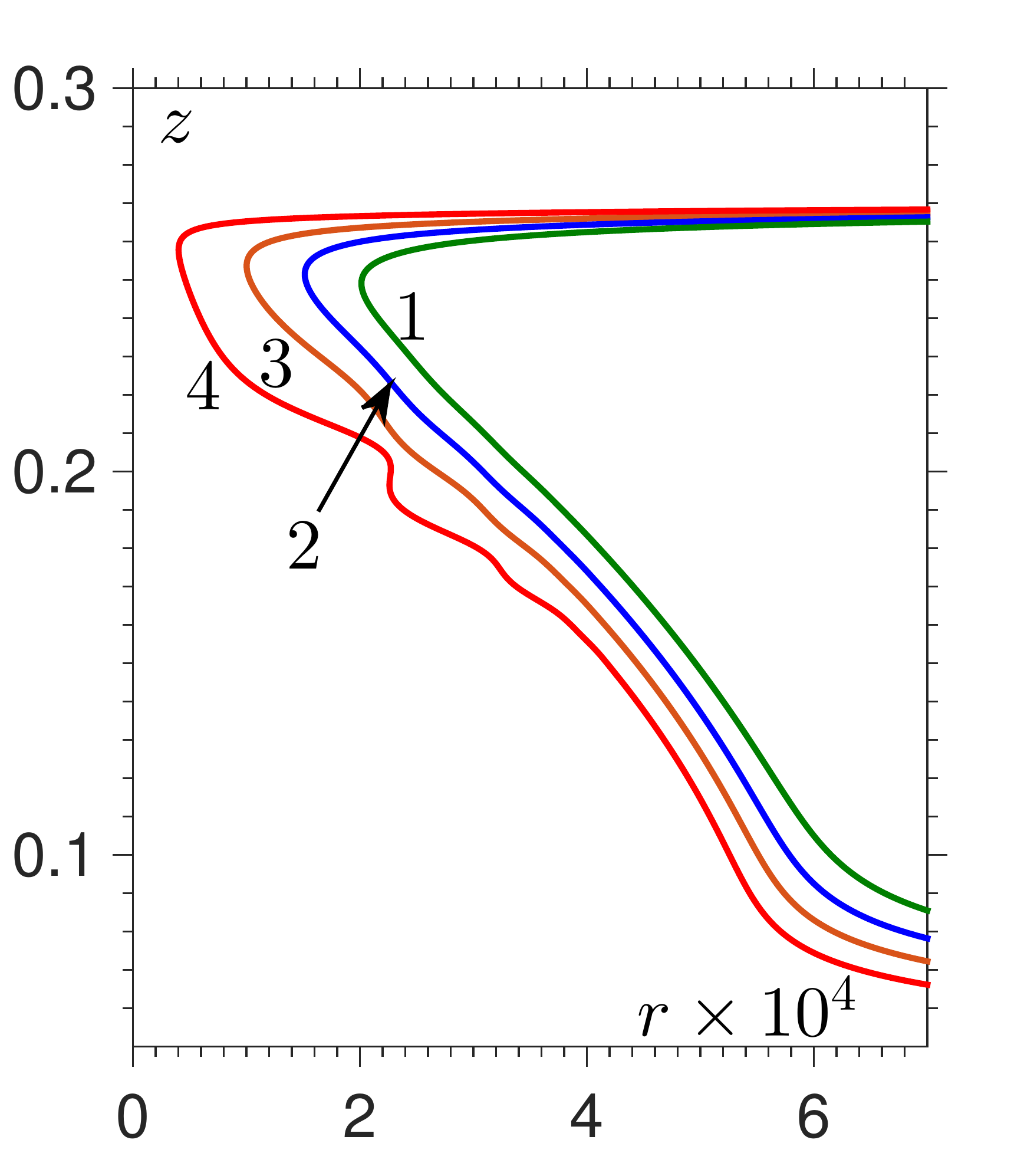}\label{F:bumps}}
\subfigure[Axial velocity distribution]{\includegraphics[scale=0.3]{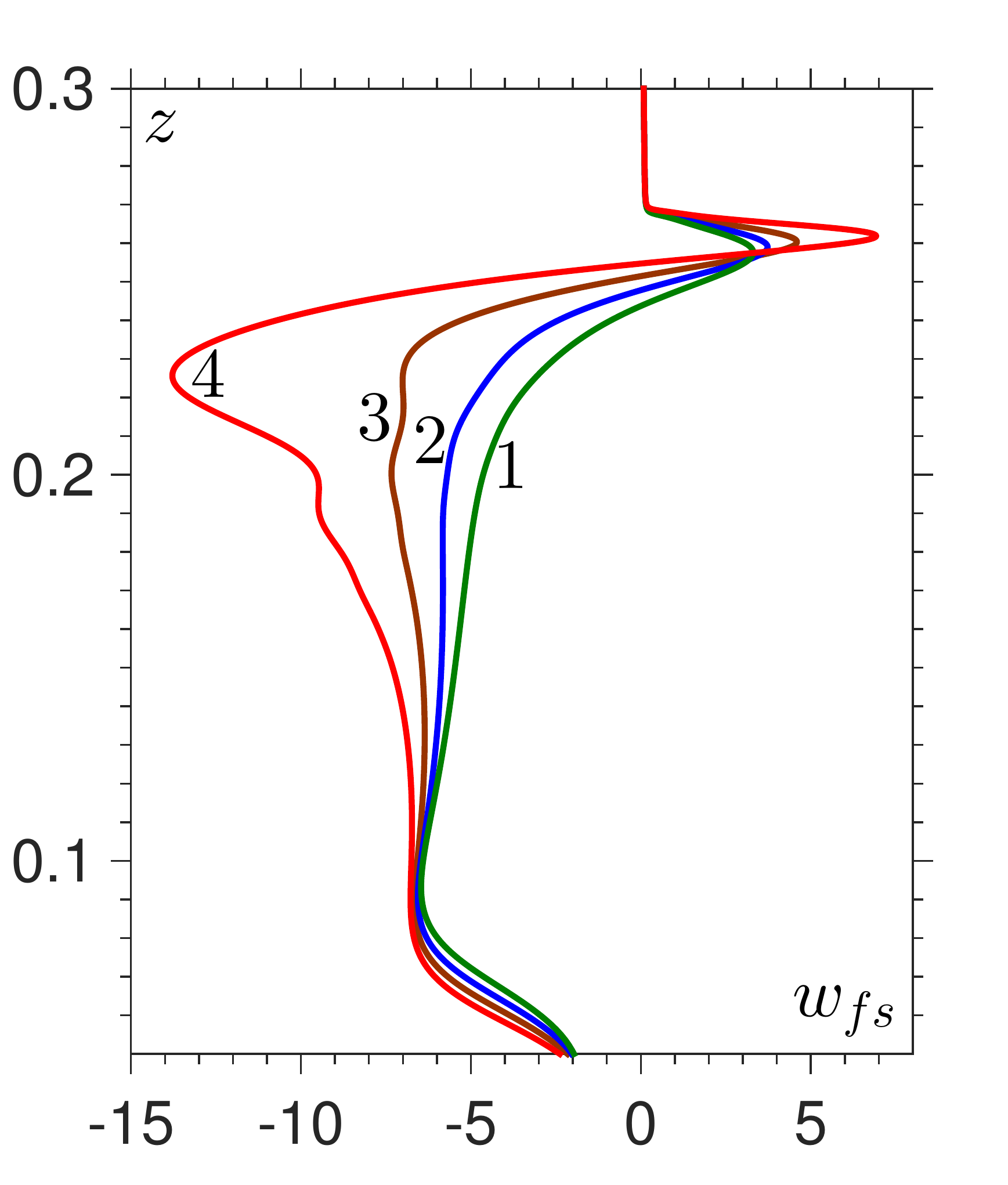}\label{F:bumps_w}}
\caption{Curves are for $Oh=0.16$ at 1: $r_{min}=2\times10^{-4}$, 2: $r_{min}=1.5\times10^{-4}$, 3: $r_{min}=10^{-4}$ and 4: $r_{min}=5\times10^{-5}$. }
\end{figure}

As can be seen from Figure~\ref{F:bumps} for $Oh=0.16$, at around $r_{min}=5\times10^{-5}$ the free-surface has developed waves close to the pinch point.  These feature in plots of axial velocity in Figure~\ref{F:bumps_w} where one can see that the maximum absolute velocity (which will have $w<0$) shifts from around $z=0.08$ (in curve 1) to $z=0.23$ (curve 4).  The velocity profile in curve 4, with the maximum absolute velocity near the pinch point, is closer to that predicted by the similarity solution in \cite{eggers93}. 

It is possible these bumps could be related to the iterated instabilities observed experimentally in \cite{brenner94} and interpreted as successive instabilities of the similarity solution of \cite{eggers93}.   In this work, the instability was attributed to thermal fluctuations which were added to the lubrication formulation. However, it was pointed out that because this formulation only approximates the full Navier-Stokes system, such external noise may be unnecessary. The iterated instabilities have been seen computationally in \cite{mcgough06} for a surfactant covered jet, but our results suggest that the instabilities can be generated without requiring additional effects such as noise or surfactants.

There could also be a relation to the destabilisation of similarity solutions due to finite amplitude perturbations observed in \cite{brenner96}, with intriguing similarities between our Figure~\ref{F:bumps} and their Figure~5 for the evolution of an unstable similarity solution.

\subsection{Overturning of the Free Surface in the I-Regime}\label{S:overturning}

Figure~\ref{F:Oh0p001} showed that at $Oh=10^{-3}$ no overturning of the free-surface was observed and, as far as we are aware, this does not contradict any previous findings in the liquid bridge geometry. Although simulations in \cite{suryo06} show overturning in the same geometry, this phenomenon only occurs for non-Newtonian cases.  In contrast, for drop formation from an orifice experimental studies show cases that do \citep{castrejon12} and do not \citep{brenner97} show overturning whilst simulations using the inviscid equations do show this feature \citep{schulkes94} and those using slender jet theory are unable to \citep{brenner97}.  

The most comprehensive computational study of overturning is given in \cite{wilkes99} where the finite element method is used to study the effect of $Oh$ and the Bond number $Bo$ (i.e. gravitational forces) on this phenomenon.  It is shown that the critical $Oh$ at which overturning is observed and the angles which the free-surface shape makes with the $z$-axis at the pinch point depend on $Bo$, which alters the geometry of the breakup.  Furthermore, the angles recovered vary, with a maximum angle of $95^\circ$ observed, in contrast to the value of $112.8^\circ$ predicted from the inviscid theory \citep{day98}.

In the drop formation geometry, threads of arbitrary large lengths can be formed whereas in our setup the liquid is confined between two stationary plates.  This appears to be the reason why no overturning has been observed even in cases of small $Oh$ which fall below the overturning limit in \cite{wilkes99}.  In particular, rather than a thread connected to a free drop, our geometry has a short thread (the satellite drop) connected to a hemispherical volume pinned to a plate.

\subsection{Implications for Breakup in CFD}

Computational approaches that `go through' the topological change of breakup all rely on cutting the thread, either manually (as in FEM) or automatically (as in Volume-of-Fluid approaches - VoF) once it reaches a specific $r_{min}$, which may be implicitly related to the grid size (as in VoF).  Therefore, it is of interest to know how accurate this approach is and what could be lost from a premature truncation.  To do so, for sake of argument consider a scheme with a fixed resolution (mesh size) of $\triangle h=5\times 10^{-3}$, giving a (relatively large) grid of $200\times200$ for our problem, that cuts the thread wherever $r_{min}<\triangle h$. 

The most obvious feature that can be lost from the cut-off is the satellite drop, as seen in \cite{fawehinmi05}.  Looking at Figure~\ref{F:zmin}, satellite drop formation, indicated by $z(r=r_{min})$ becoming non-zero, can clearly be seen for both $Oh=0.16$ (curve 2) and $Oh=0.5$ (curve 3).  This occurs when the symmetric V-regime transitions into the asymmetric VI one, as can be seen clearly from free surface profiles at $Oh=0.16$ in Figure~\ref{F:Oh0p16}.  For $Oh=0.16$ the pinch point moves when $r_{min}\approx 10^{-2}$ whilst for $Oh=0.5$ this does not happen until $r_{min}\approx 10^{-3}$.  Therefore, using our cut-off, at $Oh=0.16$ the thread would be cut near to the correct pinch point whilst for $Oh=0.5$ the thread would be severed at $z=0$ so that the  formation of a satellite drop would be missed.  Most worryingly, in the latter case the post-breakup state would be entirely wrong as instead of having two breakup points a distance $\triangle z=0.35$ apart (giving 3 distinct volumes) there would be just one at the centre of the drops (and hence 2 volumes).  Notably, for $R=1$~mm the minimum radius $r_{min}=10^{-3}$ corresponds to a dimensional radius of $1\mu$m, so that these are truly macroscopic quantities.

To capture the features of breakup, such as satellite drops, one either has to (a) develop codes with huge levels of resolution/refinement, which will be extremely costly (particularly in 3D) due to the multiscale nature of this phenomenon, or (b) intelligently utilise the similarity solutions to take the codes through the topological change and up to a suitable post-breakup state.  Such a scheme would be complex, but could lead to increased accuracy at a lower computational cost.  By identifying in which regions of parameter space the different similarity solutions are accurate and devising methods to analyse in which flow regime a given breakup is occurring we have taken just some of the first steps in the development of such a scheme. Forthcoming works will extend these results to cases in which there is a strong externally-driven elongation of the thread and/or interface formation physics which creates a singularity-free breakup, see \cite{shik07}.

\section*{Appendix:  Identification of Regimes}

\begin{figure}
     \centering
\includegraphics[scale=0.26]{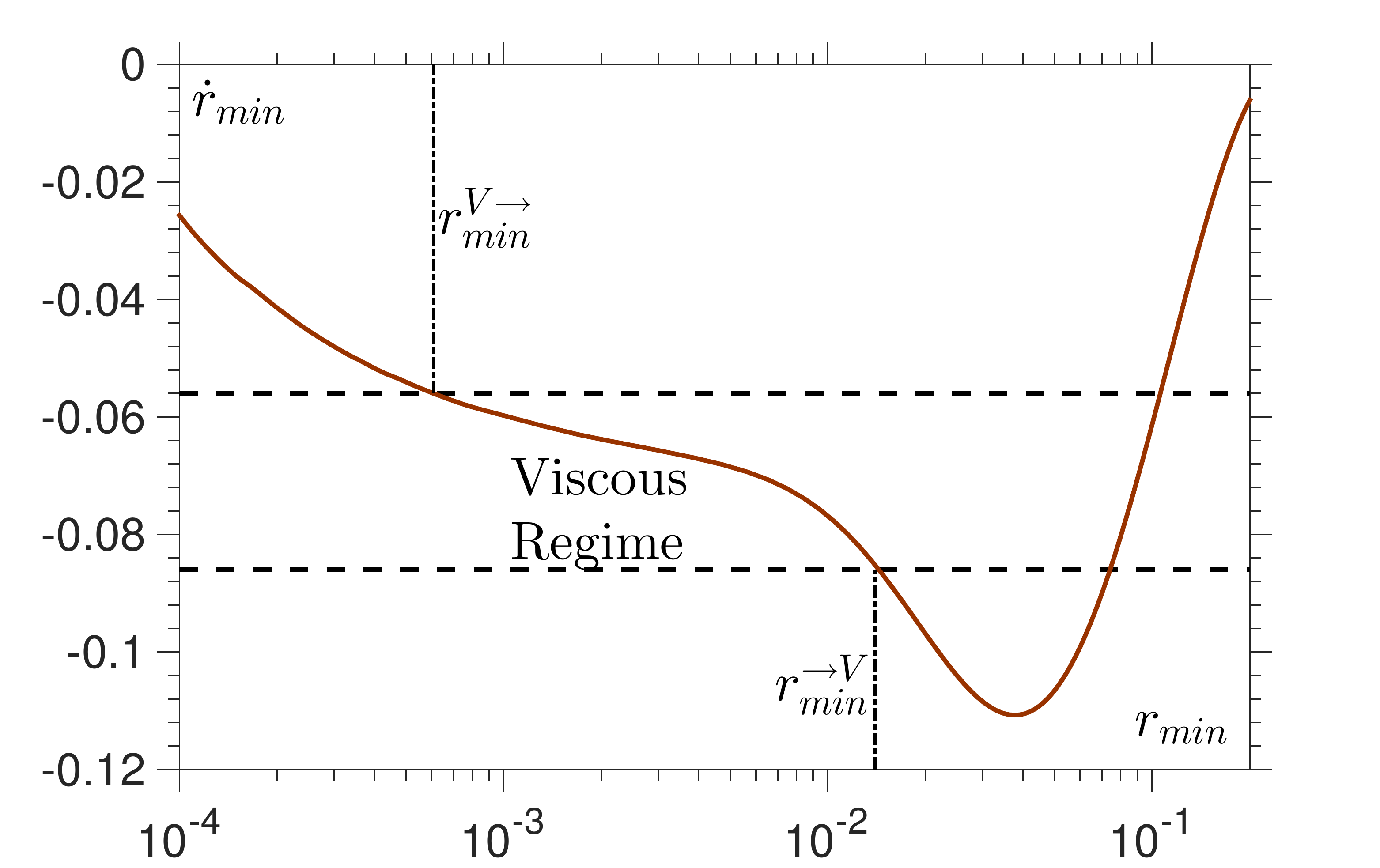}
\caption{Demonstration of method for calculating boundaries of the viscous regime for the case of $Oh=1$, where $r_{min}^{\to V}$ denotes the transition into this regime and $r_{min}^{V\to}$ is the exit.}
\label{F:rdot_vs_r_Oh1_example}
\end{figure}

In Figure~\ref{F:rdot_vs_r_Oh1_example}, the calculation of the boundaries of the viscous regime are shown for the case of $Oh=1$.  The V-regime is defined by speeds $-0.086<\dot{r}_{min}<-0.056$ (dashed lines).  The transition into this regime is found to occur when $r_{min}^{\to V}=1.4\times10^{-2}$ and the exit from this regime is when $r_{min}^{\to V}=6.1\times 10^{-4}$.  Once these values have been calculated they are placed on the phase diagram, as shown in Figure~\ref{F:highOh_transition} (as circles), to define the V-regime.  

\section*{Acknowledgements}

The authors thank Jens Eggers for his feedback on the manuscript and for providing them with the similarity solutions used in Figure~\ref{F:fs_vs_similarity} as well as the Referees for their constructive comments.  They are grateful to the John Fell Oxford University Press Research Fund and the EPSRC (grant EP/N016602/1) for supporting this research.

\bibliographystyle{jfm}
\bibliography{Bibliography}% Produces the bibliography via BibTeX.

\end{document}